\documentclass[11pt]{article}
\pdfoutput=1

\usepackage[no-natbib-sort]{jheppub}
\usepackage[T1]{fontenc} 

\usepackage{amssymb}
\usepackage{amsfonts}
\usepackage{amsbsy}
\usepackage{amsmath}
\usepackage{amsthm}
\usepackage{graphicx}
\usepackage{epstopdf}
\usepackage[vcentermath]{youngtab}
\usepackage{multirow}
\usepackage{latexsym} 
\usepackage{array}
\usepackage{color}
\usepackage{verbatim}
\usepackage{comment}
\usepackage[utf8]{inputenc}

\input cyracc.def %sha
\newfont{\twelvecyr}{wncyr10 at 12pt}
\def\Z{\mathbb{Z}}

\def\C{\mathbb{C}}

\def\P{\mathbb{P}}

\def\n3a{t}

\newcommand{\SU}[0]{\mathrm{SU}}
\newcommand{\SO}[0]{\mathrm{SO}}
\newcommand{\U}[0]{\mathrm{U}}

\newcommand*{\cy}{CY }
    \newcommand*{\gsm}{G_{\rm SM}}

\newcommand{\wati}[1]{\footnote{\textcolor{blue}{\textbf{WT:\ #1}}}}
\newcommand{\kobe}[1]{\footnote{\textcolor{red}{\textbf{KL:\ #1}}}}

\newcommand{\clean}{
\renewcommand{\kobe}[1]{}
\renewcommand{\wati}[1]{}
}
\clean

\title{Towards natural and realistic $E_7$ GUTs in F-theory}

\author{Shing Yan Li}
\author{and Washington Taylor}

\affiliation{Center for Theoretical Physics\\
Department of Physics\\
Massachusetts Institute of Technology\\
77 Massachusetts Avenue\\
Cambridge, MA 02139, USA}

\emailAdd{sykobeli at mit.edu}
\emailAdd{wati at mit.edu}

\preprint{MIT-CTP/5566}
\abstract{
We consider phenomenological aspects of a natural class of
Standard Model-like supersymmetric F-theory vacua realized through
flux breaking of rigid $E_7$ gauge factors.  
Three
generations of Standard Model matter
are realized in many of these vacua. 
We further
find that many other Standard Model-like features
are naturally %can easily be
 compatible
with these constructions. For example, dimension-4 and 5 terms associated
with proton decay are ubiquitously suppressed. Many of these features are
due to the %peculiar group theoretical and F-theory geometrical
group theoretical structure
of $E_7$ and associated F-theory geometry. In particular, a set of approximate global symmetries descends
from the $E_7$ group,
leading to exponential suppression of undesired
couplings.
}
  
\begin{document}
\maketitle
\flushbottom

%--------------------------------
\section{Introduction}
\label{sec:Intro}

String theory provides a consistent framework for a unified theory
that combines gravity with the other fundamental forces described by
quantum field theory.  To describe the real world, however,
ten-dimensional string theory must be compactified on a real
six-dimensional manifold, and various further objects like branes,
fluxes, and orientifolds must be incorporated.  Such constructions give
an enormous number (perhaps on the order of something like
$10^{272000}$ \cite{TaylorWangVacua}) of string theory vacua, known as
the string landscape. As part of the program to realize our Universe
in string theory, it has been a long-standing and primary goal
to find the  structure of the Standard Model (SM) of particle
physics within the string landscape.
%Despite decades of work, 
While many (supersymmetric) string vacua have been identified that share
 many of the principal features of the Standard Model,
there is as yet no single vacuum known in the string landscape that reproduces
all the observed phenomenological details of our world;
% can be 
%realized within
%the string landscape
 for recent reviews of work in this direction, see
\cite{Cvetic:2022fnv,Marchesano:2022qbx}.  

Beyond the simple question of the
  existence of a vacuum matching observed physics, it is
perhaps even more important to understand the extent to which the
physical features of the
Standard Model arise  \emph{naturally} in string
theory.  In other words, we would like to understand the extent to
which solutions like the Standard Model are widespread in the string
landscape or require extensive fine-tuning.  
This is a principal focus of this work and the associated research
program: we take a top-down perspective on the global set of string
vacua and attempt to identify realizations of the Standard Model that
are compatible with the most typical structures arising in string
theory.  %It is natural to 
We
use
F-theory \cite{VafaF-theory,MorrisonVafaI,MorrisonVafaII} to study
these
 questions, as this approach gives a global and nonperturbative
picture of the largest currently understood set of string vacua.
For reviews of F-theory and applications to Standard Model
constructions, see \cite{WeigandTASI,HeckmanReview}.
This paper describes some more detailed phenomenological aspects of
SM constructions
 originally presented in \cite{Li:2021eyn,Li:2022aek}
that are realized through flux breaking of
rigid $E_6$ and $E_7$ gauge factors,
which are relatively common features in F-theory geometries.

Constructing the detailed
Standard Model requires many elements such as the gauge group, the
matter content including both chiral matter and the Higgs, the Yukawa
couplings, a supersymmetry (SUSY)-breaking mechanism, values of the 19
free parameters, and possibly some room to address beyond-SM problems
as well as cosmological aspects such as the density of dark energy. 
Unfortunately, the current available string theory techniques are far
from enough to compute all these features precisely. Among the above SM 
features, string theory techniques for constructing the
gauge group and the chiral spectrum are well-established. While there is
some recent progress on the Higgs sector \cite{Bies:2021nje,Bies:2021xfh,Bies:2022wvj,Bies:2023jqg, Bies:2023sfm} and the Yukawa
couplings \cite{Cvetic:2019sgs} in a large class of F-theory models,
so far no fully precise statement on the realization of these features
in a way that matches observed physics
has been made in this context.
On the other hand, incorporating these established features
with e.g. SUSY breaking is far beyond our current techniques.  
Although at 
this moment no complete  realization of the Standard Model
has been constructed in any version of string theory, if we can identify a 
natural class of models that realize a decent portion of
the coarsest features of the SM, 
these structures may
naturally correlate with certain other features of the SM or beyond SM
physics. We will explore this philosophy in this paper.

One obvious way in which the
models studied here (and elsewhere in much of the string theory literature)
differ from observed physics is that we focus on solutions with
supersymmetry.  Supersymmetry has not yet been observed at low (TeV or
below) energies in nature, but as a theoretical tool it increases our
level of analytic control.  By studying solutions with supersymmetry,
we can gain some perspective on global aspects of the string
landscape.  Of course, eventually we need to understand
non-supersymmetric solutions to match observed physics.  One
possibility is that the physics we see is in a broken-symmetry phase
of a theory with supersymmetry at energies beyond the TeV scale.
Even if supersymmetry is broken at the Planck or string scale,
many insights gained by exploring the space of supersymmetric vacua may
be relevant to the less controlled
non-supersymmetric vacua.

\subsection{Natural vs. tuned  features}
\label{subsec:natural}

Before describing our results, it is worth clarifying the concept of
\emph{naturalness} used in this paper.
To obtain vacua with all the SM features considered in this paper, 
quite a few specific choices must be made in the
construction of vacua.  A
list of such choices in the context of the models studied in this
paper
 is summarized in \S\ref{subsec:overview}, and the mathematical
 conditions imposed for such choices are given at the beginning of
\S\ref{sec:ExplicitConstruction}. 
%Many of the choices are natural in
%different extents, 
The extent to which these different kinds of choices are natural varies,
within a hierarchy of naturalness/tuning.
Roughly speaking, each of the choices made in constructing a specific
class of string vacua can be characterized as
belonging to one of the following categories:

\begin{enumerate}
\item Physical constraints: These constraints come from string theory itself
and must be satisfied in all string compactifications. These constraints
ensure physically sensible vacua that have, e.g., Poincar\'{e} (or AdS)
invariance. Examples include tadpole cancellation and primitivity of fluxes.

\item Ubiquitous/common conditions: Let us consider a reasonably large
but presumably finite set of 
string vacua or compactification geometries, such as ${\cal N} = 1$ 4D
F-theory vacua or the associated set of topologically distinct
 complex
threefold bases that support elliptic Calabi-Yau fourfolds.
A condition is common if it holds for
an ${\cal O} (1)$ fraction
of the set of vacua or geometries, considered with a simple counting
measure. In particular, the condition is ubiquitous if it holds for a
substantial majority. As examples, the existence of rigid
$E_8$ gauge factors in (known) F-theory base geometries is ubiquitous, and
that of $E_7$ gauge factors appears to be common. (See, e.g.,
\cite{HalversonLongSungAlg,TaylorWangLandscape,twy} and discussions below)

\item Fairly likely conditions: Sometimes, there are a family of
  similar conditions, such as possible values of a discrete
  parameter. Each possible value may
only hold for a relatively small fraction of vacua
  within the above set, so that none of the conditions are
  ubiquitous. We refer to a condition as being fairly likely if the
 fraction of vacua or geometries with this property
 is considerably higher than for most of the
  other possibilities. 
As an illustration, we would say that rolling a sum of 6 on a pair of
six-sided dice is ``fairly likely,'' although rolling a sum of 7 is slightly
more likely.
%In other words, 
As another example, 
%preferred conditions are represented by
  points near the peaks in a distribution of some discrete parameters
  correspond to fairly likely conditions.  See
  \cite{Andriolo:2019gcb} for more discussions along these lines. As
%  an
a further example, \cite{Li:2021eyn,Li:2022aek} argued that three
  generations of chiral matter is fairly likely in this sense in
  our $E_7$ model (although, for example, zero generations may be  more
  likely). (See also, e.g., \cite{BraunWatariGenerations}.)

\item Natural choices: These are  choices for discrete parameters having
many possible values that are not (obviously)
preferred in any way, but 
imposing a particular chosen value does not require 
exponential amounts of tuning. Such choices may
hold at the level of, e.g., 0.1\% of the given set of vacua.
%These choices are
%necessary conditions
Such choices may be needed for obtaining some \emph{qualitative}
features of observed phenomenology in some constructions.  For
example, obtaining the SM gauge group and matter representations from
flux breaking of $E_7$ involves some choices of fluxes given by mild
linear constraints, which seem to be natural in this sense, although
they do not seem to be preferred in any particular way over other
choices that would give a variety of other possible groups and
representations.

\item Fine-tuning: These are choices involving setting one or more
continuous variables to take specific values, 
or making an exponentially rare choice among discrete possibilities.
Vacua based on such choices
 are increasingly non-generic 
in the landscape as the number of such tunings increases. In some 
constructions of string vacua, such
choices are needed to obtain  certain \emph{qualitative} features of
observed phenomenology.
For example, a tuned $\SU(5)$ or SM
gauge group in F-theory involves extensively fine-tuning many moduli to
specific values
\cite{BraunWatariGenerations} (unless these moduli are somehow
automatically tuned by a specific class of flux choices). 
Notably, it seems that
no such fine-tuning is involved in our $E_7$
models.

\item Technical choices: To facilitate analytic control of the vacua
  and make some particular calculations manageable, in some cases
  technical choices are made by restricting attention to some specific
  relatively simple choices of vacua. These choices are not necessary
  either for physical or phenomenological consistency, but are made to
  illustrate specific examples as simply as possible. The features of
  the models chosen in this way should be representative of some
  larger class of vacua or geometries.  
In some situations, technical choices can be made just to simplify
calculations that are in principle possible and expected to give
qualitatively similar results for all other choices.  In other cases,
technical choices are made where it is not clear how to do the
computation explicitly in general, and/or whether a completely general
choice will give qualitatively similar results. If not, some choices or
tuning of one of the above types may be
implicitly involved.  For the specific technical choices made here,
we have some confidence that
  similar results should also hold for a broader  class of vacua
  without those technical choice. Nevertheless, some qualitative
  simplifications occur based on these choices, thus
more
  explicit further studies are required to understand the extent to which these technical choices
  are relevant for phenomenologically interesting features. Examples of 
technical
  choices include picking some certain topological types for the
  compactification, which we do in this paper 
(specifically by choosing  models where the gauge divisor is a del
  Pezzo surface and the matter curve is a $\P^1$)
%for explicit  computations in some examples.
to facilitate and simplify the analysis; these technical choices made
here fall in the latter category above
that may implicitly involve some more or less natural choice or
tuning,
 as they may affect qualitative
aspects of the low-energy physics.
\end{enumerate}

While the term ``natural'' is used widely in many different ways in
the literature, we attempt to use the above classification to be slightly
more precise about the types of choices involved in the construction of
our models and the realization of phenomenological features.  This is a  coarse characterization, however, as choices and
tunings can occur across a broad spectrum, and
we do not attempt to make any precise
division between the gradations of ``common,'' ``fairly likely,'' and
``natural'' conditions.  In particular,
we do not have a perfect understanding of the class of string
geometries or F-theory compactifications, so any attempt at classification of
this type is necessarily quite imperfect given the current state of knowledge.
Moreover, the
measure problem on the landscape is not at all understood, so we
really do not have any good sense of the proper probability measure to
use on the landscape.  Nonetheless, in the absence of any known or
conjectured dynamical mechanism that would modify these considerations,
features that seem to require exponentially large amounts of fine-tuning under a
simple counting measure seem likely to occur less frequently in a
large string multiverse than  features that are ubiquitous,
fairly likely, or even natural in the preceding terminology.  In principle, even without solving the
measure problem, this may give us some insight into the extent to
which the Standard Model may be realized naturally in string theory,
and what BSM physics may be most naturally associated with those SM structures.

\subsection{Review of previous work}

In recent years, F-theory
 has become a particularly
promising framework for studying many aspects of
 string compactifications and
phenomenology, as it provides a global description of a large
connected class of supersymmetric string vacua. (See
\cite{WeigandTASI} for a review.) In particular, F-theory gives 4D
$\mathcal N=1$ supergravity models when compactified on 
elliptically fibered Calabi-Yau (CY) fourfolds $Y$, corresponding to
non-perturbative compactifications of type IIB string theory on general
(non-Ricci flat) complex K\"ahler threefold base manifolds $B$.  The
number of such threefold geometries $B$  seems to already be on the order
of $10^{3000}$ for toric bases $B$
\cite{TaylorWangMC,HalversonLongSungAlg,TaylorWangLandscape}, without
even considering the exponential multiplicity of fluxes possible for
each geometry, although the number of flop equivalence classes of bases is
somewhat smaller
%, on the order of $10^{50}$
\cite{twy}.  F-theory is also known to be dual to many other types
of string compactifications such as heterotic models.  Briefly,
F-theory is a strongly coupled version of type IIB string theory with
non-perturbative configurations of 7-branes balancing the curvature of
the compactification space. The non-perturbative brane physics is
encoded geometrically into the elliptically fibered manifold, which
can be analyzed using powerful tools from algebraic geometry. The
gauge groups and 
chiral
matter content supported on these branes can then be
easily determined when combined with flux data.

Applying the above
techniques, many SM-like constructions of 4D F-theory models with the gauge group $\gsm=\SU(3)\times \SU(2)\times \U(1)/\mathbb
Z_6$ have been
achieved in the literature. The early literature, starting from 
\cite{Donagi:2008ca,BeasleyHeckmanVafaI,BeasleyHeckmanVafaII,DonagiWijnholtGUTs}, focused on the breaking of GUT groups of $\SU(5)$ and its $\U(1)$
extensions
\cite{Blumenhagen:2009yv,Marsano:2009wr,Grimm:2009yu,KRAUSE20121,Braun:2013nqa}, 
while there has also been some study of $\SO(10)$ 
\cite{Chen:2010ts} and $E_6$ 
\cite{Chen:2010tg,Callaghan:2012rv,Callaghan:2013kaa} GUTs. (See \cite{WeigandTASI,HeckmanReview} for more extensive reviews.) These constructions break the GUT group using the
so-called hypercharge flux further discussed in
\cite{Mayrhofer:2013ara,Braun:2014pva}, which is a kind of ``remainder''
flux \cite{Buican:2006sn,Braun:2014xka} to be reviewed below. Some later constructions 
tried to construct $\gsm$ directly without any symmetry breaking, with the recent culmination of finding $10^{15}$ explicit solutions of
directly tuned $\gsm$ with three generations of SM chiral
matter (a ``quadrillion Standard Models'' \cite{CveticEtAlQuadrillion}), 
based on the ``$F_{11}$'' fiber in \cite{KleversEtAlToric}. These 
constructions are further generalized in \cite{Raghuram:2019efb,Jefferson:2022yya}. Although these models nicely
capture some of the most important phenomenological features, they face
one common issue: 
In terms of the notions discussed in \S\ref{subsec:natural}
the gauge groups in these constructions are
highly ``fine-tuned'', namely they are
obtained by setting specific values for many complex structure
moduli. 
Furthermore, on
most F-theory bases such tuning of $\gsm$ is forbidden due to the presence
of rigid gauge groups (to be discussed shortly). Even if the tuning is
available, it may not be compatible with moduli stabilization by fluxes and/or nonperturbative effects.

A more natural class of SM-like constructions in F-theory comes from
rigid gauge groups such as $E_7,E_6$ \cite{MorrisonTaylorClusters,MorrisonTaylor4DClusters}. These are gauge groups enforced by
strong curvature (to be more precise, very negative normal bundle) on the
base, and are present throughout the whole branch of moduli space
over that base, hence 
avoid the issue of tuning moduli. Moreover, statistical studies on (toric)
F-theory bases have suggested that these rigid gauge groups are
%ubiquitous
fairly common
in the landscape.
While the specific base naively associated with the most flux vacua
\cite{TaylorWangVacua}
does not contain $E_7$ or $E_6$ factors,
 these gauge factors arise in a substantial fraction of F-theory base
 geometries enumerated by a simple counting measure (which may or may
 not distinguish bases related by a flop).
The fraction of toric bases for 6D F-theory models that contain
rigid $E_7$ and $E_6$ factors is more than 50\% \cite{MorrisonTaylorToric}.
The statistics of $E_7$ and $E_6$ factors in threefold bases for 4D
F-theory models is less well understood; one study found
$E_7$ factors in $\sim 20$\% of a limited simple of bases
\cite{TaylorWangMC}, and a more detailed analysis of the prevalence of
such factors is currently underway \cite{twy}.
Nonetheless, breaking these gauge groups to $\gsm$ should
give us a very large set of SM-like constructions. Recently in \cite{Li:2021eyn,Li:2022aek}, we have proposed a general class of
SM-like models using rigid $E_7,E_6$ GUT groups in
F-theory, with an intermediate $\SU(5)$ group. These models 
enjoy the advantages of being natural
and involving little or no fine-tuning.
  Specifically, a combination of ``vertical''
and ``remainder'' fluxes can be used to break the rigid gauge groups
in a way that is not
transparent in the low-energy field theory, but
gives the correct SM gauge group and some chiral matter.
Although in many cases the breaking leads to exotic chiral
matter, there are large families of models in which the
correct SM chiral matter representations are obtained through
an intermediate $\SU(5)$.  The number of generations
can easily be small and we have demonstrated that three
generations are fairly likely in many of these models. In particular,
a fully global explicit construction of such an $E_6$ model has been given
in \cite{Li:2022aek}.

\subsection{Overview of results}
\label{subsec:overview}

As discussed before, 
%we may hope to naturally find more SM features in this
%large set of SM-like constructions. 
one might hope that
there are string realizations of the Standard Model in which most or
all of the features observed in nature arise in a relatively natural
way.
In this paper, we show that apart from
the SM gauge group and chiral spectrum, several
additional
 important SM features can be easily obtained in the $E_7$ (but not 
$E_6$) models, with some additional but mild tuning on the geometry and 
flux background. Specifically, we obtain the following features, 
%by making 
%the following choices,
each of which depends upon making choices with
 various extents of naturalness:
\begin{itemize}
    \item As mentioned above, rigid $E_7$ factors are 
%ubiquitous
quite common in the F-theory landscape, and may be natural or likely
depending on the proper vacuum measure. 
For generic (non-toric) bases,
    the $E_7$ gauge group can be broken down to $\gsm$ by some natural choices
    of vertical and remainder fluxes.
\item 
For the models with flux breaking of $E_7 \rightarrow\gsm$, a set of
approximate global $\U(1)$ symmetries descend from the $E_7$, leading to
exponential suppression of certain couplings.
    \item With the $E_7 \rightarrow\gsm$ gauge-breaking fluxes, it 
appears to be fairly likely to have 3 generations of SM chiral
matter (although 0, 1, or 2 generations may be more likely), and
fairly likely that the
    exotic $(\mathbf 3,\mathbf 2)_{-5/6}$ representation is removed
from the spectrum.
    \item Due to the use of $E_7$, there are always candidate Higgs 
    sectors with a string
    theory origin different from that of chiral matter. Such a structure
    automatically leads to distinct dynamics between the Higgs and chiral
    matter, and gives rise to unsuppressed SM Yukawa couplings.
    \item Under this setup, dimension-4 and 5 proton decay is
      ubiquitously suppressed to phenomenologically safe levels.
\end{itemize}

The distinction between the Higgs and chiral matter, the appearance of the approximate global $\U(1)$ symmetries, and the
ubiquitous suppression of dimension-4 and 5 proton decay are the
strongest features of these constructions, in which desirable
properties associated with observed physics arise essentially
automatically.  Most of
the remaining features we explore generally require
small amounts of discrete tuning. They may involve common, fairly
likely or natural choices and do not arise automatically,
but do not seem to require extensive fine-tuning.

\begin{itemize}
    \item There is some automatic splitting between the doublet and triplet masses, although the amount of splitting and the exact masses are unknown.
    \item 
%Despite the presence of additional fields, 
%we can realize the correct set of Yukawa couplings with some further
%natural choices of fluxes. 
Although there are extra charged vector-like exotics in the spectrum, the Yukawa couplings
between most of these fields (all besides the triplet Higgs) and the
SM matter are exponentially suppressed through the
above-mentioned approximate symmetries. We call these fields \emph{inert}
vector-like exotics.

\item It is plausible that there
    is some hierarchy in the SM
Yukawa couplings, but the exact values are unknown.
    \item With the setup so far, the model contains three right-handed
    neutrinos with masses lower than the string/GUT scale. It is plausible but not fully
    clear that the seesaw mechanism occurs.
    %\item Gauge coupling unification is achieved by assuming the presence of several inert
    %vector-like exotics at TeV scales, which are well within the current
    %observational bounds but may be observable in the near
    %future.
\end{itemize}

 To facilitate the discussions and calculations in this paper, we technically choose the gauge divisor to be a del Pezzo surface, and the
    matter curve to be a $\mathbb P^1$.  Although we expect similar
    results for many other choices, these choices do lead to some
    qualitative simplifications in the analysis, and further work is
    needed to determine whether  low-energy models
with similar structure arise for a broader
    class of gauge divisors and matter curves, and/or to determine how
    natural or fine-tuned these geometric choices may be.

As an 
example, we  work out an explicit global construction of the $E_7$ models that 
%explicitly
 realizes all of the above SM features. We emphasize that many
of these phenomenological advantages are specific for the $E_7$ models,
and may be (much) harder to realize in other types of SM-like constructions
in F-theory. Some of the above features are inherited from the group 
structure of $E_7$ itself, regardless of the string theory physics. To the
authors' knowledge, however, these group theoretical features
have not been noticed in the field theory literature, probably because $E_7$
itself does not support any chiral matter, if there is no additional input
like fluxes from the UV.

While
 the $E_7$ models considered in this paper
 have quite a few phenomenological advantages
 over some other stringy realizations of the Standard Model, 
we
note
that these models potentially still suffer from the following
issues, in light of which extra care must be taken when interpreting  the
results presented here. First, these models contain many vector-like exotics that cannot be
removed by fluxes (except the most dangerous $(\mathbf 3,\mathbf
2)_{-5/6}$, which is fairly likely
to be absent).  In particular, these exotics
include other copies of the Higgs field. From the effective field theory
perspective, we generically expect these exotics to get heavy masses
near the GUT/string scale such that they do not affect the low-energy
phenomenology. On the other hand, this expectation in general may not
be true in string theory, and it is important to develop further
techniques to ensure the right masses.  Although it may be possible,
we do not see any reason in these models why one of the Higgs doublet
pairs
should get much
lower masses than the other copies. In other words, there is no
totally clear solution to the $\mu$-problem in our setup. Next, these
$E_7$ models have codimension-3 $(4,6)$ singularities on the base,
which correspond to an extra family of flux and may be associated to
an extra sector of strongly coupled superconformal and chiral matter
\cite{Candelas:2000nc,Lawrie:2012gg,Achmed-Zade:2018idx,Jefferson:2021bid,46}. We
can easily control the flux such that this sector is non-chiral, but
since we understand very little about this sector, further studies are
needed to ensure that this sector does not affect phenomenology.

\subsection{Outline of paper}

This paper is organized as follows. In Section \ref{sec:Fluxes}, we review fluxes in 
F-theory, which are the central tools in our construction of $E_7$ models.
We first discuss the notion of vertical and remainder fluxes, and the 
constraints satisfied by these fluxes. We then describe how these fluxes 
break a nonabelian gauge group (known as ``flux breaking''), determine the
chiral spectrum, and tell us something about the vector-like spectrum.
Despite the difficulty of computing the vector-like spectrum in general,
the vector-like spectrum can be completely determined in some special 
cases. We discuss how our $E_7$ models easily fit into these cases, so that
we can fully compute the matter spectrum in our models.

To initiate our discussions on semi-realistic $E_7$ GUTs in F-theory, we
first review our previous work on the $E_7$ models 
\cite{Li:2021eyn,Li:2022aek} in Section \ref{sec:E7Review}. We describe the geometry and 
fluxes needed to get the SM gauge group and three generations of SM chiral
matter from a rigid $E_7$.

In Section \ref{sec:Pheno}, we discuss various phenomenological aspects of the $E_7$ 
models, namely the vector-like matter, Yukawa couplings, proton decay, the Higgs sector, the neutrino sector, and
gauge coupling unification. One of our main tools is the St\"uckelberg mechanism \cite{DonagiWijnholtModelBuilding,Grimm:2010ks,Grimm:2011tb} used
in our flux breaking of $E_7$ to $\gsm$, which leaves several approximate
$\U(1)$ global symmetries. We discuss these symmetries in detail and 
study how they constrain the couplings and mass 
terms\footnote{Throughout the paper, ``mass terms'' refer
to the $\mu$-term and other similar terms in the superpotential, which involve two different fields.} in the low-energy 
theory. These constraints, plus some additional tuning on the fluxes, lead
to many of the above phenomenological advantages, especially the 
ubiquitous suppression of proton decay. We also discuss the vector-like 
matter that appears in the spectrum. We demonstrate how to easily remove the
most dangerous $(\mathbf 3,\mathbf 2)_{-5/6}$ vector-like exotic, and 
discuss why most other vector-like exotics are \emph{inert}. Although we 
cannot make
any precise statements, we discuss various possible origins of the 
vector-like (including the Higgs) masses. Based on such discussions, we
make some brief comments about the Higgs sector, the neutrino sector, and
gauge coupling unification.
%and suggest some estimates on
%the mass scales based on a rough calculation on gauge coupling unification.
%In this process, we will see that it is unavoidable to have several inert
%vector-like exotics at TeV scales, although we do not explain their 
%origins. These exotics include both the Higgs and the other SM 
%representations, but they do not violate current experimental bounds.

After describing the recipe of getting semi-realistic $E_7$ GUTs F-theory,
in Section \ref{sec:ExplicitConstruction} we write down an explicit global construction of a $E_7$ model
that achieves all the above phenomenological features. This example
demonstrates the fact that these features can indeed be obtained through 
some mild tuning on the geometry and the fluxes, but %not any
without the necessity of fine-tuning any
 moduli.
Therefore, it is reasonable to regard these features as being natural in
the string landscape. To emphasize various advantages and disadvantages of
the $E_7$ models, in Section \ref{sec:Comparison} we briefly compare our models with other
SM-like F-theory constructions in the literature. We finally conclude in Section
\ref{sec:Conclusion}. In Appendices \ref{appendix:Hypersurfaces} and \ref{appendix:Quantization}, we discuss several technical tools that are useful
in the construction in Section \ref{sec:ExplicitConstruction}.

\section{Fluxes in 4D F-theory models}
\label{sec:Fluxes}

In this section, we review vertical and remainder fluxes in 4D F-theory 
models, and how these fluxes determine the gauge group and matter 
spectrum. Except in \S\ref{subsec:Vectorlikematter} on vector-like matter, all the content in 
this section has been discussed in depth in \cite{Li:2022aek}. Here
we only recap the essential facts for our
construction of $E_7$ models and set up the notation. Interested readers
can refer to \cite{Li:2022aek} for more background information.

\subsection{Vertical and remainder fluxes}
\label{subsec:FluxesIntro}

To describe the flux backgrounds, we first need some basic geometric
facts about the compactifications. As mentioned in \S\ref{sec:Intro}, we consider F-theory compactified on a \cy fourfold
$Y$, which is an elliptic fibration on a threefold base
$B$. Nonabelian gauge groups arise when sufficiently high degrees of
singularities are developed in the elliptic fibers over divisors on $B$ (denoted by $D_\alpha$), called gauge
divisors $\Sigma$. When this happens, $Y$ itself is also singular and
we need to consider its resolution $\hat Y$ such that we can study
cohomology and intersection theory. Let the total gauge
group be $G$, where $G$ has no
$\U(1)$ factors before flux breaking. 
In the main part of this paper, we always study the simple case where 
$G=E_7$, but for generality we assume any simple Lie group $G$ in this section.
The nonabelian group
 $G$ is supported on a gauge divisor $\Sigma$, and
the resolution
results in exceptional divisors $D_{1\leq{i}\leq\mathrm{rank}(G)}$
in $\hat Y$.  For ADE (or simply-laced) groups, the intersection structure of these divisors matches the Dynkin
diagram of $G$, where each exceptional divisor corresponds to a
Dynkin node \cite{Kodaira,Neron}. By the Shioda-Tate-Wazir theorem
\cite{shioda1972,Wazir}, the divisors $D_I$ on $\hat Y$ are spanned by
the zero section $D_0$ of the elliptic fibration, pullbacks of base
divisors $\pi^* D_\alpha$ (which we also call $D_\alpha$ depending on
context), and the exceptional divisors $D_{i}$.\footnote{If $G$ has
  $\U(1)$ factors, there are also divisors associated with these
  factors coming from the Mordell-Weil group of rational sections with
  nonzero rank.} Although the choice of resolution is not unique, our
analysis and results are clearly resolution-independent
\cite{Jefferson:2021bid}.

Now we are ready to understand fluxes. These are
most easily understood by considering the dual
M-theory picture of the F-theory models, that is,
M-theory compactified on the resolved fourfold
$\hat Y$ (reviewed in \cite{WeigandTASI}). In the M-theory perspective, fluxes are
characterized by the three-form potential $C_3$ and
its field strength $G_4=dC_3$. The data of $G_4$
flux, which can be studied with well-established
tools, is sufficient for constructing our $E_7$ models.

In general, $G_4$ is a discrete flux that takes values in the fourth
cohomology $H^4(\hat Y,\mathbb R)$.
The quantization condition on $G_4$ is slightly subtle and is given by
\cite{Witten:1996md}
\begin{equation} \label{quantization}
    G_4+\frac{1}{2}c_2(\hat Y) \in H^4(\hat Y,\mathbb Z)\,,
\end{equation}
where $c_2(\hat Y)$ is the second Chern class of $\hat Y$. In
general (and particularly for $E_7$ models), $c_2(\hat Y)$ can be odd (i.e., non-even), in which case the discrete
quantization of $G_4$ contains a half-integer shift.
When we construct an $E_7$ model explicitly in \S\ref{sec:ExplicitConstruction}, we will make use of
an odd $c_2(\hat Y)$ and half-integer fluxes. More details will be discussed in
that section.

Next, to preserve the minimal amount of SUSY in 4D, $G_4$
must live in the middle cohomology i.e. $G_4\in H^{2,2}(\hat Y,\mathbb
R)\cap H^4(\hat Y,\mathbb Z/2)$. 
Supersymmetry also imposes the condition of primitivity \cite{Becker:1996gj,Gukov:1999ya}:
\begin{equation}
\label{eq:primitivity}
    J\wedge G_4=0\,,
\end{equation}
where $J$ is the K\"ahler form of $\hat Y$. This is
automatically satisfied when the geometric gauge group is not broken, but
%not obviously satisfied
 when the gauge group is broken by \emph{vertical} flux (to be
 discussed below),
%. The interpretation of
 this condition
% is that it
 stabilizes some (but not all)
K\"ahler moduli; stabilizing these moduli within the K\"ahler cone
imposes additional flux constraints. 

We also have the D3-tadpole condition \cite{Sethi:1996es} that must be
satisfied for a consistent vacuum solution:
\begin{equation} \label{tadpole}
    \frac{\chi(\hat Y)}{24}-\frac{1}{2}\int_{\hat Y} G_4\wedge G_4=N_{D3}\in \mathbb Z_{\geq 0}\,,
\end{equation}
where $\chi(\hat Y)$ is the Euler characteristic of $\hat Y$,
and $N_{D3}$ is the number of D3-branes, or M2-branes in the
dual M-theory. To preserve SUSY and stability, we
forbid the presence of anti-D3-branes i.e. $N_{D3}\geq 0$. The
integrality of $N_{D3}$ is guaranteed by
Eq.\ (\ref{quantization}). This condition has an immediate
consequence on the sizes of fluxes. Since in general
$h^{2,2}>2\chi/3\gg \chi/24$, if we randomly turn on
flux in the whole middle cohomology such that the tadpole constraint
is satisfied, a generic flux
configuration vanishes or has small magnitude in most of the
$h^{2,2}$ independent directions. In this sense, the tadpole contributed
by fluxes along some particular directions can be treated as a rough
estimate on the amount of fine-tuning on fluxes. We leave a more
precise and detailed analysis of these considerations to future work.

We now consider
the orthogonal decomposition of the middle cohomology
\cite{Braun:2014xka}:
\begin{equation}
    H^{4}(\hat Y,\mathbb C)=H^{4}_\mathrm{hor}(\hat Y,\mathbb C)\oplus H^{2,2}_\mathrm{vert}(\hat Y,\mathbb C)\oplus H^{2,2}_\mathrm{rem}(\hat Y,\mathbb C)\,.
\end{equation}
The horizontal subspace comes from the complex structure
variation of the holomorphic 4-form $\Omega$,
% associated with
%the CY fourfold. 
%Flux in these directions has the effect of inducing a superpotential and
%stabilizing complex structure moduli \cite{Gukov:1999ya}. 
while 
the vertical subspace is spanned by products of harmonic
$(1,1)$-forms (which are Poincar\'e dual to divisors, denoted
by $[D_I]$)
\begin{equation}
    H^{2,2}_\mathrm{vert}(\hat Y,\mathbb C)=\mathrm{span}\left(
    H^{1,1}(\hat Y,\mathbb C)\wedge H^{1,1}(\hat Y,\mathbb C)\right)\,.
\label{eq:vertical-c}
\end{equation}
%Finally, there may be
 Components that do not belong to the
horizontal or vertical subspaces
%; these 
are referred to as
remainder flux.  
%While there are various types of remainder
%flux, we will need the following type in
%the analysis below. 
In the construction here we use a specific type of remainder flux.
Consider a
curve $C_{\mathrm{rem}}\in H_{1,1}(\Sigma,\mathbb Z)$  in
$\Sigma$, such that its pushforward $\iota_*C_\mathrm{rem}\in H_{1,1}(B,\mathbb Z)$ is trivial, where
$\iota:\Sigma\rightarrow B$ is the inclusion map.
While such a
curve  cannot be realized on toric divisors on toric bases, it has
been suggested that such curves do exist on ``typical'' bases
\cite{Braun:2014xka}, so that toric geometry may be insufficiently
generic for this class of constructions; understanding this question
of typicality is an important problem for further study.
In any case, we
now restrict each $D_i$ 
(considered as a fibration over $\Sigma$)
onto $C_\mathrm{rem}$, giving a surface in $\hat{Y}$. Its Poincar\'e
 dual
(in $\hat{Y}$)
 $[D_i|_{C_\mathrm{rem}}]$ is a $(2,2)$-form, but since $C_\mathrm{rem}$ cannot be obtained by intersections of base divisors, we must have
\begin{equation}
    \left[D_i|_{C_\mathrm{rem}}\right]\in H^{2,2}_\mathrm{rem}(\hat Y,\mathbb C)\,.
\end{equation}

Here we explain more about vertical flux (denoted by $G_4^\mathrm{vert}$), as there are more
constraints specifically on vertical flux such that $G_4$ dualizes to a consistent F-theory
background that preserves Poincar\'{e} invariance. Combining Eqs. (\ref{eq:vertical-c}) and (\ref{quantization}) gives the
integral vertical subspace $H^{2,2}_\mathrm{vert}(\hat Y,\mathbb
R)\cap H^4(\hat Y,\mathbb Z)$.\footnote{We remind readers that the
quantity $G_4^\mathrm{vert}+c_2(\hat Y)/2$ instead of $G_4^\mathrm{vert}$ lives in this subspace.
Note that $c_2(\hat Y)$ is always vertical.}
We focus primarily here on the vertical subspace spanned by integer
multiples of forms $[D_{I}] \wedge[D_J]$
\begin{equation}
    H^{2,2}_\mathrm{vert}(\hat Y,\mathbb Z):=\mathrm{span}_{\mathbb Z}\left(
    H^{1,1}(\hat Y,\mathbb Z)\wedge H^{1,1}(\hat Y,\mathbb
    Z)\right)\,.
\label{eq:vertical-z}
\end{equation}
While this subspace does not necessarily include all lattice points in
the full vertical cohomology $H^{2, 2}_\mathrm{vert} (\hat{Y},\C)\cap
H^4 (\hat{Y},\Z)$ of the same dimension, 
this subspace is sufficient for us to construct the $E_7$ models, and we
leave the analysis of the full vertical subspace to future work.

Now we set up some notations for vertical flux. We expand
\begin{equation}
    G_4^\mathrm{vert}=\phi_{IJ} [D_I]\wedge [D_J]\,,
\label{eq:g-phi}
\end{equation}
and work with integer (or half-integer if $c_2$ is
odd) flux parameters $\phi_{IJ}$. 
%Note that the
%expansion depends on the choice of basis of base divisors,
%which we will specify depending on context. 
We denote the
integrated flux as \cite{Grimm:2011fx}
\begin{equation}
    \Theta_{\Lambda\Gamma}=\int_{\hat Y} G_4\wedge [\Lambda]\wedge
          [\Gamma]\,,
\label{eq:integrated-flux}
\end{equation}
where $\Lambda,\Gamma$ are arbitrary linear combinations of $D_I$; subscripts $0, i, \alpha, \ldots$ refer to the basis divisors $D_0, D_i, D_\alpha, \ldots$. In this paper, we use the
following
resolution-independent
 formula to relate
$\Theta_{i \alpha}$ to $\phi_{i \alpha}$
\cite{Grimm:2011sk,Jefferson:2021bid}:\footnote{Indices appearing twice are
summed over, while other
  summations are indicated explicitly.}
\begin{equation} \label{eq:phitoTheta}
    \Theta_{i \alpha}=-\kappa^{i j} \Sigma\cdot D_\alpha\cdot D_\beta \phi_{j\beta}\,,
\end{equation}
where $\kappa^{i j}$ is the inverse Killing
metric of $G$, and ``dots'' denote the
intersection product.
%\footnote{We will not mention
%explicitly the space where
%the products are taken in such formulae, as the space ($\hat{Y}$ or
%$B$, the latter in this case)
%is already clear from context.}
%This is the same as
%the Cartan matrix $C^{ij}$ of $G$ for ADE groups.
In the $E_7$ models, the only vertical flux parameters we turn on have
indices of type $\phi_{i\alpha}$; although it is also possible to turn on
nontrivial $\phi_{ij}$, we always turn them off for reasons to be discussed below.

Now we write down the extra constraints for vertical flux.
To preserve Poincar\'e symmetry after dualizing, we require
\cite{Dasgupta:1999ss}
\begin{equation} \label{PoincareSym}
    \Theta_{0\alpha}=\Theta_{\alpha\beta}=0\,.
\end{equation}
%(Recall that Greek  indices $\alpha, \ldots$ correspond to divisors that are
%pullbacks from the base, while Roman indices $i$ correspond to Cartan
%divisors, and the index 0 refers to the global zero section of the elliptic
%fibration.)
%Next, 
If the whole geometric gauge symmetry is preserved, a necessary condition is that
\begin{equation} \label{gaugeSym}
    \Theta_{i\alpha}=0\,,
\end{equation}
for all $i,\alpha$, 
otherwise flux breaking occurs. This condition is not
sufficient when there is nontrivial remainder flux, which
will be discussed more in \S\ref{subsec:FluxBreaking}.
When flux breaking occurs, i.e. Eq.\ (\ref{gaugeSym}) is violated, the 
condition (\ref{PoincareSym}) for Poincar\'{e} symmetry
is unchanged, but there are extra constraints from primitivity, which
are demonstrated in later sections.

Much of the above discussion on vertical flux extends naturally to the
type of remainder flux $G_4^\mathrm{rem}$ we need. Similarly, we expand
\begin{equation}
    G_4^\mathrm{rem}=\phi_{ir} \left[D_i|_{C_\mathrm{rem}}\right]\,,
\end{equation}
and work with integer $\phi_{ir}$. In this paper, we always turn on 
remainder flux with a single $C_\mathrm{rem}$ only, so we do not specify
the choice of $C_\mathrm{rem}$ in the flux parameters; instead we just 
label them by ``$r$''. Eq.\ (\ref{eq:phitoTheta}) straightforwardly generalizes
to remainder flux by replacing the triple intersection on the base with
the intersection of two $C_\mathrm{rem}$'s on $\Sigma$.

\subsection{Flux breaking}
\label{subsec:FluxBreaking}

With the knowledge of vertical and remainder fluxes, we now
describe the breaking of geometric gauge groups
with these fluxes, a.k.a. flux breaking. This
kind of breaking has been used as early as \cite{BeasleyHeckmanVafaI} (see also \cite{WeigandTASI}), and
is recently developed in depth in \cite{Li:2022aek}. In this
paper, we only list the results essential for our
analysis on the $E_7$ models, and we refer readers to
\cite{Li:2022aek} for full technical details.

Let us first study vertical flux. Recall that we need $\Theta_{i \alpha}=0$ for all
$i,\alpha$ to preserve the whole $G$. Now we
break $G$ into a smaller group $G'$ by turning on
some nonzero $\phi_{i\alpha}$. Such flux breaks
some of the roots in $G$. It also induces masses
for some Cartan gauge bosons by the St\"uckelberg
mechanism \cite{Grimm:2010ks,Grimm:2011tb}, hence breaks some combinations of Cartan
$\U(1)$'s in $G$. 
Let $\alpha_{i}$ be the simple
roots of $G$, and $T_{i}$ be the Cartan generators
associated with $\alpha_{i}$ i.e. in the co-root
basis. The root $b_{i} \alpha_{i}$ is preserved under the breaking if
\begin{equation} \label{eq:fluxbreakingcondition}
    \sum_{i} b_{i}\left<\alpha_{i},\alpha_{i}\right>\Theta_{i \alpha}=0\,, 
\end{equation}
for all $\alpha$. Here $\left<.,\,.\right>$ denotes
the inner product of root vectors. Moreover, the
corresponding linear combination of Cartan generators
\begin{equation}
    \sum_{i} b_{i}\left<\alpha_{i},\alpha_{i}\right>T_{i}\,, 
\end{equation}
is preserved. These generators form a nonabelian
gauge group $G'\subset G$ after breaking. Note that for ADE groups like
$E_7$, $\left<\alpha_{i},\alpha_{i}\right>$ is the same for all $i$, and
we will drop this factor in the above conditions. Below we will use
a simple version of the breaking: we turn on $\Theta_{i'\alpha}\neq 0$
for some set of Dynkin indices $i'\in I'$ and some $\alpha$, in a
generic way such that Eq.\ (\ref{eq:fluxbreakingcondition}) is satisfied only when $b_{i'}=0$ for
all $i' \in I'$. Then
$G'$ is given by removing the corresponding
nodes in the Dynkin diagram of $G$. The simple roots of $G'$ are
directly descended from $G$ and are given by $\alpha_{i\notin I'}$.
In the $E_7$ models, one can show that all flux breaking routes to $\gsm$
(or $\SU(5)$ before including remainder flux) are related to this simple
version of breaking by automorphisms.

There are additional constraints on vertical flux
breaking coming from primitivity, since Eq.\ 
(\ref{eq:primitivity}) is not automatically
satisfied when there is vertical flux breaking and
$\Theta_{i\alpha}\neq 0$ for some $i,\alpha$.
To understand these constraints, we consider the F-theory limit where
the fibers shrink to zero volume. Hence we can expand $J$ of $\hat Y$ as
$J\rightarrow\pi^* J_B=t^\alpha[D_\alpha]$, where the K\"ahler moduli 
$t^\alpha$ are restricted to the positive K\"ahler cone.
Now primitivity requires that
\begin{equation} \label{eq:intprimitivity}
    \int_{\hat Y}[D_{i}]\wedge J\wedge G_4=t^\alpha \Theta_{i\alpha}=0\,,
\end{equation}
which is true only for specific choices of $J$ when
there is vertical flux breaking. The condition of
primitivity then stabilizes some but not all K\"ahler
moduli in $J$; consistently stabilizing these moduli within the K\"ahler
cone imposes additional flux constraints.

As discussed in \cite{Li:2022aek}, 
the above flux constraints lead to an important necessary condition for
consistent vertical flux breaking:
Let $r$ be the number of
%$\alpha$'s
linearly independent $D_\alpha$'s appearing
 in the set of all homologically
independent surfaces in the form of $S_{i \alpha}=D_{i}\cdot
D_\alpha$
(for any $i$ of the given $G$). Now consider the ($r$  $\times$
$\mathrm{rank}(G)$)
matrix $\Theta_{(\alpha_{a})(i)}$ (where $a$ and $i$
are the indices for rows and columns respectively). 
The condition (\ref{eq:intprimitivity}) asserts that $t^\alpha
\Theta_{\alpha i} = 0$.
Since the solution
to primitivity thus requires a nontrivial left null space of the matrix,
the rank of the matrix is at most $r-1$. Moreover from Eq.
(\ref{eq:fluxbreakingcondition}), the rank of the
matrix is also the change in rank of $G$ during flux breaking. Therefore, we require
\begin{equation} \label{eq:rs}
    r\geq\mathrm{rank}(G)-\mathrm{rank}(G')+1\,.
\end{equation}
Note that when remainder flux breaking is not
available,
and all divisors in $\Sigma$ descend
from intersections in $B$, we have
$r=h^{1,1}(\Sigma)$. In the $E_7$ models, however, we require the
presence of remainder flux and $r$ is smaller than $h^{1,1}(\Sigma)$.
This condition means that we must have a sufficiently large $r$ in order
to get a desired $G'$, hence imposing constraints on
the possible geometries that support a given vertical flux breaking.

So far we have focused on the nonabelian part of the broken
gauge group, but there may also be $\U(1)$ factors in the broken gauge
group like $\gsm$. In the formalism of flux breaking, there are two ways
to get $\U(1)$ factors: for $\U(1)$ factors not along any roots, we can
use vertical flux to get these factors using the St\"uckelberg mechanism;
for a recent application see \cite{Li:2022vfj}. On the other hand,
$\U(1)$ factors like the hypercharge in $\gsm$ are along some roots of 
a higher gauge group ($\SU(5)$ in the case of hypercharge), and remainder
flux is necessary for obtaining these factors. Therefore, we proceed as
follows: If we turn on
\begin{equation}
    G_4^\mathrm{rem}=\phi_{ir}[D_i|_{C_\mathrm{rem}}]\,,
\end{equation}
for some $C_\mathrm{rem}$ satisfying the property mentioned
in \S\ref{subsec:FluxesIntro}, $G$ is broken into the commutant of
$T=\phi_{ir} T_i$ within $G$. The difference is that the remainder
flux does not turn on any $\Theta_{i\alpha}$, so there is no
St\"uckelberg mechanism and all the $\U(1)$ factors in 
the commutant are preserved. In other words, breaking using
remainder flux never decreases the rank of the gauge group,
while breaking using vertical flux always decreases the rank.
Note that
when $G$ is a rigid gauge group, $\Sigma$ is a rigid divisor
and supports remainder flux breaking only when embedded into
a non-toric base.
This follows because for a toric base $B$, toric divisors span the
cone of effective divisors, so any rigid effective divisor $\Sigma$ is
toric, and toric curves in a toric $\Sigma$ span $h^{1,1} (\Sigma)$.

\subsection{Chiral matter}
\label{subsec:ChiralMatter}

It is well known that vertical flux induces some chiral matter, and the
same is true for vertical flux that breaks the gauge group. Perhaps more
strikingly, even if the unbroken gauge group $G$ does not have any complex
representations and does not support chiral matter, there may still be some
chiral matter after \emph{vertical} flux breaking \cite{Li:2022aek}. The famous
index formula states that for a weight $\beta$ in
representation $R$, its chiral index $\chi_\beta$ is \cite{Braun_2012,Marsano_2011,KRAUSE20121,Grimm:2011fx}
\begin{equation} \label{ordinarychi}
    \chi_\beta=\int_{S(\beta)}G_4^\mathrm{vert}\,,
\end{equation}
where $S(\beta)$ is called the matter surface of $\beta$. When
$R$ is localized on a matter curve $C_R$, $S(\beta)$ is the
fibration of the blowup $\mathbb P^1$ corresponding to $\beta$
over $C_R$. Here we recall that a matter curve is a curve on $\Sigma$
where the fiber singularity is enhanced, resulting in additional fibral 
curves in the resolution, which corresponds to matter multiplets in the
4D theory.

Since weights differ by roots,
given a weight $\beta$ in $R$ of $G$, it is useful to
expand
$\beta=-b_i\alpha_i$. Hence we can decompose its matter surface
$S(\beta)$ as \cite{WeigandTASI}
\begin{equation} \label{Sdecompose}
    S(\beta)=S_0(R)+b_i\left.D_i\right|_{C_R}\,,
\end{equation}
where $S_0$ only depends on $R$ but not $\beta$. $S_0(R)$ corresponds to
the flux that gives chiral matter without breaking $G$, when $G$ supports
chiral matter. In contrast, when $G$ itself does not support chiral
matter as in the $E_7$ models, $S_0(R)$ is trivial in homology. From
now on we will ignore $S_0(R)$ and focus on the
second term of Eq.\ (\ref{Sdecompose}). Matter curves in general
can be written as $C_R=\Sigma\cdot D_R$ for some divisor $D_R$. Then,
\begin{equation} \label{partialchiR'}
    \int_{S(\beta)}G_4^\mathrm{vert}=b_{i}\Theta_{iD_R}\,.
\end{equation}
We can replace the $i$ summation with $i' \in I'$ since the other terms vanish. When $G$ is broken to $G'$, $R$ decomposes into
different irreducible representations $R'$ in $G'$, which can be labelled
by $b_{i'}$. In
general, different $b_{i'}$ and different $R$ can give rise to the same irreducible representation
$R'$. The total chiral index $\chi_{R'}$ for $R'$ is then
\begin{equation} \label{fullchiR'}
    \chi_{R'}=\sum_R \sum_{b_{i'}} b_{i'}\Theta_{i'D_R}\,.
\end{equation}
Note that this expression is nontrivial even when $R$ is not complex.

There are also adjoint chiral multiplets (apart from the vector multiplet
of the gauge fields) living on the bulk of $\Sigma$, and matter curves or surfaces for this representation are not well-defined.
Nevertheless, it has been shown that adjoint matter can also
become chiral after flux breaking, and the chiral indices are
given by setting $S_0(\mathrm{Adj})=0$ and replacing $C_R$ by
$K_\Sigma$ \cite{Bies:2017fam}. Here $K$ denotes the canonical class.
By the adjunction formula,
$K_\Sigma=\Sigma\cdot(K_B+\Sigma)$ and we should set
$D_R=K_B+\Sigma$.

\subsection{Vector-like matter}
\label{subsec:Vectorlikematter}

To fully understand the phenomenology of these models, it is important to study
the vector-like spectrum 
%apart from
in addition to the chiral spectrum. One of the main reasons 
for this is that in
a realistic model we need to realize the Higgs sector, while avoiding
dangerous vector-like exotics.  While the techniques for computing the
chiral spectrum are already at hand as above, computing the
vector-like spectrum in general requires not only the $G_{4}$ flux,
but the full information of $C_{3}$ in terms of line bundles on $\Sigma$
and $C$.\footnote{In general, these are described by sheaves when
  there are more severe singularities on $\Sigma$ and/or $C$. In this
  paper, we only consider completely smooth geometry on the bases, so
  the description by line bundles is sufficient.} In many cases, these
things are hard to compute and some of the relevant technology has
only been developed fairly recently
\cite{Bies:2021nje,Bies:2021xfh,Bies:2022wvj,Bies:2023jqg}. Fortunately,
our models admit several important simplifications such that the
$G_{4}$ flux itself already determines the vector-like spectrum.

Let us first focus on vector-like matter that lives on the bulk of $\Sigma$.
We follow the formalism in \cite{Bies:2014sra,Bies:2017fam}. At least in most cases, the full $C_3$ can be captured by an
algebraic complex 2-cycle $\mathcal{A}$ in the Chow group $\mathrm{CH}^{2}(\hat{Y})$
(algebraic cycles modulo  rational equivalence
instead of homological equivalence),
with homology class is $\left[G_{4}\right]$ \cite{Braun_2012}. We consider the restriction of $\mathcal{A}$
onto a Cartan divisor $D_{i}$, given by the intersection product
$\mathcal{A}\cdot D_{i}\in\mathrm{CH}^{2}\left(D_{i}\right)$. Its
projection onto $\Sigma$, given by $\pi_{*}\left(\mathcal{A}\cdot D_{i}\right)\in\mathrm{CH}^{1}\left(\Sigma\right)$,
is a curve on $\Sigma$ associated with the line bundle
\begin{equation}
L_{i}=\mathcal{O}_{\Sigma}\left(\pi_{*}\left(\mathcal{A}\cdot D_{i}\right)\right)\,.
\end{equation}
Now for each weight $\beta=-b_{i}\alpha_{i}$ of the adjoint,
we define the line bundle
\begin{equation}
L_{\beta}=\otimes_{i}L_{i}^{b_{i}}\,.
\end{equation}
Then the chiral and anti-chiral multiplicities for $\beta$ are counted
by the following sheaf cohomologies \cite{Donagi:2008ca,BeasleyHeckmanVafaI}
\begin{align}
\mathrm{chiral}&:\quad H^{0}\left(\Sigma,L_{\beta}\otimes K_{\Sigma}\right)\oplus H^{1}\left(\Sigma,L_{\beta}\right)\oplus H^{2}\left(\Sigma,L_{\beta}\otimes K_{\Sigma}\right)\,, \\
\mathrm{anti-chiral}&:\quad H^{0}\left(\Sigma,L_{\beta}\right)\oplus H^{1}\left(\Sigma,L_{\beta}\otimes K_{\Sigma}\right)\oplus H^{2}\left(\Sigma,L_{\beta}\right)\,.
\end{align}

Notice that the sheaf cohomologies for chiral and anti-chiral
matter are related by Serre duality. To calculate their dimensions,
we apply the following two simplifications \cite{BeasleyHeckmanVafaI}. First, for
fluxes with \emph{nontrivial} $L_\beta$ and satisfying primitivity
(i.e. preserving SUSY), we have $H^{0}\left(\Sigma,L_{\beta}\right)=H^{2}\left(\Sigma,L_{\beta}\otimes K_{\Sigma}\right)=0$.
Next, we %restrict
assume that
 $\Sigma$ is a rational surface with effective
$-K_{\Sigma}$,\footnote{As shown below, the condition that
$-K_\Sigma$ is effective is a reasonable simplifying assumption in the
context of rigid gauge groups. All toric surfaces are rational and
have effective $-K_\Sigma$.
We assume these conditions on $\Sigma$ in the
rest of the paper; in much of the paper we restrict attention to the
special case where $\Sigma$ is a (generally non-toric)
del Pezzo surface.
Further work would be needed to understand the detailed structure of
the resulting models when these technical conditions are relaxed.} %then
in this case, we have $H^{2}\left(\Sigma,L_{\beta}\right)=H^{0}\left(\Sigma,L_{\beta}\otimes K_{\Sigma}\right)=0$.
Therefore, the exact multiplicities $n_{\beta}$ and $n_{-\beta}$ are fully
determined by $h^{1}\left(\Sigma,L_{\beta}\right)$ and $h^{1}\left(\Sigma,L_{\beta}\otimes K_{\Sigma}\right)$
respectively. Since only $H^1$ is nontrivial, the multiplicities are also captured by the topological Euler
characteristics $\chi\left(\Sigma,L_{\beta}\right)$ and $\chi\left(\Sigma,L_{\beta}\otimes K_{\Sigma}\right)$.
These are fully determined by $c_{1}\left(L_{\beta}\right)$, given
by the Hirzebruch-Riemann-Roch theorem: \cite{Donagi:2008ca,BeasleyHeckmanVafaI,Blumenhagen:2008zz}
\begin{align}
\label{eq:HirzebruchRiemannRoch}
n_{\beta} & = -\chi\left(\Sigma,L_{\beta}\right) \nonumber \\
& =\frac{1}{2}\left[c_{1}\left(L_{\beta}\right)\right]\cdot K_{\Sigma}-\frac{1}{2}\left[c_{1}\left(L_{\beta}\right)\right]^{2}-1\nonumber \\
 & =\frac{1}{2}\chi_{\beta}-\frac{1}{2}\left[c_{1}\left(L_{\beta}\right)\right]^{2}-1\,,
\end{align}
where $\chi_{\beta}$ is the chiral index for $\beta$ given in \S\ref{subsec:FluxBreaking}, and
the Poincar\'e dual is taken with respect to $\Sigma$.
The expression for $n_{-\beta}$ is the same except that the sign of the
first term is flipped; indeed we get back $n_\beta-n_{-\beta}=\chi_\beta$.
In our models where only gauge-breaking fluxes are turned on, we can
read off $c_{1}\left(L_{\beta}\right)$ from Eq.\ (\ref{eq:phitoTheta}):
\begin{equation}
\label{eq:c1ofL}
\left[c_{1}\left(L_{\beta}\right)\right]=-b_{i}\kappa^{ij}\left(\phi_{j\alpha}\Sigma\cdot D_\alpha+\phi_{jr}C_{\mathrm{rem}}\right)\,.
\end{equation}
Combining Eqs. (\ref{eq:HirzebruchRiemannRoch}) and (\ref{eq:c1ofL}), 
this gives us a formula to compute the exact matter multiplicities from
the bulk of $\Sigma$ in terms of the vertical and remainder flux
parameters.

Now we turn to vector-like matter localized on matter curves. Similarly for a weight $\beta\in R$ supported on $C_R$,
we consider the pullback of $\mathcal A$ onto a matter surface $S(\beta)$
given by $\mathcal A\cdot S(\beta)\in \mathrm{CH}^{2}\left(S(\beta)\right)$. 
Its projection onto $C_R$, given by $\pi_*(A\cdot S(\beta))\in \mathrm{CH}^{1}\left(C_R\right)$ defines a line bundle for $\beta$:
\begin{equation}
L_\beta=\mathcal{O}_{C_R}\left(\pi_{*}\left(\mathcal{A}\cdot S(\beta)\right)\right)\,.
\end{equation}
Then the chiral and anti-chiral multiplicities for $\beta$ are counted
by the following sheaf cohomologies
\begin{align}
\label{eq:mattercurvecohomologies}
\mathrm{chiral}&:\quad H^{0}\left(C_R,L_{\beta}\otimes \sqrt{K_{C_R}}\right)\,, \\
\mathrm{anti-chiral}&:\quad H^{1}\left(C_R,L_{\beta}\otimes \sqrt{K_{C_R}}\right)\,,
\end{align}
where $\sqrt{K_{C_R}}$ is the spin bundle on $C_R$. These sheaf 
cohomologies are more subtle than those for the bulk of $\Sigma$. While
they are well understood when the matter curve has genus 0 or 1, for
irreducible curves with higher genus, these cohomologies have complicated dependence on moduli,
and their dimensions can jump at special points in the moduli space. For
reducible curves, there can also be vector-like pairs between 
different irreducible components of the curves, if the total chiral index
is split into different components accordingly. Instead of running into
all these subtleties, below we just focus on a special case where the
matter curve is simply a $\mathbb P^1$. In this case, there can never be
any vector-like pairs from the matter curve, since for $\mathbb P^1$ only
one of the $H^0,H^1$ is nontrivial, depending on the sign of the line 
bundle. 
This means in particular that the vector-like spectrum is independent of the choice
of spin bundle, significantly simplifying the analysis.
As shown below, this geometry is 
not hard to achieve in our $E_7$ models, and we leave the generalizations
to more complicated matter curves in future work.

So far, we have discussed the vector-like multiplicities for each
weight separately. On the other hand, we recall that weights from 
different $b_{i'}$ and $R$ can contribute to the same chiral $R'$ in Eq.\ 
(\ref{fullchiR'}) during flux breaking. Similarly, these different weights
can form vector-like pairs after flux breaking, even if each weight is 
purely chiral. This effect
can occur on both matter curves and the bulk of $\Sigma$. Such vector-like
matter has qualitatively different behavior from that obtained from sheaf
cohomologies, and has interesting phenomenological implications. More 
details will be discussed in later sections.

\section{Review of $E_7$ GUTs in F-theory}
\label{sec:E7Review}

With the above background knowledge, now we are ready to describe the
$E_7$ models. For completeness, first we briefly review our previous work
\cite{Li:2021eyn,Li:2022aek} on $E_7$ models, namely
%getting
describing how the SM gauge
group and chiral spectrum can be realized in a
 natural way through flux breaking of a rigid $E_7$ factor. We refer readers to those
two papers for more details.

\subsection{Flux breaking of rigid $E_7$ factors}

Recall that gauge groups in F-theory arise from sufficiently high degrees
of fiber singularities on a gauge divisor $\Sigma$. For an $E_7$ gauge 
group, the (singular) elliptic \cy fourfold $Y$
is described by a certain form of Weierstrass model 
\cite{Kodaira,Neron,BershadskyEtAlSingularities}. Treating the elliptic curve as the
\cy hypersurface in $\mathbb{P}^{2,3,1}$ with
homogeneous coordinates $[x:y:z]$, $Y$ is given by the
locus of
\begin{equation}
y^2 = x^3 + s^3 f_3xz^4 + s^5 g_5 z^6 \,,
\label{eq:e7}
\end{equation}
where $s, f_3, g_5$ are sections of line bundles ${\cal O} (\Sigma),{\cal O} (-4K_B-3
\Sigma),{\cal O} (-6K_B-5 \Sigma)$ on the base $B$,
and the gauge divisor
$\Sigma$
 supporting the $E_7$ factor  is
given by $s=0$. There is adjoint matter $\mathbf{133}$ arising from
excitations localized around the bulk of
$\Sigma$. There is also fundamental matter $\mathbf{56}$ localized on
the curve $s= f_3=0$, or $C_{\mathbf{56}}=-\Sigma \cdot (4K_B+3\Sigma)$
in terms of the intersection product, when the curve is nontrivial in
homology. When $\Sigma$ has a sufficiently negative normal bundle 
$N_\Sigma$, singularities of the elliptic fibration are enforced on $\Sigma$, and the 
Weierstrass model for $Y$ is automatically restricted to the form 
 (\ref{eq:e7}). A rigid $E_7$ is then realized on $\Sigma$. To be precise, we can consider the following divisors
on $\Sigma$ (not on $B$) \cite{MorrisonTaylor4DClusters}:
\begin{align}
    F_k &= -4K_\Sigma+(4-k)N_\Sigma\,, \nonumber \\
    G_l &= -6K_\Sigma+(6-l)N_\Sigma\,,
\label{eq:nonHiggsable}
\end{align}
where $k,l$ are integers. Then there is a rigid $E_7$ on $\Sigma$ if 
$F_k,G_l$ are effective for $k\geq 3,l\geq 5$ only. A simple way to 
satisfy this condition is to consider effective $-K_\Sigma$ and
$-N_\Sigma$, such that $-3K_\Sigma+N_\Sigma$ is not effective but 
$-4K_\Sigma+N_\Sigma$ is effective. As discussed
% before
previously, the natural choice of effective $-K_\Sigma$ 
matches nicely with the simplification we made in \S\ref{subsec:Vectorlikematter} for computing
vector-like spectrum, and we assume that this condition holds.

\begin{figure}[t]
\centering
\includegraphics[width=0.5\columnwidth]{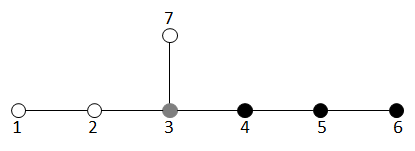}
\caption{The Dynkin diagram of $E_7$. The Dynkin node labelled
$i$ corresponds to the exceptional divisor $D_i$. The solid
nodes are the ones we break to get the Standard Model gauge
group and chiral matter. Node 3 (in gray) is broken by
remainder flux while the others are broken by vertical flux.}
\label{dynkine7}
\end{figure}

After setting up the geometry, we now turn on the flux background.
We break $E_7$ to $\gsm$ in two steps. Since remainder flux preserves
$\U(1)$'s along the roots but vertical flux does not, we first break 
$E_7$ to $\SU(5)$ with vertical flux, then break $\SU(5)$ to $\gsm$ with
remainder flux. The latter flux is very similar to the hypercharge flux
in traditional $\SU(5)$ GUTs in F-theory. To perform the first step
of breaking, we turn on nonzero $\Theta_{i'\alpha}$ for some $\alpha$ and
$i'=4,5,6$ subject to the flux constraints listed in \S\ref{subsec:FluxesIntro}, see Figure \ref{dynkine7}. In terms of flux parameters $\phi_{i\alpha}$, we turn on
\begin{equation}
    \phi_{1\alpha}=2n_\alpha\,,\quad\phi_{2\alpha}=4n_\alpha\,,\quad\phi_{3\alpha}=6n_\alpha\,,\quad
    \phi_{4\alpha}=5n_\alpha\,,\quad\phi_{7\alpha}=3n_\alpha\,.
    \label{eq:phi-n}
\end{equation}
The values of $\phi_{5 \alpha}, \phi_{6 \alpha}$, if sufficiently 
generic, do not affect the gauge group, but they will be fixed by other
flux and phenomenological constraints. Here we define a new set of flux parameters
$n_\alpha$, which can be integers or half-integers depending on the parity
of $c_2(\hat Y)$. 
At this point the gauge group has been broken to SU(5).
To perform the second step of breaking, we similarly
turn on the remainder flux
\begin{equation}
    \phi_{1r}=2n_r\,,\quad\phi_{2r}=4n_r\,,\quad\phi_{3r}=6n_r\,,\quad
    \phi_{7r}=3n_r\,,
    \label{eq:phi-r}
\end{equation}
and $\phi_{4r},\phi_{5r},\phi_{6r}$ plays the same role as $\phi_{5 \alpha}, \phi_{6 \alpha}$. Here $n_r$ is always integer. Under the
construction of rigid $E_7$, we require a non-toric base $B$ to ensure
the existence of $C_\mathrm{rem}$, hence this remainder flux.
After the remainder flux breaking, the remaining unbroken gauge group
is
\begin{equation}
\gsm=\SU(3)\times \SU(2)\times \U(1)/\mathbb
Z_6\,.
\label{eq:SMgaugegroup}
\end{equation}

\subsection{Chiral spectrum in flux-broken $E_7$ models}

It is straightforward to calculate the chiral spectrum given the above
fluxes. Since only the vertical flux induces chiral matter, we can analyze
the matter content by breaking $E_7 \rightarrow SU(5)$, where
the $\mathbf{56}$ breaks into a combination of $\mathbf{5},
\mathbf{10}$, uncharged singlets and conjugate representations, and $\mathbf{133}$
includes these as well as the adjoint $\mathbf{24}$.  Since the
adjoint
of SU(5)
 is non-chiral, the only chiral representations we expect for
$\gsm$ after the whole breaking
are the Standard Model representations,
which descend from the  $\mathbf{5},
\mathbf{10}$ of SU(5),
\begin{equation}
\label{eq:SMreps}
    Q=\left(\mathbf{3},\mathbf{2}\right)_{1/6}\,,\quad
    \bar U=\left(\bar{\mathbf{3}},\mathbf{1}\right)_{-2/3}\,,\quad
    \bar D=\left(\bar{\mathbf{3}},\mathbf{1}\right)_{1/3}\,,\quad
    L=\left(\mathbf{1},\mathbf{2}\right)_{-1/2}\,,\quad
    \bar E=\left(\mathbf{1},\mathbf{1}\right)_{1}\,.
\end{equation}
Using 
Eq.\ (\ref{fullchiR'}), we
indeed get the anomaly-free combination of SM chiral matter from vertical
flux. It will be useful to separate the contributions from $\mathbf{56}$
and $\mathbf{133}$ to the total chiral index, i.e. $\chi_{\left(\mathbf{3},\mathbf{2}\right)_{1/6}}=\chi_{\left(\mathbf{3},\mathbf{2}\right)_{1/6}}^{\mathbf{56}}+\chi_{\left(\mathbf{3},\mathbf{2}\right)_{1/6}}^{\mathbf{133}}$, where each contribution is anomaly-free by itself. The fundamental $\mathbf{56}$ gives
\begin{equation} \label{chi56}
    \chi_{\left(\mathbf{3},\mathbf{2}\right)_{1/6}}^{\mathbf{56}}=\Sigma\cdot (4K_B+3\Sigma)\cdot D_\alpha n_\alpha\,,
\end{equation}
and the adjoint $\mathbf{133}$ gives
\begin{equation} \label{chi133}
    \chi_{\left(\mathbf{3},\mathbf{2}\right)_{1/6}}^{\mathbf{133}}=2\Sigma\cdot (K_B+\Sigma)\cdot D_\alpha n_\alpha\,.
\end{equation}
Note that the total chiral indices only depend on $n_\alpha$ but not
$\phi_{5 \alpha}, \phi_{6 \alpha}$. This is no longer true when
we look at the chiral indices for each weight in the phenomenological
analysis below. An important feature of these chiral indices is that
they have a linear Diophantine structure in the quantized flux parameters,
with the coefficients not being very large.
If we randomly pick some small values of $n_\alpha$ (bounded by the 
tadpole constraint (\ref{tadpole})), 
%most probably
generically different terms in the
chiral indices will cancel each other, resulting in small chiral indices.
Therefore, small chiral indices are preferred in these models,
and it is not hard to achieve three generations of SM chiral matter.

In most cases, the above Weierstrass model also has codimension-3 singularities at the 
locus $s=f_3=g_5=0$. Traditionally, codimension-3 singularities are
interpreted as Yukawa couplings in the low-energy theory. In $E_7$ models,
however, these are so-called non-minimal singularities (with degrees 
$(4,6)$ or higher) where the fiber becomes non-flat, i.e. its dimension
jumps. Such singularities can no longer be interpreted as Yukawa 
couplings;\footnote{It was pointed out in \cite{Achmed-Zade:2018idx} that
these singularities may give rise to quartic or higher order couplings. 
Nevertheless, the singularities in \cite{Achmed-Zade:2018idx} have an 
unusual local geometry where a curve intersects another curve three times.
We do not see any such intersections or any evidence of such higher order 
couplings in our models.} this is also manifest by noticing that $\mathbf{56}^3$ does not
contain any singlets, hence cannot form any gauge-invariant 
couplings.
%\footnote{From the perspective after the breaking of $E_7$,
%Eq.\ (\ref{eq:branchingrules56}) tells us that all fields descended from 
%$\mathbf{56}$ carry half-integer additional $\U(1)$ charges. Hence we 
%cannot form invariant couplings from any three of those fields.}
Instead, there is an extra family of vertical flux associating to the 
non-flat fiber with nontrivial $\phi_{ij}$ components 
\cite{Jefferson:2021bid}. Analogous to codimension-2 $(4,6)$ 
singularities in 6D F-theory models \cite{HeckmanMorrisonVafa,Apruzzi:2018oge}, there has been
evidence that this flux switches on an extra sector of strongly coupled
superconformal \emph{chiral} matter, given by M2-branes wrapping curves
on the non-flat fiber \cite{46}. For our phenomenological purpose, we can always
set this flux to zero i.e. $\phi_{ij}=0$ for all $i,j$, such that the
extra sector becomes non-chiral and probably does not affect the Standard
Model sector. We should warn readers, however, that without further
studies on these extra sectors, we cannot precisely rule out the 
possibility that these sectors ruin the desired phenomenology.

\section{Phenomenology of $E_7$ GUTs}
\label{sec:Pheno}

So far, we have studied the gauge group and the chiral spectrum in the above class of $E_7$ models. In this
section, we start to analyze the phenomenological aspects of these models
in more detail. 
 The presence of approximate global symmetries descending from the
 underlying $E_7$ group suppresses certain couplings, with significant
 implications for phenomenology of these models;
in particular, we show that proton decay is automatically suppressed.
%Specifically, we
We focus further on the Higgs sector and the
interactions in these models.
We also discuss vector-like exotics in these models and %argue how
%they can still
the circumstances under which they can
 be phenomenologically safe. We consider the extent
to which the various features of the $E_7$ flux-broken models possibly, or even
naturally, match with observed phenomenology. The analysis in this
section, together with an explicit
construction of an example model in \S\ref{sec:ExplicitConstruction}, are the main
results of this paper. 
These results provide evidence
%They provide strong evidence towards
that natural and realistic $E_7$ GUTs can be realized in F-theory. 
On the other hand, due to %lack of
limits on existing
technologies %to compute
for computing detailed aspects of % continuous data in
 F-theory models (such as the specific values of couplings), most of the
analysis in this section is purely qualitative.

\subsection{Approximate global symmetries}
\label{subsec:approxu1}

Our starting point is 
based on considering
 approximate $\U(1)$ global symmetries that arise in the $E_7$  flux-broken
models.  
These approximate 
symmetries directly originate from vertical flux breaking, and
 control the structure of interactions and mass terms in the
low-energy theory,  as well as leading to many of the phenomenologically attractive
features of these models such as suppression of proton decay. 
 Recall that during vertical
flux breaking from $E_7$ to $\gsm$, the Cartan gauge bosons along the generators $T_4,T_5,T_6$ get masses from the St\"uckelberg
mechanism \cite{DonagiWijnholtModelBuilding,Grimm:2010ks,Grimm:2011tb}. 
These masses explicitly break the corresponding Cartan $U(1)$'s of the 
nonabelian gauge  symmetry,
but the matter interactions descending from the unbroken
gauge group still respect the global parts of the broken $U(1)$ gauge
symmetries (at least for the symmetries without mixed anomalies with the remaining gauge 
group).  These
$U(1)$ global symmetries are broken only by D3/M5-instanton
effects, which are exponentially suppressed in K\"ahler moduli \cite{Witten:1996bn, Blumenhagen:2006xt,Ibanez:2006da,Blumenhagen:2007zk}. In the low-energy theory, these effects turn on exponentially
suppressed mass terms and interactions that violate the global symmetries.
Since these effects are small, the symmetries still remain as approximate
global symmetries in the theory. This scenario is consistent with
the No Global Symmetries Conjecture \cite{Banks:2010zn,Harlow:2018tng}. Note that without more details of
the model, we cannot quantitatively specify %
the extent to which
 a certain quantity is
%exponentially
 suppressed by these symmetries. Throughout this paper, we 
only adopt the qualitative picture of exponential suppression, and leave
efforts towards explicit calculations of these quantities
for future work.

To study the implications of these $U(1)$ symmetries, it is important to understand
the branching rules from $E_7$ to $\gsm$  in the presence of these additional $U(1)$
charges. Below we use the basis $\left(Y,b_{4},b_{5},b_{6}\right)$
for the $U(1)$ charges, where $Y$ is the SM hypercharge.\footnote{One 
direction of the additional $\U(1)$ symmetries actually has mixed 
anomalies with the SM gauge group (see also, e.g.\ \cite{Ibanez:2006da},
for a similar situation). In the example below, the anomaly-free
directions are spanned by $-4b_4/5+b_5$ and $2b_4/5+b_6$. It turns out 
that excluding the anomalous symmetry does not affect the  selection
rules below at
all, so for completeness we still use all the $\U(1)$ charges 
%as the 
%basis
in labeling the charges.} The
branching rules are
\begin{align}
\label{eq:branchingrules56}
\mathbf{56} & \rightarrow\left(\mathbf{1},\mathbf{1}\right)_{0,5/2,2,3/2}+\left(\mathbf{1},\mathbf{1}\right)_{0,5/2,2,1/2}+\left(\mathbf{1},\mathbf{1}\right)_{0,5/2,1,1/2}+\left(\mathbf{1},\mathbf{1}\right)_{1,3/2,1,1/2}\nonumber \\
 & +\left(\mathbf{3},\mathbf{2}\right)_{1/6,3/2,1,1/2}+\left(\bar{\mathbf{3}},\mathbf{1}\right)_{-2/3,3/2,1,1/2}+\left(\bar{\mathbf{3}},\mathbf{1}\right)_{1/3,1/2,1,1/2}+\left(\bar{\mathbf{3}},\mathbf{1}\right)_{1/3,1/2,0,1/2}\nonumber \\
 &
+\left(\bar{\mathbf{3}},\mathbf{1}\right)_{1/3,1/2,0,-1/2}+\left(\mathbf{1},\mathbf{2}\right)_{-1/2,1/2,1,1/2}+\left(\mathbf{1},\mathbf{2}\right)_{-1/2,1/2,0,1/2}+\left(\mathbf{1},\mathbf{2}\right)_{-1/2,1/2,0,-1/2}
\nonumber \\
&+\mathrm{conjugates}\,,
\end{align}
\begin{align}
\label{eq:branchingrules133}
\mathbf{133} & \rightarrow\left(\mathbf{8},\mathbf{1}\right)_{0,0,0,0}+\left(\mathbf{1},\mathbf{3}\right)_{0,0,0,0}+4\times\left(\mathbf{1},\mathbf{1}\right)_{0,0,0,0}\nonumber \\
 &
+\left[\left(\mathbf{1},\mathbf{1}\right)_{0,0,0,1}+\left(\mathbf{1},\mathbf{1}\right)_{0,0,1,0}+\left(\mathbf{1},\mathbf{1}\right)_{0,0,1,1}+\left(\mathbf{1},\mathbf{1}\right)_{1,-1,0,0}+\left(\mathbf{1},\mathbf{1}\right)_{1,-1,-1,0}
\right.\nonumber \\
 &\hspace*{0.1in} +\left(\mathbf{1},\mathbf{1}\right)_{1,-1,-1,-1}+\left(\mathbf{3},\mathbf{2}\right)_{-5/6,0,0,0}+\left(\mathbf{3},\mathbf{2}\right)_{1/6,-1,0,0}+\left(\mathbf{3},\mathbf{2}\right)_{1/6,-1,-1,0}+\left(\mathbf{3},\mathbf{2}\right)_{1/6,-1,-1,-1}\nonumber \\
&\hspace*{0.1in}  +\left(\bar{\mathbf{3}},\mathbf{1}\right)_{-2/3,-1,0,0}+\left(\bar{\mathbf{3}},\mathbf{1}\right)_{-2/3,-1,-1,0}+\left(\bar{\mathbf{3}},\mathbf{1}\right)_{-2/3,-1,-1,-1}+\left(\bar{\mathbf{3}},\mathbf{1}\right)_{1/3,-2,-1,0}\nonumber \\
&\hspace*{0.1in}  +\left(\bar{\mathbf{3}},\mathbf{1}\right)_{1/3,-2,-1,-1}+\left(\bar{\mathbf{3}},\mathbf{1}\right)_{1/3,-2,-2,-1}+\left(\bar{\mathbf{3}},\mathbf{1}\right)_{1/3,3,2,1}+\left(\mathbf{1},\mathbf{2}\right)_{-1/2,-2,-1,0}\nonumber \\
&\hspace*{0.1in} \left. +\left(\mathbf{1},\mathbf{2}\right)_{-1/2,-2,-1,-1}+\left(\mathbf{1},\mathbf{2}\right)_{-1/2,-2,-2,-1}+\left(\mathbf{1},\mathbf{2}\right)_{-1/2,3,2,1}+\mathrm{conjugates}\right]\,.
\end{align}
It is then straightforward to apply the rule
that only terms  with all net
$U(1)$ charges vanishing are %not
unsuppressed in the superpotential
of the low-energy theory.
Notice that there are three copies of 
$\bar{D} =\left(\bar{\mathbf 3}, \mathbf 1\right)_{1/3}$
and $L = \left(\mathbf 1,\mathbf 2\right)_{-1/2}$ (or the $\SU(5)$ 
fundamental before remainder flux breaking) in each of
the decompositions $\mathbf{56},\mathbf{133}$, which are distinguished
by having different
%with different additional
$U(1)$ charges.
Without further information or inputs, 
 three families of SM chiral matter arising in a given model
 in general may be distributed
within the three copies. As shown below, such a distribution may lead to 
phenomenological inconsistencies, and some extra tuning must be done
% in the
%model
 to avoid those issues.

\subsection{Vector-like exotics}
\label{subsec:vectorexotics}
 
As in all GUT models, the $E_7$ models face the issue of having (many)
vector-like exotics that have not been observed in experiments. Although
we have chosen a $\mathbb P^1$ matter curve to ensure
that there are
 no vector-like 
pairs on the matter curve, generically there are many vector-like
matter
fields
on the bulk of $\Sigma$. Therefore, all representations in Eq. 
(\ref{eq:branchingrules133}), except the SM adjoint representations, 
generically have nontrivial vector-like matter multiplicities. 
Interestingly, these bulk vector-like fields involve the usual MSSM
 Higgs 
fields $H_u, H_d$, which indeed play the role of
a SM Higgs sector in the discussion below.
This feature does not seem to happen in models with smaller GUT groups.
On the other hand, there are also inert Higgs fields $H_u', H_d'$, which
have the same representation $(\mathbf 1,\mathbf 2)_{\pm1/2}$ under
the SM
gauge group, but do not have the right additional $\U(1)$ charges to form
unsuppressed Yukawa couplings with SM matter (see 
\S\ref{subsec:Yukawalist}). There are also similar sets of fields for the
triplet Higgs $T_u,T_d,T_u',T_d'$, as well as vector-like fields in other
exotic representations, namely $(\mathbf 3,\mathbf 2)_{-5/6},(\mathbf 3,\mathbf 2)_{1/6},(\bar{\mathbf 3},\mathbf 1)_{-2/3},(\mathbf 1,\mathbf 1)_{1}$.
In particular, the exotic $(\mathbf 3,\mathbf 2)_{-5/6}$ ruins
phenomenology by causing proton decay and spoiling gauge coupling unification (see also \S\ref{subsec:unification}), and must be removed
from the spectrum. We will also discuss the phenomenological safety of
other vector-like exotics as we proceed in later sections.

As seen
in \S\ref{subsec:Vectorlikematter}, the multiplicities of these vector-like fields are controlled
by the fluxes. Unfortunately, it has been shown in 
\cite{BeasleyHeckmanVafaII} that for GUT groups higher or equal to $\SO(10)$,
it is impossible to remove all the vector-like exotics by tuning the 
fluxes. Nevertheless, it was pointed out in \cite{Li:2022aek} that it is easy to remove the most dangerous
$(\mathbf 3,\mathbf 2)_{-5/6}$.

Now we show that in the $E_7$
models, this representation is %preferred
reasonably likely
to be removed from the spectrum, at least for certain kinds of gauge
divisor $\Sigma$.
First, we notice that the representation and its conjugate have
$(b_3,b_4,b_5,b_6)=(\pm1,0,0,0)$. Recall that the vertical flux we turn
on breaks directions $4,5,6$. We then see that $\chi_{(\mathbf
  3,\mathbf 2)_{-5/6}}=0$,
since $\Theta_{i \alpha} \neq 0$ only for $i = 4, 5, 6$ in
(\ref{eq:fluxbreakingcondition}),
 and the vertical flux does not contribute in Eq.
(\ref{eq:c1ofL}). In other words, the multiplicity is purely controlled
by the remainder flux in Eq.\ (\ref{eq:phi-r}), given by\footnote{When $\phi_{4r}=5n_r$, $L_\beta$ becomes trivial and Eq.\ (\ref{eq:HirzebruchRiemannRoch}) no longer applies.}
\begin{equation}
    n_{(\mathbf 3,\mathbf 2)_{-5/6}}=-\frac{1}{2}(5n_r-\phi_{4r})^2 C_{\mathrm{rem}}^2-1\,.
\end{equation}
Therefore, $n_{(\mathbf 3,\mathbf 2)_{-5/6}}=0$ if $5n_r-\phi_{4r}=\pm1$
and $C_\mathrm{rem}^2=-2$. Interestingly, some choices of the remainder flux with the 
smallest tadpole %satisfies
satisfy these conditions. Consider the tadpole
\begin{equation}
    \frac{1}{2}[G_4^\mathrm{rem}]^2=-\frac{1}{2}C_\mathrm{rem}^2\kappa^{ij}\phi_{ir}\phi_{jr}\,.
\end{equation}
As demonstrated in \S\ref{sec:ExplicitConstruction}, in many cases $\Sigma$ is a del Pezzo surface
and the available $C_\mathrm{rem}$ with the least negative
self-intersection has $C_\mathrm{rem}^2=-2$. 
Going forward we assume these technical conditions, which also imply the
condition discussed earlier that $-K_\Sigma$ is effective.
Further work would be needed to generalize the analysis to the
situation when  these technical conditions are relaxed.
In this situation, $\kappa^{ij}\phi_{ir}\phi_{jr}$ is minimized by e.g.
\begin{equation}
\label{eq:remainderfluxchoice}
    \phi_{ir}=(2,4,6,4,2,1,3)\,,
\end{equation}
which indeed leads to $n_{(\mathbf 3,\mathbf 2)_{-5/6}}=0$. From now on,
we always assume this choice of remainder flux, with tadpole
\begin{equation}
    \frac{1}{2}[G_4^\mathrm{rem}]^2=4\,.
\end{equation}

How about other vector-like exotics? Unlike the above, vertical flux also
contributes to the multiplicities
of other vector-like exotics. Comparing to remainder flux, vertical flux
satisfies more constraints like primitivity. We also need vertical flux
to get the right chiral spectrum and, to be discussed below, the right 
interactions. After fulfilling these more important requirements, 
we find no more room to remove the remaining vector-like exotics; i.e.,
generically there is a nontrivial or even large contribution to $n_\beta$ from vertical
flux. Since remainder flux is orthogonal to vertical flux, there is
no remainder flux that can cancel the contribution from
vertical flux. Therefore, we expect that all the other vector-like
exotics are present in our models. Fortunately, below we will show that
these vector-like exotics, including the triplet Higgs $(\bar{\mathbf 3},\mathbf 1)_{1/3}$, which potentially mediates dangerous proton decay, 
can still be phenomenologically acceptable if some additional assumptions
and tunings are made.

\subsection{List of Yukawa couplings}
\label{subsec:Yukawalist}

With the selection rules \S\ref{subsec:approxu1}, we can now list the Yukawa couplings that
are not suppressed by the approximate global symmetries. Recall that in
a general 4D F-theory model, there are three types
of Yukawa couplings on a gauge divisor $\Sigma$ \cite{BeasleyHeckmanVafaI}. First, there are Yukawa
couplings between three fields on the bulk of $\Sigma$ (denoted by 
$\Sigma\Sigma\Sigma$), but it has been shown in 
\cite{BeasleyHeckmanVafaI} that these couplings are all absent when 
$-K_\Sigma$ is effective, which is assumed in our $E_7$ models
as discussed in \S\ref{subsec:Vectorlikematter}. The 
second type of Yukawa couplings are between one field on the bulk of $\Sigma$ and two fields on matter curves (denoted by $\Sigma CC$). These
couplings are generically present, and we will simply assume that
all 
 couplings of this type that satisfy the symmetry constraints are present. Finally, there
are Yukawa couplings between three fields on matter curves (denoted by
$CCC$). These couplings are characterized by codimension-3 singularities
of the fibration. Nevertheless as discussed in last section, the 
codimension-3 singularities in the $E_7$ models cannot be interpreted as
Yukawa couplings. In conclusion, there are only $\Sigma CC$-type
couplings in our theory.\footnote{This coupling structure, from the low-energy perspective, can also be understood as a kind of R-symmetry, where
the fields on the bulk of $\Sigma$ have R-charge $1$, and those on matter
curves have R-charge $1/2$. We thank Jesse Thaler for this comment.} Note
that this UV structure of Yukawa couplings
%are vastly
is quite different from that of other % any
 previous SM-like constructions in 
F-theory, and we expect new phenomenological features in the IR to arise from this structure.

We now 
investigate how
%try to reproduce
 the SM Yukawa couplings can arise from
% with
 $\Sigma CC$-type couplings. In principle, we can
localize the vector-like Higgs on matter 
curves more general than $\mathbb P^1$. If we make such a choice, however,
the general (including both diagonal and off-diagonal) Yukawa couplings 
will require the non-existent $CCC$- and/or $\Sigma\Sigma C$-type 
couplings apart from $\Sigma CC$-type couplings. Therefore, to reproduce the SM Yukawa couplings
with mixing between all three generations, it is necessary to localize the Higgs on the bulk of $\Sigma$, and
all SM chiral matter on the matter 
curve $C_\mathbf{56}$.
 This choice of localization also matches with the fact 
that, from the discussion of \S\ref{subsec:Vectorlikematter}, 
generically
there are  many 
vector-like fields on the bulk of $\Sigma$, but no such pairs on $\mathbb P^1$ matter
curves.
We also notice that 
this choice of Higgs is only 
available when the
GUT group is as large as $E_7$, such that the adjoint includes the Higgs
after breaking. On the other hand, the chiral matter induced
by vertical flux breaking descends from both $\mathbf{133}$ and 
$\mathbf{56}$, and can be localized on both $\Sigma$ and $C_\mathbf{56}$. Therefore to reproduce
the SM Yukawa couplings, we need to impose the flux constraint
\begin{equation}
\label{eq:nochifrom133}
     \chi_{\left(\mathbf{3},\mathbf{2}\right)_{1/6}}^{\mathbf{133}}=2\Sigma\cdot (K_B+\Sigma)\cdot D_\alpha n_\alpha=0\,.
\end{equation}
Below we
will see that this constraint can be easily satisfied. It is worth
emphasizing that this choice of localization automatically implies very
different low-energy physics between the Higgs and chiral matter, due to
their distinct geometric origins.

Now, assuming that
the SM chiral spectrum is supported on $C_{\mathbf{56}}$, we can easily
list all couplings that do not violate the approximate global
symmetries. For simplicity, here we 
first ignore the couplings involving
uncharged singlets under the SM gauge group; %while
 these singlets may play the
role of right-handed neutrinos and will be studied in \S\ref{subsec:neutrino}.
We then have the SM Yukawa couplings:\footnote{Similar to mass terms, the couplings described here are 
terms in the superpotential $W$; Yukawa couplings between one boson
and two fermions come as usual from the contributions to the
potential $V$ of the form $ (\partial^2 W/\partial \phi_i \partial
\phi_j) \psi_i \psi_j$ and its conjugate.}
\begin{align}
H_uQ\bar U&:\quad\left(\mathbf 1,\mathbf 2\right)_{1/2,-3,-2,-1}\times\left(\mathbf 3,\mathbf 2\right)_{1/6,3/2,1,1/2}\times\left(\bar{\mathbf 3},\mathbf 1\right)_{-2/3,3/2,1,1/2}\,,\nonumber \\
H_dQ\bar D&:\quad
\begin{cases}
\left(\mathbf 1,\mathbf 2\right)_{-1/2,-2,-2,-1}\times\left(\mathbf 3,\mathbf 2\right)_{1/6,3/2,1,1/2}\times\left(\bar{\mathbf 3},\mathbf 1\right)_{1/3,1/2,1,1/2}\,, \\
\left(\mathbf 1,\mathbf 2\right)_{-1/2,-2,-1,-1}\times\left(\mathbf 3,\mathbf 2\right)_{1/6,3/2,1,1/2}\times\left(\bar{\mathbf 3},\mathbf 1\right)_{1/3,1/2,0,1/2}\,, \\
\left(\mathbf 1,\mathbf 2\right)_{-1/2,-2,-1,0}\times\left(\mathbf 3,\mathbf 2\right)_{1/6,3/2,1,1/2}\times\left(\bar{\mathbf 3},\mathbf 1\right)_{1/3,1/2,0,-1/2}\,,
\end{cases} \nonumber \\
H_d L\bar E&:\quad
\begin{cases}
\left(\mathbf 1,\mathbf 2\right)_{-1/2,-2,-2,-1}\times\left(\mathbf 1,\mathbf 2\right)_{-1/2,1/2,1,1/2}\times\left(\mathbf 1,\mathbf 1\right)_{1,3/2,1,1/2}\,, \\
\left(\mathbf 1,\mathbf 2\right)_{-1/2,-2,-1,-1}\times\left(\mathbf 1,\mathbf 2\right)_{-1/2,1/2,0,1/2}\times\left(\mathbf 1,\mathbf 1\right)_{1,3/2,1,1/2}\,, \\
\left(\mathbf 1,\mathbf 2\right)_{-1/2,-2,-1,0}\times\left(\mathbf 1,\mathbf 2\right)_{-1/2,1/2,0,-1/2}\times\left(\mathbf 1,\mathbf 1\right)_{1,3/2,1,1/2}\,,
\end{cases}
\label{eq:SMYukawa}
\end{align}
where the first representation in each product is the up and down
Higgs $H_u,H_d$. We also have %the following
a number of other exotic couplings. 
There are %also
 couplings involving the triplet Higgs $\left(\mathbf 3,\mathbf 1\right)_{-1/3}$:
\begin{align}
T_uQQ&:\quad\left(\mathbf 3,\mathbf 1\right)_{-1/3,-3,-2,-1}\times\left(\mathbf 3,\mathbf 2\right)_{1/6,3/2,1,1/2}\times\left(\mathbf 3,\mathbf 2\right)_{1/6,3/2,1,1/2}\,,\nonumber \\
T_u\bar U\bar E&:\quad\left(\mathbf 3,\mathbf 1\right)_{-1/3,-3,-2,-1}\times\left(\bar{\mathbf 3},\mathbf 1\right)_{-2/3,3/2,1,1/2}\times\left(\mathbf 1,\mathbf 1\right)_{1,3/2,1,1/2}\,,\nonumber \\
T_dQL&:\quad
\begin{cases}
\left(\bar{\mathbf 3},\mathbf 1\right)_{1/3,-2,-2,-1}\times\left(\mathbf 3,\mathbf 2\right)_{1/6,3/2,1,1/2}\times\left(\mathbf 1,\mathbf 2\right)_{-1/2,1/2,1,1/2}\,, \\
\left(\bar{\mathbf 3},\mathbf 1\right)_{1/3,-2,-1,-1}\times\left(\mathbf 3,\mathbf 2\right)_{1/6,3/2,1,1/2}\times\left(\mathbf 1,\mathbf 2\right)_{-1/2,1/2,0,1/2}\,, \\
\left(\bar{\mathbf 3},\mathbf 1\right)_{1/3,-2,-1,0}\times\left(\mathbf 3,\mathbf 2\right)_{1/6,3/2,1,1/2}\times\left(\mathbf 1,\mathbf 2\right)_{-1/2,1/2,0,-1/2}\,,
\end{cases} \nonumber \\
T_d\bar U\bar D&:\quad
\begin{cases}
\left(\bar{\mathbf 3},\mathbf 1\right)_{1/3,-2,-2,-1}\times\left(\bar{\mathbf 3},\mathbf 1\right)_{-2/3,3/2,1,1/2}\times\left(\bar{\mathbf 3},\mathbf 1\right)_{1/3,1/2,1,1/2}\,,\\
\left(\bar{\mathbf 3},\mathbf 1\right)_{1/3,-2,-1,-1}\times\left(\bar{\mathbf 3},\mathbf 1\right)_{-2/3,3/2,1,1/2}\times\left(\bar{\mathbf 3},\mathbf 1\right)_{1/3,1/2,0,1/2}\,,\\
\left(\bar{\mathbf 3},\mathbf 1\right)_{1/3,-2,-1,0}\times\left(\bar{\mathbf 3},\mathbf 1\right)_{-2/3,3/2,1,1/2}\times\left(\bar{\mathbf 3},\mathbf 1\right)_{1/3,1/2,0,-1/2}\,.
\end{cases}
\label{eq:tripletHiggsYukawa}
\end{align}
These
couplings are always present together with the SM Yukawa couplings, but
the ones with triplet Higgs mediate dimension-5 proton decay and need
extra attention. Note that there are unique sets of additional $\U(1)$ 
charges for $H_u,T_u$.
This uniquely identifies these fields in the decomposition
 (\ref{eq:branchingrules133}) as
\begin{equation}
H_u=\left(\mathbf{1},\mathbf{2}\right)_{1/2, -3, -2, -1}\,,\quad
T_u=\left(\bar{\mathbf{3}},\mathbf{1}\right)_{-1/3, -3, -2, -1}\,.
\label{eq:Higgsrepresentations}
\end{equation}
On the other hand, there are three possible %sets
fields with distinct approximate U(1) charges for each of $H_d,T_d$,
each of which couples to $\bar{D}, L$ in one of the three possible
copies of the SU(5) fundamentals. The
choices of these charges will be discussed below.

There are also couplings involving
other types of vector-like exotics, namely $(\mathbf 3,\mathbf 2)_{1/6},(\mathbf 3,\mathbf 1)_{2/3},(\mathbf 1,\mathbf 1)_1$:
\begin{gather}
\left(\mathbf 3,\mathbf 2\right)_{1/6,-1,-1,-1}\times\left(\bar{\mathbf 3},\mathbf 1\right)_{1/3,1/2,1,1/2}\times\left(\mathbf 1,\mathbf 2\right)_{-1/2,1/2,0,1/2}\,,\nonumber \\
\left(\mathbf 3,\mathbf 2\right)_{1/6,-1,-1,-1}\times\left(\bar{\mathbf 3},\mathbf 1\right)_{1/3,1/2,0,1/2}\times\left(\mathbf 1,\mathbf 2\right)_{-1/2,1/2,1,1/2}\,,\nonumber \\
\left(\mathbf 3,\mathbf 2\right)_{1/6,-1,-1,0}\times\left(\bar{\mathbf 3},\mathbf 1\right)_{1/3,1/2,1,1/2}\times\left(\mathbf 1,\mathbf 2\right)_{-1/2,1/2,0,-1/2}\,,\nonumber \\
\left(\mathbf 3,\mathbf 2\right)_{1/6,-1,-1,0}\times\left(\bar{\mathbf 3},\mathbf 1\right)_{1/3,1/2,0,-1/2}\times\left(\mathbf 1,\mathbf 2\right)_{-1/2,1/2,1,1/2}\,,\nonumber \\
\left(\mathbf 3,\mathbf 2\right)_{1/6,-1,0,0}\times\left(\bar{\mathbf 3},\mathbf 1\right)_{1/3,1/2,0,1/2}\times\left(\mathbf 1,\mathbf 2\right)_{-1/2,1/2,0,-1/2}\,,\nonumber \\
\left(\mathbf 3,\mathbf 2\right)_{1/6,-1,0,0}\times\left(\bar{\mathbf 3},\mathbf 1\right)_{1/3,1/2,0,-1/2}\times\left(\mathbf 1,\mathbf 2\right)_{-1/2,1/2,0,1/2}\,,\nonumber \\
\left(\bar{\mathbf 3},\mathbf 1\right)_{-2/3,-1,-1,-1}\times\left(\bar{\mathbf 3},\mathbf 1\right)_{1/3,1/2,1,1/2}\times\left(\bar{\mathbf 3},\mathbf 1\right)_{1/3,1/2,0,1/2}\,,\nonumber \\
\left(\bar{\mathbf 3},\mathbf 1\right)_{-2/3,-1,-1,0}\times\left(\bar{\mathbf 3},\mathbf 1\right)_{1/3,1/2,1,1/2}\times\left(\bar{\mathbf 3},\mathbf 1\right)_{1/3,1/2,0,-1/2}\,,\nonumber \\
\left(\bar{\mathbf 3},\mathbf 1\right)_{-2/3,-1,0,0}\times\left(\bar{\mathbf 3},\mathbf 1\right)_{1/3,1/2,0,1/2}\times\left(\bar{\mathbf 3},\mathbf 1\right)_{1/3,1/2,0,-1/2}\,,\nonumber \\
\left(\mathbf 1,\mathbf 1\right)_{1,-1,-1,-1}\times\left(\mathbf 1,\mathbf 2\right)_{-1/2,1/2,1,1/2}\times\left(\mathbf 1,\mathbf 2\right)_{-1/2,1/2,0,1/2}\,,\nonumber \\
\left(\mathbf 1,\mathbf 1\right)_{1,-1,-1,0}\times\left(\mathbf 1,\mathbf 2\right)_{-1/2,1/2,1,1/2}\times\left(\mathbf 1,\mathbf 2\right)_{-1/2,1/2,0,-1/2}\,,\nonumber \\
\left(\mathbf 1,\mathbf 1\right)_{1,-1,0,0}\times\left(\mathbf 1,\mathbf 2\right)_{-1/2,1/2,0,1/2}\times\left(\mathbf 1,\mathbf 2\right)_{-1/2,1/2,0,-1/2}\,.
\label{eq:exoticYukawa}
\end{gather}
These couplings induce, e.g., additional proton decay and may not be 
compatible with phenomenology. On the other hand, all these couplings
%exchange
mix
distinct copies of $(\bar{\mathbf 3},\mathbf 1)_{1/3}$ and/or 
$(\mathbf 1,\mathbf 2)_{-1/2}$ with different U(1) charges, while 
%each
the couplings in Eqs. 
(\ref{eq:SMYukawa}) and (\ref{eq:tripletHiggsYukawa}) 
relate $H_d, T_d$ in a given copy with matter fields within the same corresponding
copy.  As
shown in \S\ref{subsubsec:Yukawastructure}, extra tuning of the model is available such that the
couplings in Eq.\ (\ref{eq:exoticYukawa}) are absent, and we assume
such absence throughout the paper. All the other couplings without
uncharged singlets and not listed above, if present, are exponentially
suppressed by the approximate global symmetries.

\subsection{Proton decay}
\label{subsec:protondecay}

Before building semi-realistic Higgs sector and Yukawa couplings, there
is an important issue to be resolved: as in every GUT model, 
%there must
%be a mechanism to suppress proton decay. 
there is a possibility of couplings that give a proton decay rate that
exceeds experimental limits.
In particular, since the triplet Higgs cannot
be removed from the spectrum, the couplings in last subsection naively seem to
suggest that the $E_7$ models will suffer from
excess %too large
 proton decay mediated by
 dimension-5 operators.  We
now show that, fortunately, both dimension-4 and 5 proton decay 
are ubiquitously suppressed in the $E_7$ models.
This feature in some sense ``comes for free'' with the construction of
these models.

First, dimension-4 proton decay in the MSSM is driven by the R-parity
violating terms in the superpotential:
\begin{equation}
\label{eq:dim4protondecay}
    W\supset \alpha_1QL\bar D+\alpha_2LL\bar E+\alpha_3\bar D\bar D\bar U\,,
\end{equation}
where we have used the notation in Eq.\ (\ref{eq:SMreps}).
These are
couplings between three chiral fields, which all descend from 
$\mathbf{56}$ under the assumption of Eq.\ (\ref{eq:nochifrom133}). Hence
from the geometric perspective, dimension-4 proton decay requires 
$CCC$-type couplings, which are absent in the $E_7$ models. 
The absence of these couplings is also %complemented
natural from
% by
 the form of the fields in the
symmetry-broken theory: since the approximate global charges $b_4,b_6$ of
all fields (including all copies of $\bar D,L$) appearing in Eq.\ 
(\ref{eq:dim4protondecay}) are half-integers, none of the interactions
of this type
have vanishing net $b_4,b_6$ charges, so all such interactions violate
one of the approximate
global symmetries.
%On the other
%hand, it is
%not clear whether flux breaking would induce new but exponentially small
%$CCC$-type couplings; the interplay between codimension-3 $(4,6)$ 
%singularities and flux breaking will be studied in the future. 
%This
Thus, dimension-4 proton decay is automatically absent in the $E_7$ 
models (with the caveat that we do not completely understand the $(4,6)$
singularities; it is not fully clear whether the interplay between these
singularities and flux breaking would modify this conclusion). In fact, it was already pointed out in \cite{Tatar:2006dc} that
dimension-4 proton decay can be eliminated in this way only when the
GUT group is $E_7$ or $E_8$; %such  elimination was also realized in
this suppression arises in
heterotic and M-theory
models as well as in F-theory.

Now we turn to dimension-5 proton decay. Such decay in conventional
supersymmetric GUTs comes from the following terms
\begin{equation}
\label{eq:conventionaldim5}
W\supset \lambda_1 T_uQQ+\lambda_2 T_dQL+MT_{u}T_{d}\,,
\end{equation}
where $T_u,T_d$ are the triplet Higgs, and $M$ is a large mass close to 
the GUT scale $M_\mathrm{GUT}$. Integrating
out $T_{u},T_{d}$, we then get the dimension-5 operator $QQQL/M$, which is
only suppressed by $1/M$ and 
leads to
an unacceptable rate of proton decay. Nevertheless, the $E_7$ models are different
from %the
 conventional GUTs in the following sense: although the first two terms
in Eq.\ (\ref{eq:conventionaldim5}) are not suppressed, we see from Eq.\ 
(\ref{eq:tripletHiggsYukawa}) that $T_u,T_d$ never have opposite additional $\U(1)$
charges, hence the mass term in Eq.\ (\ref{eq:conventionaldim5}) is 
exponentially suppressed. Instead, $T_u,T_d$ have their
own vector-like partners, denoted by $T_d',T_u'$, with opposite
additional $\U(1)$ charges. These ``primed'' fields
are \emph{inert}, i.e.\ their Yukawa interactions with SM chiral matter are
exponentially suppressed, but they give $T_u,T_d$ large masses by the
conventional mass terms. Now the superpotential schematically has the form of
\begin{equation}
W\supset \lambda_1 T_uQQ+\lambda_2 T_dQL+MT_{u}T_{d}'+MT_{u}'T_{d}+mT_{u}T_{d}+mT_{u}'T_{d}'\,,
\end{equation}
where $m$ is exponentially suppressed compared to $M$.
The operator we get by integrating out the triplet Higgs and the vector-like partners is roughly
$\left(m/M\right)QQQL/M$, which is indeed further exponentially 
suppressed by the factor $m/M$. Therefore, as long as $M$ is
sufficiently
large (probably close to $M_{GUT}$) and $m/M$ is sufficiently small, the
$E_7$ models are safe from overly dangerous proton decay.

Regarding dimension-6 proton decay mediated by gauge bosons along the 
broken directions of the gauge group, we do not see an obvious mechanism
of suppression.
 The gauge bosons typically have masses around the KK/GUT scale, %and 
which
may be 
sufficiently high to evade the current experimental bounds \cite{Workman:2022ynf}. Moreover,
fluxes on $\Sigma$ can make the ground state wavefunction of the gauge
bosons more localized, and suppress its wavefunction overlap
(hence the coupling) to the chiral matter on $C_\mathbf{56}$
\cite{Ibanez:2012zg}. Note that such suppression may not be exponential 
\cite{Hebecker:2014uaa}, but may already be sufficient for our purposes due
to the high GUT scale.
Despite all these heuristic arguments, more techniques and explicit
calculations are still required to %figure
determine the exact rate of proton decay,
which is essential for realistic model building.

\subsection{Higgs and Yukawa sectors}
\label{subsec:HiggsAndYukawa}

Now we turn to the Higgs and Yukawa sectors in the $E_7$ models. Even
with the Yukawa couplings in Eq.\ (\ref{eq:SMYukawa}) that have the right 
representations, it is not
guaranteed that those couplings resemble the structure of Higgs and Yukawa sectors
in the Standard Model, due to various differences of the $E_7$ models
from the conventional Standard Model. Below we explain each of these
differences, and write down the necessary conditions for realizing the
SM Higgs and Yukawa sectors.

\subsubsection{Doublet-triplet splitting and the Higgs masses}
\label{subsubsec:Higgs}

Apart from proton decay, one of the important questions in general GUT
models is the doublet-triplet splitting problem, or why the masses of 
doublet and triplet Higgs are separated %or
by many orders of magnitude.
In F-theory GUTs,  this splitting in principle can be explained by
the presence of hypercharge flux \cite{BeasleyHeckmanVafaII}. For
the tuned $\SU(5)$ GUTs, however, the explicit realization of such 
splitting can be difficult; see \S\ref{sec:Comparison} for further
discussion.  In contrast, the doublet and triplet
Higgs in the $E_7$ model live on the bulk of $\Sigma$ and always receive
mass splitting from hypercharge flux, which is also localized on a
(remainder) surface on $\Sigma$. Therefore, the doublet and triplet
masses are automatically split once we break $E_7$ to $\gsm$, although
the amount of splitting is still unknown and new techniques
must be developed for finding out the Higgs mass spectrum.

\begin{comment}
Although
the amount of splitting is still unknown, in principle the hypercharge 
flux can induce exponential
separation between the two masses as follows.
It was demonstrated in \cite{BeasleyHeckmanVafaII} that when $\Sigma$ is
a del Pezzo surface as in our models, depending on
the choice of flux, the wavefunctions involved in the mass terms can repel
one another, hence give exponentially suppressed mass terms. The ranges 
of such choice are different for doublet and triplet Higgs. Therefore, 
there may be a choice of hypercharge flux such that the mass term for
doublet Higgs is exponentially suppressed by wavefunction repulsion, 
while that for triplet Higgs is not. Without explicit
geometric and flux curvature data of
the model, however, it is hard to check whether this scenario can be 
realized.
\end{comment}

Still, what controls the mass terms before the splitting by
hypercharge flux? Similar to the triplet Higgs in
\S\ref{subsec:protondecay}, the conventional $\mu$-term
i.e. $\mu H_u H_d$ is exponentially suppressed. This suppression, however,
does not mean that the $\mu$-problem is solved, since the Higgs can still
get large masses from terms $H_u H_d', H_d H_u'$, when 
$H_u,H_d$ have their own vector-like partners $H_d',H_u'$. On the other 
hand, it means that there is some vector-like matter with light masses
when such vector-like partners do not exist. 
%Such 
Indeed, such a scenario generically 
%happens
arises for $H_d$, and the essence of this effect lies in
vertical flux breaking: although we have imposed that the total chiral
index of fields arising
from $\mathbf{133}$ vanishes, there can still be nontrivial chiral
surpluses for each of the three copies of doublet and triplet Higgs in
this representation
(see \S\ref{subsec:approxu1}). Suppose we have the following spectrum
for the three copies of Higgs fields:
\begin{align}
    (\mathbf 1,\mathbf 2)_{-1/2,-2,-2,-1}&:\,n_1\,,\quad(\mathbf 1,\mathbf 2)_{1/2,2,2,1}:\,n_1'\,,\nonumber\\
    (\mathbf 1,\mathbf 2)_{-1/2,-2,-1,-1}&:\,n_2\,,\quad(\mathbf 1,\mathbf 2)_{1/2,2,1,1}:\,n_2'\,,\nonumber\\
    (\mathbf 1,\mathbf 2)_{-1/2,-2,-1,0}&:\,n_3\,,\quad(\mathbf 1,\mathbf 2)_{1/2,2,1,0}:\,n_3'\,,\label{eq:Higgs-n-spectrum}
\end{align}
where $n_i,n_i'$ denote the multiplicities. Eq. (\ref{eq:nochifrom133})
implies that $n_1+n_2+n_3=n_1'+n_2'+n_3'$, hence the spectrum is non-chiral
under the SM gauge group. On the other hand, generically we have 
$n_i\neq n_i'$ and the spectrum would be
% is
 chiral if the additional 
$\U(1)$'s were gauge symmetries, i.e. the St\"uckelberg mechanism was absent.
If $n_i\neq n_i'$ for some $i$, there must be a
field direction in the $i$-th copy that cannot acquire any mass terms 
within the same copy. It can only get mass terms from fields in other 
copies. Since they do not have opposite additional $\U(1)$ charges,
the resulting mass terms are exponentially suppressed, leading
to a light doublet Higgs $H_d$.

It is tempting to use the above mechanism to solve the $\mu$-problem.
Unfortunately, the $\mu$-problem cannot be solved in this way %because
                                %of
for
two reasons. First, only $H_d$, but not $H_u$, has three copies in the
branching rule. In other words, %the same
this mechanism for producing light
$H_d$ cannot produce a light $H_u$. 
Second, the above mechanism relies on vertical flux breaking,
which only breaks $E_7$ to $\SU(5)$ instead of $\gsm$. This means that
whenever a light $H_d$ is produced in this way, there must also be a
light $T_d$. Although
there may still be doublet-triplet splitting from hypercharge flux, the
mass of the $T_d$ is still exponentially suppressed. Such $T_d$ directly interacts with SM chiral matter and
ruins the argument in \S\ref{subsec:protondecay}, i.e. there is still too much dimension-5
proton decay even with the exponential suppression in \S\ref{subsec:protondecay}. In this sense, we should even avoid
any light $H_d$ or $T_d$ produced in this way. As discussed in \S\ref{subsubsec:Yukawastructure}, we will
arrange the fluxes such that only one copy of $H_d$ interacts with SM
chiral matter. Without loss of generality, let us pick the copy with
$(b_4,b_5,b_6)=(2,2,1)$. Then avoiding light $H_d$ and $T_d$ coming from the above mechanism is 
achieved by the flux constraint
\begin{equation}
\label{eq:133samecopy}
    \chi^\mathbf{133}_{(\bar{\mathbf 3},\mathbf 1)_{1/3,-2,-2,-1}}=0\,.
\end{equation}
This is another linear constraint on the flux parameters, similar to the
ones for breaking the gauge group or inducing three generations of chiral
matter. This new constraint, however, is the first constraint that
involves the previously unused flux parameters $\phi_{5\alpha},\phi_{6\alpha}$. Therefore, given the gauge group and total chiral spectrum, 
there is always still some room in the $E_7$ models for
satisfying this new constraint.

The above still does not explain the origin of light masses in the SM
Higgs sector. Sadly, in the current construction of our $E_7$ models,
there is still no obvious solution to the $\mu$-problem. This is 
understandable, however, since the Higgs masses in F-theory are very 
complicated quantities to calculate. Traditionally, the Higgs masses come from the vevs of some fields
localized on divisors other than $\Sigma$ but intersecting with $\Sigma$.
These fields behave as singlets and couple to the vector-like matter
on $\Sigma$. Nevertheless, the vevs or potential of these fields depends on
many factors, including %by
but not limited to the detailed couplings between
these fields, the D-term potential, the nonperturbative superpotential,
and most importantly, soft SUSY breaking \cite{Palti:2016kew}. Therefore without understanding
more basic issues like moduli stabilization and SUSY breaking in F-theory, no precise
statements on these vector-like masses can be made. On the other hand,
given such a complicated origin of the Higgs masses, it is reasonable to
expect that some hierarchy is generated and brings some of the Higgs
to light scales. At the same time, we should
not allow more than one pair
of Higgs to be at the electroweak scale, although generically there are
many vector-like fields with the same representation. This is because
when more than one Higgs field couples to SM chiral matter in the same
way, the flavor basis generically does not align with the Higgs mass
basis. Such misalignment produces tree-level flavor-changing neutral
currents (FCNCs), which are not observed in experiments.\footnote{We
thank Jesse Thaler for pointing out this issue.} In conclusion, to
reproduce the SM Higgs sector, it is far from clear how to realize
exactly one pair of light Higgs doublets among all the Higgs fields. This is a major
%disadvantage 
shortcoming
 of the $E_7$ models, and we hope to give a better explanation
for this Higgs hierarchy in the future.

As a remark, there are still many \emph{inert} Higgs fields in the other
two copies. In particular, there can be multiple light inert Higgs fields,
coming from pairing chiral surpluses between the copies or other ways. 
Fortunately since they are inert, there is no tight constraint on these
fields. We note that the 
current experimental lower bound on $H_u',H_d'$ masses is around 100 GeV \cite{Workman:2022ynf}.

\subsubsection{Structure of Yukawa couplings}
\label{subsubsec:Yukawastructure}

One of the most dangerous features in the $E_7$ model is that there are three copies
of $(\bar{\mathbf 3},\mathbf 1)_{1/3}$ and $(\mathbf 1,\mathbf 2)_{-1/2}$
in the branching rules in \S\ref{subsec:approxu1}, with different
additional $\U(1)$ charges. The
generic case where the three generations of chiral matter are distributed
in all the copies is not phenomenologically acceptable for various 
reasons. First, generically there are chiral %surpluses
differences for each of the
copies,
as in (\ref{eq:Higgs-n-spectrum}).
While
these add up to the three generations in the total chiral spectrum, as
demonstrated in \S\ref{subsubsec:Higgs} they can also form light vector-like exotics between
different copies. Next, having different copies in the chiral spectrum
turns on an unsuppressed set of exotic couplings in Eq.\ (\ref{eq:exoticYukawa}),
including additional proton decay. More seriously, multiple light Higgs fields are
required to generate unsuppressed SM Yukawa couplings for all the copies, see Eq.\ (\ref{eq:SMYukawa}).\footnote{While there is indeed some hierarchy between
Yukawa couplings in the observed Standard Model, we do not expect
the hierarchy to be as large as the exponential suppression from the 
approximate global symmetries. Therefore, if we only use one Higgs for 
multiple copies with exponentially suppressed couplings, very probably it will not give
the right flavor structure.} As discussed in \S\ref{subsubsec:Higgs},
such a Higgs sector
again leads to unacceptable FCNCs. Therefore to avoid all the above 
issues, we must arrange all three generations of chiral matter to be
within the same copy. For the choice of light Higgs with
$(b_4,b_5,b_6)=(2,2,1)$ in \S\ref{subsubsec:Higgs}, for example,
we should
choose the corresponding copy of chiral matter with $(b_4,b_5,b_6)=(1/2,1,1/2)$ and impose the following
flux constraints:
\begin{equation}
\label{eq:56samecopy}
    \chi^\mathbf{56}_{(\bar{\mathbf 3},\mathbf 1)_{1/3,1/2,0,1/2}}=\chi^\mathbf{56}_{(\bar{\mathbf 3},\mathbf 1)_{1/3,1/2,0,-1/2}}=0\,.
\end{equation}
Again, these constraints are mild tuning on the remaining flux parameters
$\phi_{5\alpha},\phi_{6\alpha}$, which is
generically achievable. Nevertheless, it will be interesting to see
whether there is a more fundamental reason that leads to such choice of
fluxes.

After this choice of chiral matter, the remaining couplings in the 
low-energy theory are just the SM Yukawa couplings and their counterparts
with the triplet Higgs. It would be even more informative if we can get 
the values of the Yukawa couplings. Although calculating those
values is beyond our current F-theory technologies, we can gather some of
their qualitative features. Unlike the conventional F-theory models with
$CCC$-type couplings, the use of $\Sigma CC$-type couplings means that the
Yukawa couplings are supported on the whole $C_\mathbf{56}$ instead of
points on it. If the Higgs wavefunction is nearly uniform on 
$C_\mathbf{56}$, the Higgs will interact with all three 
generations of chiral matter in the same way, thus the Yukawa couplings
will be undesiredly close to an identity matrix.\footnote{We thank Jonathan Heckman for pointing this out.}
Nevertheless, especially with the presence of bulk fluxes, we expect 
the
Higgs wavefunction to be non-uniform and peak in some smaller
region. A simple %generic
 scenario would be that the region intersects with  
$C_\mathbf{56}$ in a connected small but finite range.
% in.
This scenario is then similar to
the case of a single Yukawa point studied in e.g. \cite{Heckman:2008qa,Cecotti:2009zf}, where the small nonperturbative
correction is now due to the finite size of the interaction region.
In this way, the 
Yukawa hierarchy is generated as in the $\SU(5)$ F-theory
GUTs. On the other hand, to really compute the Yukawa couplings, we first need to
understand the Higgs wavefunction profile and its possible
correlations with the exponentially low Higgs mass. Once we understand
these issues, we may be able to use the ultra-local approach developed in
\cite{Font:2012wq,Font:2013ida,Marchesano:2015dfa,Carta:2015eoh} to
computing Yukawa couplings within the intersecting region, but understanding those 
issues remains very challenging. In %less generic
more complicated scenarios where there are
multiple disconnected interaction regions, they are similar to the case of
multiple Yukawa points. The arguments in \cite{Cvetic:2019sgs} then 
suggest that there is also some Yukawa hierarchy, although the methods in
\cite{Cvetic:2019sgs} do not straightforwardly generalize to our models due to
the use of flux breaking. In summary, it is possible that there is
 some hierarchy between the Yukawa couplings.
This hierarchy may match with the observed Yukawa hierarchy, but 
explicitly computing the Yukawa matrix in our models will be an important
future step for realistic model building.

\subsection{Neutrino sector}
\label{subsec:neutrino}

Here we turn to the neutrino sector and make some brief comments. From Eq.
(\ref{eq:branchingrules56}), we see that $\mathbf{56}$ gives three copies
of singlets that can be right-handed neutrinos. Since in the
above we have restricted the leptons into one copy, only one copy of the
singlets $(\mathbf 1,\mathbf 1)_{0,5/2,1,1/2}$ have unsuppressed Yukawa
couplings with SM chiral matter:
\begin{gather}
    (\mathbf 1,\mathbf 2)_{1/2,-3,-2,-1}\times (\mathbf 1,\mathbf 2)_{-1/2,1/2,1,1/2}\times (\mathbf 1,\mathbf 1)_{0,5/2,1,1/2}\,,\nonumber \\
    (\mathbf 3,\mathbf 1)_{-1/3,-3,-2,-1}\times (\bar{\mathbf 3},\mathbf 1)_{1/3,1/2,1,1/2}\times (\mathbf 1,\mathbf 1)_{0,5/2,1,1/2}\,.
\end{gather}
The other two copies of singlets have nontrivial multiplicities but belong
to inert matter.

We can obtain the multiplicity of right-handed neutrinos from its ``chiral
index''. It sounds strange to calculate a ``chiral index'' for non-chiral
matter; the correct interpretation is that the right-handed neutrinos
carry additional $\U(1)$ charges, hence ``would be'' chiral matter if we ignore the St\"uckelberg
mechanism.
Since there are no vector-like exotics on the matter curve, the chiral
index is the same as the exact multiplicity, which remains unchanged
under the symmetry breaking. Now from the flux constraints we imposed in
previous sections, we see that the chiral index is fixed to be
\begin{equation}
    \chi_{(\mathbf 1,\mathbf 1)_{0,5/2,1,1/2}}=3\,.
\end{equation}
Therefore, we have three right-handed neutrinos, which favorably combine
with the left-handed ones to give three Dirac neutrinos and a square PMNS
matrix. The above Yukawa couplings then give the usual Dirac mass terms
after electroweak symmetry breaking. This scenario is more or less the
same as conventional GUTs with $\SO(10)$ gauge group or above.

There are also Majorana mass terms from the $\Sigma CC$-type couplings
involving two right-handed neutrinos and a bulk singlet. Since
all right-handed neutrinos have the same additional $\U(1)$ charges,
the Majorana masses are always exponentially suppressed (compared
to string/GUT scale) by the additional $\U(1)$ symmetries. In fact, similar suppression was already used in some 
early type II \cite{Blumenhagen:2006xt,Ibanez:2006da} and F-theory 
\cite{BeasleyHeckmanVafaII} SM-like model. It was estimated in those 
references that the exponential suppression factor might be around 
$10^{-6}$ to $10^{-4}$, which is much more mild than the electroweak 
hierarchy. This is not incompatible
 %Therefore, this 
%suppression %seems to match 
with the observational constraints on the 
seesaw mechanism. We emphasize that, however, these numerical estimates
are very crude, and without more explicit
computations of the masses and couplings, we cannot make fully precise
statements on how the left-handed neutrinos get very small masses.

\subsection{Gauge coupling unification}
\label{subsec:unification}

Here we briefly comment on the possibility of gauge coupling unification
in our models. Despite the use of $E_7$ in the construction of models,
whether %the gauge coupling unification is preserved 
gauge coupling unification is present
in any useful sense
is far from obvious.
From the point of view of the world-volume theory on the
IIB 7-branes supporting the $E_7$ gauge
theory (as in, e.g., \cite{BeasleyHeckmanVafaI}), it should be possible to find a classical %(though
%nonperturbative)
 description of flux breaking through turning on flux (T-dual to
 turning on an adjoint scalar as in, e.g., \cite{Taylor:1996ik}).
From this perspective, at sufficiently high energies
 the world-volume
$E_7$ gauge symmetry would be effectively restored, and the expected extra gauge bosons %should
would
become  relatively light, so there is some sense in which gauge coupling unification
might be expected.  
Note, however, that the quantization of flux means that the background
flux will give a mass scale $m_\mathrm{KK}=1/l_{\rm KK}$, where $l_{{\rm KK}}$ is
the compactification scale, so that this unification only occurs much above
the KK scale.  Furthermore,
in the nonperturbative F-theory regime, where there is no
weakly coupled description, it is not clear that the 7-brane
world-volume theory can be meaningfully separated from string theory
in the bulk space.
Thus, we do not necessarily expect unification even at the
compactification scale.
To understand some of the issues, we first clarify the meaning of gauge coupling
unification in our string theory context.

There are two separate aspects. First at the GUT scale
$M_\mathrm{GUT}$,\footnote{In the string theory context, $M_\mathrm{GUT}$ may be around the KK scale or string scale depending on model details.} the gauge
couplings in our models are clearly unified if flux breaking is absent.
The coupling is given by the volume of the gauge divisor:
\begin{equation}
\label{eq:divisorvol}
\frac{1}{g^{2}}\simeq\mathrm{vol}\left(\Sigma\right)\,.
\end{equation}
It is estimated from observations that $1/\alpha\simeq 24$ at 
$M_\mathrm{GUT}$ \cite{Martin:1997ns}; we simply assume that the divisor volume is stabilized 
to this particular value by certain mechanisms. On the other hand,
the remainder flux breaks $\SU(5)$ to $\gsm$ and 
induces some splitting of gauge couplings at $M_\mathrm{KK}$. 
It is 
then important to understand such splitting and determine its
size. Such
splitting has been understood in type IIB models \cite{Blumenhagen:2008aw,Mayrhofer:2013ara}:
the splitting between the $\SU(3)$ and $\SU(2)$ gauge couplings is
\begin{equation}
\label{eq:splitting}
\frac{1}{\alpha_{2}\left(M_\mathrm{GUT}\right)}-\frac{1}{\alpha_{3}\left(M_\mathrm{GUT}\right)}\simeq-\frac{1}{10g_s}\left[c_{1}\left(L_{3}\right)\right]^{2}=\frac{1}{5g_s}(n_{Q'}+1)\,,
\end{equation}
where $g_s$ is the string coupling, and $n_{Q'}$ is the number of vector-like pairs in the exotic 
representation $(\mathbf 3,\mathbf 2)_{-5/6}$; recall that we have set $n_{Q'}=0$ in previous sections. There is also a unification-like relation
\begin{equation}
\label{eq:unificationlike}
    \frac{1}{\alpha_{Y}\left(M_\mathrm{GUT}\right)}=\frac{5}{3}\frac{1}{\alpha_{1}\left(M_\mathrm{GUT}\right)}=\frac{1}{\alpha_{2}\left(M_\mathrm{GUT}\right)}+\frac{2}{3}\frac{1}{\alpha_{3}\left(M_\mathrm{GUT}\right)}\,.
\end{equation}
All the above, however, cannot be directly applied to F-theory models,
especially when the models, like our $E_7$ models, are intrinsically 
strongly coupled and have no type IIB limit.
This is because the axio-dilaton varies over the internal space and the
meaning of the $1/g_s$ correction is no longer clear. The worldvolume 
theory, which was used to derive the type IIB result, also needs to be
reconsidered in F-theory setups. In addition, there may be large
stringy threshold corrections to the gauge kinetic functions
due to the strong coupling nature of these models. All these subtleties imply that the splitting at 
$M_\mathrm{GUT}$ may not be small even if we set $n_{Q'}=0$.

Next, at scales lower than $M_\mathrm{GUT}$, the RG flow
of the SM gauge couplings are affected by the vector-like
exotics. The RG flow depends on both the representations
and the masses of the vector-like exotics. Since the remainder flux
already breaks the GUT group at $M_\mathrm{GUT}$, it is possible that
some vector-like exotics are light and do not form GUT multiplets. These
exotics seriously alter the RG flow and may ruin gauge coupling 
unification. The existence of such exotics, however, depends crucially on
uncontrolled aspects of the models such as SUSY breaking. It is also
possible that the presence of these exotics compensates the above splitting
at $M_\mathrm{GUT}$ and makes the couplings apparently unified from the
bottom-up perspective. Therefore, so far we cannot make any definite
statement on how the vector-like spectrum may affect the RG flow.

In conclusion, the gauge couplings in our models are affected by a
number of uncontrolled aspects, thus gauge coupling unification is not
guaranteed in our models. From this perspective, the unification of 
the observed gauge couplings in ordinary MSSM looks like an accident if
our models really describe our Universe. Nevertheless, this conclusion
mainly comes from our inability to compute non-topological details of our
models. A more careful string theory analysis in the future may reveal 
that the observed unification is in fact not an accident at all.

\section{Explicit global constructions of $E_7$ GUTs}
\label{sec:ExplicitConstruction}

In all the above sections, we have written down many necessary constraints
on the geometry and fluxes for constructing semi-realistic $E_7$ GUTs in
F-theory. It remains important to see whether all these constraints can
be satisfied simultaneously within a 4D F-theory model. In this section,
we provide an explicit global construction of such a model, using the
tools of toric hypersurfaces. The construction here is a 
generalization of that in \cite{Braun:2014pva}. It is also the first
explicit example of a rigid $E_7$ GUT
% instead of the 
 (rigid $E_6$ GUTs were presented in 
\cite{Li:2022aek}). Although we only present a single example here,
the same construction can be generalized to large class of F-theory
compactifications. Before writing down such an explicit model, it is
useful to first review the geometric and flux constraints we want to
achieve:
\begin{itemize}
\item $\Sigma$ as a del Pezzo surface supporting both rigid $E_{7}$ 
(with effective $-K_\Sigma$)
and
hypercharge flux, and $C_{\mathbf{56}}=-\Sigma\cdot\left(4K_{B}+3\Sigma\right)$
as a $\mathbb{P}^{1}$, to enable explicit computations and interesting
phenomenology. The first requirement demands that $\Sigma$ is a rigid
divisor on a non-toric base.
\item The general flux constraints in \S\ref{subsec:FluxesIntro}: flux quantization, primitivity for
vertical flux, and tadpole cancellation. In particular, we should
look for flux configurations with minimal tadpole.
\item A vertical flux breaking $E_{7}\rightarrow\SU(5)$ and a remainder
(hypercharge) flux breaking $\SU(5)\rightarrow G_{\mathrm{SM}}$.
In particular we need $r\geq4$, see \S\ref{subsec:FluxBreaking}.
\item $\chi_{\left(\mathbf{3},\mathbf{2}\right)_{1/6}}^{\mathbf{133}}=0$ and $\chi_{\left(\mathbf{3},\mathbf{2}\right)_{1/6}}^{\mathbf{56}}=3$ for the total chiral
spectrum.
\item All three families of chiral $\left(\bar{\mathbf 3},\mathbf 1\right)_{1/3}$ and $\left(\mathbf 1,\mathbf 2\right)_{-1/2}$
coming from the same copy, i.e. Eq.\ (\ref{eq:56samecopy}),
to avoid exotic vector-like spectrum and couplings.
\item The copy of bulk $\left(\bar{\mathbf 3},\mathbf 1\right)_{1/3}$ and $\left(\mathbf 1,\mathbf 2\right)_{-1/2}$
that interacts with the chiral matter
being itself non-chiral, i.e. Eq.\ (\ref{eq:133samecopy}), to avoid light vector-like exotics.
\item $\left[c_{1}\left(L_{3}\right)\right]^{2}=-2$ for hypercharge flux,
to remove the exotic $\left(\mathbf 3,\mathbf 2\right)_{-5/6}$.
\end{itemize}
As always, we set the flux associated to the non-flat fiber to zero.

Now we write
down an explicit F-theory model that satisfies all the above constraints.
As in \cite{Li:2022aek}, we choose the base $B$ through the following
procedure. We start with an auxilliary toric threefold $A$ with $h^{1,1}(A)=4$.
Then the ambient fourfold $X$ is a $\mathbb P^1$-bundle over $A$ with a
certain normal bundle, and $B$ is a certain hypersurface in $X$. The
geometry of $B$ can be analyzed using the techniques in Appendix \ref{appendix:Hypersurfaces}.
With appropriate choices in the above procedure, we can construct $B$
containing a rigid $\Sigma$ with $r=4$ and nontrivial remainder flux.

\begin{table}[t]
\centering
\begin{tabular}{|c|c|}
\hline 
Toric ray & Divisor \tabularnewline
\hline
$(1,0,0,0)$ & $F_{E_1}+F_F$ \tabularnewline
\hline
$(0,1,0,0)$ & $F_{E_2}+F_F$ \tabularnewline
\hline
$(-1,-1,-1,0)$ & $F_F$ \tabularnewline
\hline
$(-1,0,-1,-3)$ & $F_{E_1}$ \tabularnewline
\hline
$(0,-1,-1,0)$ & $F_{E_2}$ \tabularnewline
\hline
$(0,0,-1,-4)$ & $F_\sigma$ \tabularnewline
\hline
$(0,0,1,0)$ & $F_\sigma+F_{E_1}+F_{E_2}+F_F$ \tabularnewline
\hline
$(0,0,0,-1)$ & $\sigma_A$ \tabularnewline
\hline
$(0,0,0,1)$ & $\sigma_A+4F_\sigma+3F_{E_1}$ \tabularnewline
\hline 
\end{tabular}
\caption{The toric rays and the corresponding divisors in the toric
construction of the ambient fourfold $X$.}
\label{tableoftoricrays}
\end{table}

Let us first construct the ambient space $X$. 
We choose $A$ to be a 
$\mathbb{P}^{1}$-bundle over the del Pezzo surface $dP_{2}$, which has
a toric description.
Let us first introduce the notations. Within $dP_{2}$ i.e. blowup of
$\mathbb{P}^{2}$ at two generic points, let $e_{1},e_{2}$ be the
exceptional curves from the blowup, and $h=f+e_{1}+e_{2}$ be the
hyperplane. The intersection numbers are $f^{2}=e_{1}^{2}=e_{2}^{2}=-1,f\cdot e_{1}=f\cdot e_{2}=1,e_{1}\cdot e_{2}=0$.
Now on $A$, we denote $\sigma$ as the $dP_{2}$ section and $E_{1},E_{2},F$
as the $\mathbb{P}^{1}$-fibers along $e_{1},e_{2},f$ respectively.
We choose the normal bundle of the $\mathbb{P}^{1}$-bundle to be $N_{\sigma}=-h$,
and the anticanonical class is $-K_{A}=2\sigma+3E_{1}+3E_{2}+4F$.
The intersection numbers on $A$ follow straightforwardly from those
on $dP_{2}$ and the relation $\sigma\cdot\left(\sigma+F+E_{1}+E_{2}\right)=0$.
Finally we let the fourfold $X$ be a $\mathbb{P}^{1}$-bundle over $A$
with normal bundle $N_{A}=-4\sigma-3E_{1}$. We denote $\sigma_{A}$ as the section
and $F_{I}$ be the fiber along $I\in\left\{ \sigma,E_{1},E_{2},F\right\} $.
The anticanonical class is $-K_{X}=2\sigma_{A}+6F_{\sigma}+6F_{E_{1}}+3F_{E_{2}}+4F_{F}$.
Again, the intersection numbers follow from those on $A$ and the
relation $\sigma_{A}\cdot\left(\sigma_{A}+4F_{\sigma}+3F_{E_{1}}\right)=0$. Note that with these choices of normal bundles, there is
a unique triangulation such that $X$ is a 
smooth and projective toric variety. The toric rays of $X$ are listed in Table \ref{tableoftoricrays}.

We now choose the threefold base $B$ as a hypersurface in $X$ with
irreducible class $B=\sigma_{A}+5F_{\sigma}+5F_{E_{1}}+2F_{E_{2}}+3F_{F}$.
By abuse of notation, we use $B$ to denote both the base and its divisor
class in $X$.
By adjunction $-K_{B}=B\cdot\left(\sigma_{A}+F_{\sigma}+F_{E_{1}}+F_{E_{2}}+F_{F}\right)$.
Using the techniques in Appendix \ref{appendix:Hypersurfaces}, one can
check that $h^{1,1}(B)=h^{1,1}(X)=5$. 
%In other words,
In particular, in this situation
 the divisors of $B$ are spanned by intersections in $X$. The
intersection numbers
of these divisors relevant to our purpose are
\begin{equation}
B\cdot\sigma_{A}\cdot F_{I}\cdot F_{J}=\left(\begin{array}{cccc}
-2 & 1 & 1 & 0\\
1 & -1 & 0 & 1\\
1 & 0 & -1 & 1\\
0 & 1 & 1 & -1
\end{array}\right)\,.
\end{equation}

Now we consider the divisor $\Sigma=B\cdot\sigma_{A}$. It is also
the hypersurface in $A$ with class $\sigma+2E_{1}+2E_{2}+3F$. This
class is irreducible and does not have any base locus (considered as a
hypersurface in $A$),
 so it is a
well-defined %rigid
irreducible gauge divisor. We compute
\begin{align}
-K_{\Sigma}
& =
B\cdot\sigma_{A}\cdot\left(F_{\sigma}+F_{E_{1}}+F_{E_{2}}+F_{F}\right)\nonumber \\
 & =\sigma_{A}\cdot\left(2F_{\sigma}\cdot F_{E_{1}}+2F_{\sigma}\cdot F_{E_{2}}+3F_{\sigma}\cdot F_{F}+3F_{E_{1}}\cdot F_{F}\right)\,,\\
N_{\Sigma} & =B\cdot\sigma_{A}^{2}\nonumber \\
 & =-\sigma_{A}\cdot\left(7F_{\sigma}\cdot F_{E_{1}}+4F_{\sigma}\cdot F_{E_{2}}+8F_{\sigma}\cdot F_{F}+3F_{E_{1}}\cdot F_{F}\right)\,.
\end{align}
Therefore by Eq.\ (\ref{eq:nonHiggsable}), we see that $\Sigma$ is indeed a rigid divisor supporting
$E_{7}$. The matter curve is
\begin{align}
C_{\mathbf{56}}  & =\Sigma \cdot (-4K_B - 3 \Sigma)\\
& =B\cdot\sigma_{A}\cdot\left(\sigma_{A}+4F_{\sigma}+4F_{E_{1}}+4F_{E_{2}}+4F_{F}\right)\nonumber \\
 & =B\cdot\sigma_{A}\cdot\left(F_{E_{1}}+4F_{E_{2}}+4F_{F}\right)\,.
\end{align}
Notice that the divisor $E_{1}+4E_{2}+4F$ in $A$ is also
irreducible and does not have any base locus. Therefore, $C_{\mathbf{56}}$
is also irreducible with genus
\begin{equation}
g=1+\frac{1}{2}C_{\mathbf{56}}\cdot\left(C_{\mathbf{56}}+K_{\Sigma}\right)=0\,,
\end{equation}
which means that the matter curve is simply a $\mathbb{P}^{1}$.
%On $\Sigma$, this matter curve self-intersects at $C_\mathbf{56}^2=7$ points,
%so it can support a rank-3 Yukawa matrix without nonperturbative effects.

We can now study the constraints on vertical flux. First, we study
primitivity by expanding the K\"ahler
form of $B$ using a basis of base divisors:
\begin{align}
\left[J_{B}\right]= & B\cdot(t_{1}\left(F_{E_{1}}+F_{F}\right)+t_{2}\left(F_{E_{2}}+F_{F}\right)+t_{3}\left(F_{E_{1}}+F_{E_{2}}+F_{F}\right)\nonumber \\
 & +t_{4}\left(F_{\sigma}+F_{E_{1}}+F_{E_{2}}+F_{F}\right)+t_{5}\left(\sigma_{A}+4F_{\sigma}+4F_{E_{1}}+4F_{E_{2}}+4F_{F}\right)\,,
\end{align}
where $t_{1},t_{2},t_{3},t_{4},t_5$ are linear
combinations of K\"ahler moduli,
 and may be negative
inside the K\"ahler cone of $B$
in general. While determining the
exact K\"ahler
cone of a hypersurface in a toric variety can be subtle, the K\"ahler cone of $B$ must contain that of $X$
 \cite{Demirtas:2018akl}.
For simplicity, we look for a solution of the primitivity constraints in
the K\"ahler cone of $X$ only.
By a direct toric computation, one can check that the K\"ahler cone of
$X$ is
given by $t_{1},t_{2},t_{3},t_{4},t_{5}>0$. The independent $S_{i\alpha}$
are $S_{i\sigma},S_{iE_{1}},S_{iE_{2}},S_{iF}$, where we have simplified
the notation and denoted $S_{i\left(B\cdot F_{I}\right)}$ as $S_{iI}$.
The primitivity condition is then
\begin{equation}
\label{eq:explicitprimitivity}
t_{1}\left(\phi_{iE_{2}}+\phi_{i\sigma}\right)+\left(t_{2}+3t_{5}\right)\left(\phi_{iE_{1}}+\phi_{i\sigma}\right)+\left(t_{3}+t_{5}\right)\left(\phi_{iF}+2\phi_{i\sigma}\right)+t_{4}\left(\phi_{iF}+\phi_{iE_{1}}+\phi_{iE_{2}}\right)=0\,,
\end{equation}
for all $i$. A necessary but not sufficient condition for satisfying
primitivity is
that there must be
some coefficients in Eq.\ (\ref{eq:explicitprimitivity}) with
opposite signs for each $i$.
Below we will find explicit solutions to primitivity and check that the
solutions are within the K\"ahler cone.

Next, we require the total chiral indices to be
\begin{gather}
\chi^{\mathbf{133}}_{(\mathbf 3,\mathbf 2)_{1/6}}=-2\left(n_{F}+n_{E_{1}}+n_{E_{2}}\right)=0\,,\\
\chi^{\mathbf{56}}_{(\mathbf 3,\mathbf 2)_{1/6}}=-n_{F}-3n_{E_{1}}-5n_{\sigma}=3\,.
\end{gather}
Here the $n_I$ parameterize the fluxes through Eq. (\ref{eq:phi-n}).
To understand what values of $n_{I}$ we should turn on to get the right
total chiral spectrum,
we should first look at flux quantization, since for $E_{7}$ models
$c_{2}(\hat{Y})$ is not necessarily even. We can calculate $c_2(\hat Y)$
using the techniques in \cite{Jefferson:2021bid}, which involve picking
a particular resolution of the $E_7$ models, but the parity of 
$c_2(\hat Y)$ is expected to be resolution-independent. We outline the
procedure in Appendix \ref{appendix:Quantization}, while here we only apply the result, which
tells us that we can turn on half-integers $n_{E_{1}},n_{F}$ and integers
$n_{E_{2}},n_{\sigma}$ (similarly for $\phi_{6\alpha}$) to guarantee flux
quantization. Note that these may not be the only choices of $n_I$, 
since the structure of $H^4(\hat Y,\mathbb Z)$ is subtle and may 
include elements with fractional coefficients. Also note that these choices of $n_I$
do not necessarily mean that the vertical flux does not belong to $H^{4}(\hat{Y},\mathbb{Z})$,
since as we will see its tadpole is still integer. Now with these choices
of $n_I$, we see that
an almost minimal flux configuration $\left(n_{\sigma},n_{E_{1}},n_{E_{2}},n_{F}\right)=\left(0,-3/2,0,3/2\right)$
already gives the above two chiral indices and is consistent with primitivity.

There are more flux conditions that constraint the values of $\phi_{5\alpha},\phi_{6\alpha}$.
First, by flux quantization we should turn on half-integers $\phi_{6E_{1}},\phi_{6F}$
and integers for the remaining parameters. Primitivity still constraints
their values nontrivially. Moreover, to put the chiral matter into
the copy $\left(b_{4},b_{5},b_{6}\right)=\left(1/2,1,1/2\right)$,
we need to impose Eq.\ (\ref{eq:56samecopy}), or in terms of flux parameters
\begin{equation}
-\phi_{5F}-3\phi_{5E_{1}}-5\phi_{5\sigma}=12\,,\quad-\phi_{6F}-3\phi_{6E_{1}}-5\phi_{6\sigma}=6\,.
\end{equation}
We also need to avoid light vector-like exotics in the bulk
copy $\left(b_{4},b_{5},b_{6}\right)=\left(2,2,1\right)$. Eq.\ (\ref{eq:133samecopy}) then leads
to
\begin{equation}
\phi_{5F}+\phi_{5E_{1}}+\phi_{5E_{2}}=0\,.
\end{equation}
These are mild linear constraints on the flux parameters $\phi_{5\alpha},\phi_{6\alpha}$.
Although there are multiple solutions to these linear constraints, we should seek for solutions with minimal
tadpole. By a brute force search, we find that one of the optimal
solutions is

\begin{equation}
\left(\phi_{5\sigma},\phi_{5E_{1}},\phi_{5E_{2}},\phi_{5F},\phi_{6\sigma},\phi_{6E_{1}},\phi_{6E_{2}},\phi_{6F}\right)=\left(0,-5,2,3,1,-\frac{7}{2},1,-\frac{1}{2}\right)\,,
\end{equation}
which consistently stabilizes the K\"ahler moduli at $t_{1}=t_{2}+3t_{5}=t_{3}+t_{5}=3t_{4}$.
Together with $n_{I}$, this vertical flux gives a tadpole
\begin{equation}
\frac{1}{2}\left[G_{4}^{\mathrm{vert}}\right]\cdot\left[G_{4}^{\mathrm{vert}}\right]=32\,.
\end{equation}
As a comparison, if we do not impose Eqs. 
(\ref{eq:133samecopy}) and (\ref{eq:56samecopy}) i.e. primitivity is the only constraint on $\phi_{5\alpha},\phi_{6\alpha}$,
the minimal tadpole is
\begin{equation}
\frac{1}{2}\left[G_{4}^{\mathrm{vert}}\right]\cdot\left[G_{4}^{\mathrm{vert}}\right]=20\,,
\end{equation}
given by e.g.
\begin{equation}
\left(\phi_{5\sigma},\phi_{5E_{1}},\phi_{5E_{2}},\phi_{5F},\phi_{6\sigma},\phi_{6E_{1}},\phi_{6E_{2}},\phi_{6F}\right)=\left(0,-4,-1,5,1,-\frac{5}{2},-1,\frac{3}{2}\right)\,.
\end{equation}
Therefore, we see that the vertical flux we need to turn on is slightly
non-generic.

Let us now turn to remainder flux. It can be shown that $\Sigma$ is a del
Pezzo surface $dP_6$ and supports remainder flux. First recall that
$\Sigma$ is a hypersurface in $A$
with class $\sigma+2E_1+2E_2+3F$. In other words, $\Sigma$ is the vanishing
locus
\begin{equation} \label{Sigmalocus}
    xP+yP'=0\,,
\end{equation}
in $A$, where $P,P'$ are sections of $\mathcal O_A(2E_1+2E_2+3F),\mathcal O_A(E_1+E_2+2F)$ respectively, and $x,y$
are the homogeneous coordinates of the $\mathbb P^1$ in $A$. For generic
points in the $dP_2$, Eq.\ (\ref{Sigmalocus}) has a unique
solution, representing a single point in $\mathbb P^1$. On the other hand,
there are $\left(2e_{1}+2e_{2}+3f\right)\cdot\left(e_{1}+e_{2}+2f\right)=4$ points in $dP_2$ such that $P=P'=0$,
and Eq.\ (\ref{Sigmalocus}) represents the whole $\mathbb P^1$.
Therefore, the geometry of $\Sigma$ is $dP_2$ blown up in 4
generic points i.e. a $dP_6$.

To construct the remainder flux, notice that the four exceptional
curves on $\Sigma$ from blowing up $dP_2$ (denoted by
$e_3$ to $e_6$) are all $\mathbb P^1$
fibers in $A$, hence all have the same class in $B$.  Therefore, we can
choose e.g. $C_{\mathrm{rem}}=e_3-e_4$ (or any difference $e_i-e_j$ for 
distinct $i,j=3,4,5,6$) with $C_{\mathrm{rem}}^2=-2$, and turn on the 
remainder flux specified in \S\ref{subsec:vectorexotics}. Therefore, we need a total tadpole of
36 to satisfy all the
flux constraints. Using the techniques in \cite{Jefferson:2021bid}, we 
find that $\chi(\hat{Y})=1176$
and $\chi(\hat{Y})/24=49>36$, so tadpole cancellation
is satisfied. Unfortunately, there seems to be not much room to achieve
full moduli stabilization, but the situation should improve if we
consider more complicated geometries.

Having the full flux configuration, it is now straightforward to also
compute the vector-like spectrum. For simplicity again we ignore the
uncharged singlets. Since $C_{\mathbf{56}}$ is a $\mathbb{P}^{1}$, all vector-like
pairs comes from the bulk of $\Sigma$. Using the
formula for $n_{\beta}$ in \S\ref{subsec:Vectorlikematter}, we get the vector-like spectrum as in
Table \ref{tableofvector}. It is clear that there are too many doublet
and triplet Higgs that are not inert, and it is important to understand
how the mass hierarchy is produced, such that we only see one doublet
Higgs (pair) at the electroweak scale. There are also a number of light and inert vector-like exotics from the table.

\begin{table}[t]
\centering
\begin{tabular}{|c|c|c|c|}
\hline 
$\left(\mathbf 1,\mathbf 2\right)_{-1/2,3,2,1}$ & 18 & $\left(\mathbf 1,\mathbf 2\right)_{1/2,-3,-2,-1}$ & $\mathbf{18}$\tabularnewline
\hline 
$\left(\mathbf 1,\mathbf 2\right)_{-1/2,-2,-2,-1}$ & $\mathbf{4}$ & $\left(\mathbf 1,\mathbf 2\right)_{1/2,2,2,1}$ & 4\tabularnewline
\hline 
$\left(\mathbf 1,\mathbf 2\right)_{-1/2,-2,-1,-1}$ & 19 & $\left(\mathbf 1,\mathbf 2\right)_{1/2,2,1,1}$ & 16\tabularnewline
\hline 
$\left(\mathbf 1,\mathbf 2\right)_{-1/2,-2,-1,0}$ & 24 & $\left(\mathbf 1,\mathbf 2\right)_{1/2,2,1,0}$ & 27\tabularnewline
\hline 
$\left(\bar{\mathbf 3},\mathbf 1\right)_{1/3,3,2,1}$ & 21 & $\left(\mathbf 3,\mathbf 1\right)_{-1/3,-3,-2,-1}$ & $\mathbf{21}$\tabularnewline
\hline 
$\left(\bar{\mathbf 3},\mathbf 1\right)_{1/3,-2,-2,-1}$ & $\mathbf{3}$ & $\left(\mathbf 3,\mathbf 1\right)_{-1/3,2,2,1}$ & 3\tabularnewline
\hline 
$\left(\bar{\mathbf 3},\mathbf 1\right)_{1/3,-2,-1,-1}$ & 16 & $\left(\mathbf 3,\mathbf 1\right)_{-1/3,2,1,1}$ & 13\tabularnewline
\hline 
$\left(\bar{\mathbf 3},\mathbf 1\right)_{1/3,-2,-1,0}$ & 21 & $\left(\mathbf 3,\mathbf 1\right)_{-1/3,2,1,0}$ & 24\tabularnewline
\hline 
$\left(\bar{\mathbf 3},\mathbf 1\right)_{-2/3,-1,-1,-1}$ & 6 & $\left(\mathbf 3,\mathbf 1\right)_{2/3,1,1,1}$ & 3\tabularnewline
\hline 
$\left(\bar{\mathbf 3},\mathbf 1\right)_{-2/3,-1,-1,0}$ & 1 & $\left(\mathbf 3,\mathbf 1\right)_{2/3,1,1,0}$ & 4\tabularnewline
\hline 
$\left(\bar{\mathbf 3},\mathbf 1\right)_{-2/3,-1,0,0}$ & 19 & $\left(\mathbf 3,\mathbf 1\right)_{2/3,1,0,0}$ & 19\tabularnewline
\hline 
$\left(\mathbf 3,\mathbf 2\right)_{1/6,-1,-1,-1}$ & 5 & $\left(\bar{\mathbf 3},\mathbf 2\right)_{-1/6,1,1,1}$ & 2\tabularnewline
\hline 
$\left(\mathbf 3,\mathbf 2\right)_{1/6,-1,-1,0}$ & 0 & $\left(\bar{\mathbf 3},\mathbf 2\right)_{-1/6,1,1,0}$ & 3\tabularnewline
\hline 
$\left(\mathbf 3,\mathbf 2\right)_{1/6,-1,0,0}$ & 16 & $\left(\bar{\mathbf 3},\mathbf 2\right)_{-1/6,1,0,0}$ & 16\tabularnewline
\hline 
$\left(\mathbf 3,\mathbf 2\right)_{-5/6,0,0,0}$ & 0 & $\left(\bar{\mathbf 3},\mathbf 2\right)_{5/6,0,0,0}$ & 0\tabularnewline
\hline 
$\left(\mathbf 1,\mathbf 1\right)_{1,-1,-1,-1}$ & 6 & $\left(\mathbf 1,\mathbf 1\right)_{-1,1,1,1}$ & 3\tabularnewline
\hline 
$\left(\mathbf 1,\mathbf 1\right)_{1,-1,-1,0}$ & 1 & $\left(\mathbf 1,\mathbf 1\right)_{-1,1,1,0}$ & 4\tabularnewline
\hline 
$\left(\mathbf 1,\mathbf 1\right)_{1,-1,0,0}$ & 15 & $\left(\mathbf 1,\mathbf 1\right)_{-1,1,0,0}$ & 15\tabularnewline
\hline 
\end{tabular}
\caption{The representations and multiplicities of vector-like matter
originated from the adjoint $\mathbf{133}$ on the bulk of gauge divisor.
Only the bold multiplicities correspond to fields interacting with the
SM chiral matter without exponential suppression. All the other fields
are inert vector-like exotics. Note that there are 
nontrivial linear relations between these numbers implied
by the formulas in \S\ref{subsec:Vectorlikematter}.}
\label{tableofvector}
\end{table}

In conclusion, we have obtained an explicit F-theory model with the SM gauge
group from rigid $E_{7}$, three families of SM chiral
matter with qualitatively standard Yukawa couplings and suppressed proton decay,
and excess numbers of heavy Higgs with some doublet-triplet splitting.
The flux configuration requires some but not too much fine-tuning. We
emphasize again that most analysis in this section depends on the local
geometry only. We expect that many of the F-theory threefold bases 
contain local geometries that are the same or similar
to the above, so this explicit
 construction can be easily generalized to large class of 4D F-theory
compactifications.

\section{Comparing with other F-theory constructions in the literature}
\label{sec:Comparison}

As we have pointed out a number of times in the previous sections, the
$E_7$ models have many features that are distinct from previous SM-like
constructions in F-theory. This distinction makes the $E_7$ models a
new interesting class of models to be studied in depth in the future.
In this section, we %elaborate the distinction in details. We 
explain in  more detail
some of the specific differences between
the models presented here and the F-theory models with tuned $\gsm$ or
$\SU(5)$ reviewed briefly in \S\ref{sec:Intro}, as well as the rigid $E_6$
GUTs.

\subsection{Tuned models}
\label{subsec:comparetuned}

There have been many SM-like F-theory constructions using tuned gauge
groups such as $\gsm$ or $\SU(5)$ (again, for reviews see
\cite{WeigandTASI,HeckmanReview,Cvetic:2022fnv,Marchesano:2022qbx}). The
most obvious difference between those
constructions and the $E_7$ models presented here has been discussed
in \S\ref{sec:Intro}: namely, fine-tuning of many complex structure
moduli is required to obtain $\gsm$ or $\SU(5)$ geometrically, while
the presence of rigid $E_7$ only depends on the normal bundle of
$\Sigma$ instead of any moduli.  It seems that rigid $E_7$ factors are relatively abundant
in the landscape. Although the measure on the landscape has never been
clear, a naive counting measure on the (singular) geometries suggests
a large exponential dominance of geometries supporting rigid $E_7$ factors over those supporting tuned gauge
factors.
Note, however, that it is possible that certain flux choices may in
some situations force
complex structure moduli to a tuned locus with an enhanced gauge
group; further investigation of this possibility is needed to clarify
the level of tuning really involved in geometrically tuned
constructions beyond the level suggested by the analysis of e.g., \cite{BraunWatariGenerations}.
Due to the  moduli-independence of the relevant gauge group, it may also be easier to
incorporate  a full analysis of moduli stabilization
 in the rigid models than the tuned
ones.

Another significant difference regards the Yukawa couplings. In many
SM-like F-theory constructions, some selection rules 
are required to get rid of exotic couplings. The Yukawa couplings
in tuned $\gsm$ or $\SU(5)$ models come from $CCC$-type couplings, and all the
matter fields are localized on matter curves. Therefore, the selection
rules are usually obtained by engineering a set of multiple matter curves
where different types of matter localize separately, or some additional
$U(1)$'s by tuning the global geometry. In the $E_7$ models,  selection
rules that remove exotic couplings
 automatically follow from the use of $\Sigma CC$-type couplings,
and easily separate the chiral matter from vector-like matter including the Higgs. There
are also approximate $\U(1)$'s from the St\"uckelberg mechanism that
arrive
without
additional tuning. Therefore,  the
selection rules needed to match expectations from observed physics
 are more easily
realized in the $E_7$ models than in other constructions. An example is the proton decay suppression
described in \S\ref{subsec:protondecay}.

The means of realizing the Higgs in the two classes of models is also
qualitatively different. In tuned $\gsm$ or $\SU(5)$, the Higgs comes
from some
vector-like matter on matter curves. Such a construction requires explicit
specification of the sheaf cohomology groups in Eq.\ 
(\ref{eq:mattercurvecohomologies}), which are in general very hard to
compute since they are moduli-dependent quantities. More exotic tools 
like root bundles \cite{Bies:2021nje,Bies:2021xfh,Bies:2022wvj,Bies:2023jqg, Bies:2023sfm} may also be needed in the construction.
 In many cases, there
is no Higgs in the low-energy theory unless some further tuning is done.
In $\SU(5)$, we also need the Higgs matter surfaces to have remainder
components, such that the hypercharge flux is present on the Higgs curves
and doublet-triplet splitting can be achieved. Generic matter surfaces,
however, are purely vertical 
unless further tuning on moduli is done, and
global examples of such scenarios are
 rare in the literature (see e.g. \cite{Braun:2014pva}). In contrast, in the $E_7$ models we can instead
realize the Higgs as bulk vector-like matter, which generically has
nonzero multiplicities that are easily calculated from the fluxes. 
In this situation,
there are 
already Higgs fields with some doublet-triplet splitting without any further tuning, but the issue becomes
having too many instead of too few Higgs fields. It is less clear how to
make one pair of the Higgs exponentially lighter in the $E_7$ models, 
while  in the tuned models there can be exactly one pair of Higgs, and thus
the way to obtain the Higgs hierarchy may be clearer.

Because of the use of flux breaking and $E_7$, there are many further
differences between these two classes of models in terms of computational abilities. First, in tuned $\gsm$ 
or $\SU(5)$ the total chiral spectrum is controlled by one flux parameter
only. In many cases the chiral indices contain large prefactors, and 
three generations of chiral matter cannot easily be obtained using integer 
fluxes, unless more nontrivial (and less completely
understood)
quantization conditions are used, as in
\cite{CveticEtAlQuadrillion,Jefferson:2022yya}. In the $E_7$ 
models, however, many flux parameters from vertical flux breaking 
contribute to the chiral indices, giving a linear Diophantine structure.
As a result, it is natural to get three generations of 
chiral matter just by generic integer fluxes. 

Specifically for $\SU(5)$, the removal of exotic $(\mathbf 3,\mathbf
2)_{-5/6}$ appears to be harder.  This is because the form of hypercharge flux
is completely fixed to be $\phi_{ir}\propto (2,4,6,3)$ and there is no
free flux parameter like $\phi_{4r}$ as in the $E_7$ models. Therefore,
there must be a factor of 5 in $c_1(L_3)$, and we need to use fractional
line bundles to satisfy the condition $[c_1(L_3)]^2=-2$ for removing the
exotic $(\mathbf 3,\mathbf 2)_{-5/6}$. In contrast, as in \S\ref{subsec:vectorexotics},
this condition in $E_7$ is already satisfied by a fairly likely choice of
integer remainder flux. On the other hand, there can be some controlled
scenario in $\SU(5)$ where all the vector-like exotics are removed, while
in $E_7$ we cannot remove most of the vector-like exotics; we can at best
arrange them into inert fields.

In conclusion, we see that the $E_7$ models are not only more natural in
the landscape, but also possess a number of phenomenological advantages
over the tuned models. These models demonstrate how using naturalness
as the guiding philosophy can help us discover more semi-realistic 
features in the landscape. On the other hand, these models still have 
their own shortcomings especially regarding the heavy mass spectrum, due
to the lack of computational technologies.

\subsection{Rigid $E_6$ GUTs}
\label{subsec:compareE6}

In \cite{Li:2021eyn,Li:2022aek}, it was proposed that the rigid 
construction of SM-like models works equally well for both $E_7$ and 
$E_6$, since these two gauge groups are similarly abundant in the 
landscape. While the two GUT groups share some features such as the 
naturalness of the gauge group and three generations of chiral matter, 
$E_6$ behaves differently when coming to Yukawa couplings. While $E_7$
models do not have any $CCC$-type couplings, in $E_6$ models the gauge 
group only gets enhanced to $E_8$ at codimension-3 singularities, which
are well-defined Yukawa points giving $CCC$-type couplings. Moreover,
the branching rules from $E_6$ to $\gsm$ including the additional $\U(1)$ charges
$(b_4,b_5)$ are
\begin{align}
\mathbf{27} & \rightarrow\left(\mathbf 1,\mathbf 1\right)_{0,5/3,4/3}+\left(\mathbf 1,\mathbf 1\right)_{0,5/3,1/3}+\left(\mathbf 1,\mathbf 1\right)_{1,2/3,1/3}\nonumber \\
 & +\left(\mathbf 3,\mathbf 2\right)_{1/6,2/3,1/3}+\left(\bar{\mathbf 3},\mathbf 1\right)_{-2/3,2/3,1/3}+\left(\bar{\mathbf 3},\mathbf 1\right)_{1/3,-1/3,1/3}+\left(\bar{\mathbf 3},\mathbf 1\right)_{1/3,-1/3,-2/3}\nonumber \\
 & +\left(\bar{\mathbf 3},\mathbf 1\right)_{1/3,4/3,2/3}+\left(\mathbf 1,\mathbf 2\right)_{-1/2,-1/3,1/3}+\left(\mathbf 1,\mathbf 2\right)_{-1/2,-1/3,-2/3}+\left(\mathbf 1,\mathbf 2\right)_{-1/2,4/3,2/3}\,,\\
\mathbf{78} & \rightarrow\left(\mathbf 8,\mathbf 1\right)_{0,0,0}+\left(\mathbf 1,\mathbf 3\right)_{0,0,0}+3\times\left(\mathbf 1,\mathbf 1\right)_{0,0,0}\nonumber \\
 & +[\left(\mathbf 1,\mathbf 1\right)_{0,0,1}+\left(\mathbf 1,\mathbf 1\right)_{1,-1,0}+\left(\mathbf 1,\mathbf 1\right)_{1,-1,-1}+\left(\mathbf 3,\mathbf 2\right)_{-5/6,0,0}+\left(\mathbf 3,\mathbf 2\right)_{1/6,-1,0}+\left(\mathbf 3,\mathbf 2\right)_{1/6,-1,-1}\nonumber \\
 & +\left(\bar{\mathbf 3},\mathbf 1\right)_{-2/3,-1,0}+\left(\bar{\mathbf 3},\mathbf 1\right)_{-2/3,-1,-1}+\left(\bar{\mathbf 3},\mathbf 1\right)_{1/3,-2,-1}+\left(\mathbf 1,\mathbf 2\right)_{-1/2,-2,-1}+\mathrm{conjugates]}\,.
\end{align}
Unlike $E_7$, one can check that there is no suitable field on the bulk
of $\Sigma$ that can play the role of Higgs, so the Higgs must be localized
on the matter curve. The Yukawa couplings in $E_6$ models are more similar to the tuned models and many results in \S\ref{sec:Pheno} do not apply to $E_6$ models, while
$E_{6}$ models also suffer from many vector-like exotics. In this
sense, $E_{7}$ models are fundamentally different from any other F-theory
GUT models.

\section{Conclusion}
\label{sec:Conclusion}

In this paper, we have studied various phenomenological aspects of $E_7$
GUTs in 4D F-theory compactifications. These models were proposed in
\cite{Li:2021eyn,Li:2022aek} as a large class of natural SM-like
constructions, since rigid $E_7$ gauge factors
are moduli independent and common in the F-theory landscape. Vertical and remainder fluxes
are used to break $E_7$ to the SM gauge group, and appear fairly likely to induce three
generations of SM chiral matter. Here we have shown that the use of $E_7$
and flux breaking also naturally implies several more phenomenologically favorable
features, including suppression of proton decay, doublet-triplet 
splitting, and Higgs candidates with the right structure of
SM Yukawa couplings due to approximate global symmetries descending
from the $E_7$ Cartan generators. 
%The model also predicts the existence of several 
%inert vector-like exotics around TeV scale, which is a valid prediction 
%on experiments. 
For the first time, we have
written down an explicit global construction of such $E_7$ models that
achieve all the above features. The construction of these features is
qualitatively distinct from other SM-like constructions in the previous 
F-theory literature. In particular, only mild tuning on the discrete data
of the geometry and the flux background is involved in the construction.
The results in this paper give us strong hints towards
SM constructions in F-theory that are both realistic and natural. In
other words, these results appear to be compatible with the hypothesis
% the philosophy
 that our Universe 
can be described as a
 \emph{natural} solution in the string landscape.

These results, on the other hand, are still far from complete in realizing
the full details of the Standard Model in string theory. Since the $E_7$
models are inherently strongly coupled and there is extensive use of 
fluxes, the machinery for computing continuous parameters, or at least
the moduli dependence of continuous parameters, is very limited. While we
can more or less fully specify the discrete data in the $E_7$ models, our
arguments are at best qualitative when it comes to questions on Yukawa
couplings, mass scales, etc. Such limitations lead to a number of
shortcomings of the $E_7$ models, especially the unavoidable presence of
(mostly inert) vector-like exotics with masses not determined. The Higgs
hierarchy problem, i.e. the $\mu$-problem,
also remains unsolved in our models.

There are many challenges to
% is a long way of studies before
 answering these important questions.
First, we need to develop new tools beyond the ultra-local approach \cite{Font:2012wq,Font:2013ida,Marchesano:2015dfa,Carta:2015eoh} to
compute the moduli dependence of various quantities. After that, we still
need to understand more fundamental questions like the realization of
moduli stabilization and SUSY breaking in F-theory, which by themselves
are essential components of realistic model building. Solutions to these questions also
involve tackling some open problems such as computing the K\"ahler
potential in F-theory. All these tasks are particularly challenging when
there is no weakly coupled type IIB limit for the $E_7$ models.
Nonetheless, some insight into aspects like gauge coupling unification may be
possible by considering the 8D world-volume theory on the $E_7$
7-branes, where flux breaking should have a classical (if
nonperturbative) description.

There are also several extensions of the $E_7$ models presented in this
paper that should be investigated further. Throughout the paper, we have made several strong assumptions on
the geometry, such as restricting the matter curve to be $\mathbb P^1$,
to enable more interesting computations on the discrete data. It will
be interesting, although technically challenging, to relax these 
assumptions and explore more behavior of the $E_7$ models. It is also
important to study the statistics of these $E_7$ models in the F-theory
landscape. By scanning through a large set of F-theory bases and flux
configurations, we can quantitatively analyze the genericity of different
features of the $E_7$ models, which further sheds light on where our
Universe sits in the landscape.

We hope to address some of these questions in future studies.

\acknowledgments{We would like to thank Lara Anderson, Martin Bies, Mirjam Cvetic, James Gray, Daniel Harlow, Jonathan Heckman, Patrick Jefferson, Manki Kim, Paul Oehlmann, Jesse Thaler, and Andrew Turner for helpful discussions.
This work was supported by the
DOE under contract \#DE-SC00012567.}

\appendix

\section{Toric hypersurfaces}
\label{appendix:Hypersurfaces}

In this Appendix, we explain how to count $h^{1,1}$, or the number of
divisors, of a threefold hypersurface in an ambient toric 
fourfold, 
following the general
approach of Danilov and Khovanskii \cite{Danilov_1987}.\footnote{We thank Manki Kim for teaching us these techniques.} This
technique is useful in \S\ref{sec:ExplicitConstruction}.
To simplify the discussion, we focus on simple cases where there is a
triangulation such that both the ambient space and the hypersurface
are smooth. We also assume that the hypersurface does
not have any base locus.

The geometry of the hypersurface can be understood from its 
stratification. First we look at the stratification of the ambient space.
A $d$-dimensional toric variety is given by a 
disjoint union of algebraic tori $(\mathbb C^*)^k$, where $0\leq k\leq d$.
These algebraic tori, called strata, are associated with the cones of a
toric (polyhedral)
fan. To be more precise,
for a toric fan $\Sigma$ with $n$-dimensional cones
$\sigma^{(n)}\in\Sigma(n)$ (where $0\leq n\leq d$), the toric variety
$\mathbb P_\Sigma$ is given by
\begin{equation}
    \mathbb P_\Sigma=\coprod_n \coprod_{\sigma^{(n)}\in\Sigma(n)}T_{\sigma^{(n)}}\,,\quad T_{\sigma^{(n)}}\cong (\mathbb C^*)^{d-n}\,.
\end{equation}
Notice that the unique $\sigma^{(0)}$ corresponds to the prime stratum
$(C^*)^d$, which is the defining feature of toric varieties. The 
one-dimensional cones $\sigma^{(1)}$ are also given by the toric rays 
$\vec v$, associated with prime toric divisors $D_{\vec v}$.

Now consider a hypersurface $Z$ as a divisor in $\mathbb P_\Sigma$. We
abuse notation and use $Z$ to also denote its divisor class:
\begin{equation}
    Z=\sum_{\vec v} a_{\vec v} D_{\vec v}\,.
\end{equation}
Note that the prime toric divisors are not all independent and there
are multiple choices of the coefficients $a_{\vec v}$ for the same $Z$;
our final results are independent of such a choice. 
We assume
 that all
the strata of $\mathbb P_\Sigma$ intersect $Z$ transversely. Then $Z$
admits the following stratification
\begin{equation}
\label{eq:Zstratasigma}
    Z=\coprod_n \coprod_{\sigma^{(n)}\in\Sigma(n)}Z_{\sigma^{(n)}}\,,\quad Z_{\sigma^{(n)}}=Z\cap T_{\sigma^{(n)}}\,.
\end{equation}
Note that the dimension of the strata $Z_{\sigma^{(n)}}$ is $d-n-1$. To 
understand the geometry of $Z_{\sigma^{(n)}}$, it is useful to 
construct the Newton polytope $\Delta$ of $Z$
\begin{equation}
    \Delta=\left\{\vec m \mid \vec m\cdot\vec v\geq -a_{\vec v},\forall \vec v\in\Sigma(1)\right\}\,.
\end{equation}
The Newton polytope encodes information about the holomorphic sections
of the line bundle $\mathcal O_{\mathbb P_\Sigma}(Z)$. Below we restrict
to the case where this line bundle is big, i.e. $\Delta$ is
also $d$-dimensional.

The faces of $\Delta$ %also
 encode
 the geometry of $Z$ in the following
way. From $\Delta$ we can construct the so-called normal fan 
$\Sigma(\Delta)$, where each $k$-dimensional face $\Theta^{(k)}$ is
associated with a $(d-k)$-dimensional cone in $\Sigma(\Delta)$.\footnote{The
explicit construction of $\Sigma(\Delta)$ is more complicated but is
not important for our purpose.} The
resulting toric variety $\mathbb P_{\Sigma(\Delta)}$ is a blowdown of
$\mathbb P_\Sigma$,
 which is singular in general. The corresponding
blowdown of $Z$ is denoted by $Z(\Delta)$. An important fact is that 
$Z(\Delta)$ is an ample divisor in $\mathbb P_{\Sigma(\Delta)}$. Now
given the one-to-one correspondence between faces of $\Delta$ and cones
of $\Sigma(\Delta)$, we can write the stratification of $Z(\Delta)$ as
(again, assuming all strata of $\mathbb P_{\Sigma(\Delta)}$ intersect
$Z(\Delta)$ transversely)
\begin{equation}
    Z(\Delta)=\coprod_k\coprod_{\Theta^{(k)}}Z_{\Theta^{(k)}}\,.
\end{equation}
Note that the dimension of the strata $Z_{\Theta^{(k)}}$ is $k-1$. Now
including the blowups from $Z(\Delta)$ back to $Z$, the stratification of
$Z$ can be written as
\begin{equation}
\label{eq:ZstrataTheta}
    Z=Z_{\Theta^{(d)}}\coprod_{\Theta^{(d-1)}}Z_{\Theta^{(d-1)}}\coprod_{k\geq 2}\coprod_{\Theta^{(d-k)}}E_{\Theta^{(d-k)}}\times Z_{\Theta^{(d-k)}}\,,
\end{equation}
where
\begin{equation}
    E_{\Theta^{(d-k)}}=\coprod^{k-1}_{i=0}\left(\coprod(C^*)^i\right)\,,
\end{equation}
is the exceptional set associated with $\Theta^{(d-k)}$ resulting from 
the blowups. The geometry of $Z_{\sigma^{(n)}}$ can then be seen by
comparing Eqs. (\ref{eq:Zstratasigma}) and (\ref{eq:ZstrataTheta}).

A great advantage of studying the stratification of $Z$ is that the Hodge
numbers of $Z$ can be computed using the Hodge-Deligne numbers together with
the stratification \cite{Danilov_1987,Braun:2014xka} (see also \cite{Jefferson:2022ssj} for more recent review and 
applications). In our case where $Z$ is smooth, the Hodge-Deligne numbers
are just certain signed combinations of the Hodge numbers, but they behave
nicely under disjoint unions and products. One can then compute the
Hodge-Deligne numbers of $Z$ by combining those of its strata, which are
easy to get. Although the formulas for general Hodge numbers are more
complicated, it can be shown that for $d\geq 4$, $h^{1,1}(Z)$ is simply
given by counting the irreducible components of $Z_{\sigma^{(1)}}$. In
terms of $\Delta$, we should look at the faces
\begin{equation}
    \Theta_i=\left\{\vec m \mid \vec m\cdot\vec v\geq -a_{\vec v},\forall \vec v\neq \vec v_i;\vec m\cdot\vec v_i=-a_{\vec v_i}\right\}\,.
\end{equation}
All $\Theta_i$'s are nontrivial when $Z$ does not have any base locus,
but they can have different dimensions and contribute differently to
$h^{1,1}(Z)$:
\begin{itemize}
    \item $\mathrm{dim}(\Theta_i)=0$: $Z_{\sigma^{(1)}}$ is given by
    the components in $E_{\Sigma^{(0)}}\times Z_{\Sigma^{(0)}}$. For generic
    moduli, however, $Z_{\Sigma^{(0)}}$ is an empty set and such $\Theta_i$
    does not contribute to $h^{1,1}(Z)$.
    \item $\mathrm{dim}(\Theta_i)=1$: $Z_{\sigma^{(1)}}$ is given by
    the components in $E_{\Sigma^{(1)}}\times Z_{\Sigma^{(1)}}$. Notice
    that $Z_{\Theta_i}$ is a degree $n=l^*(\Theta_i)+1$ hypersurface
    in $\mathbb C^*$, where $l^*$ denotes the number of interior points. For
    generic moduli, this hypersurface is a collection of $n$
    points, so there are $n$ copies of an irreducible
    component in $Z_{\sigma^{(1)}}$, contributing $n$ to $h^{1,1}(Z)$.
    \item $\mathrm{dim}(\Theta_i)=k\geq 2$: $Z_{\sigma^{(1)}}$ is given by
    the components in $E_{\Sigma^{(k)}}\times Z_{\Sigma^{(k)}}$. We see that $Z_{\sigma^{(1)}}$ is irreducible, contributing 1 to $h^{1,1}(Z)$.
\end{itemize}
Finally, the above procedure overcounts $h^{1,1}(Z)$ by $d$, since there
are $d$ linear relations between prime toric divisors in $\mathbb P_\Sigma$. One can check that
the above procedure reproduces the famous Batyrev formula for toric
hypersurface Calabi-Yau manifolds \cite{batyrev1993dual}.

For applications in \S\ref{sec:ExplicitConstruction}, it is now clear that to obtain $h^{1,1}(Z)=h^{1,1}(\mathbb P_\Sigma)$, we can pick $Z$ such that $\mathrm{dim}(\Theta_i)\geq 2$ for all rays $\vec v_i$. It is straightforward to check
this condition for the example in \S\ref{sec:ExplicitConstruction}.

\section{Flux quantization}
\label{appendix:Quantization}

In this Appendix, we compute $c_2(\hat Y)$ and determine the impact of
flux quantization on the vertical flux parameters in \S\ref{sec:ExplicitConstruction}. The 
computation of $c_2(\hat Y)$ involves an explicit choice of resolution. 
We consider the singular Weierstrass model in Eq.\ (\ref{eq:e7}), and
resolve it by performing blowups. We denote
\begin{equation}
    Y_1\stackrel{(x,y,s|e_{1})}{\longrightarrow}Y\,,
\end{equation}
as the blowup from $Y$ to $Y_1$ by the redefinition
\begin{equation}
    x\rightarrow xe_1\,,\quad y\rightarrow ye_1\,,\quad s\rightarrow se_1\,.
\end{equation}
The resulting locus $e_1=0$ is a divisor in the ambient
space, denoted by $E_1$. Using the same notation, we can then
write down the resolution as the following steps \cite{Esole:2017kyr,Bhardwaj_2019,Jefferson:2021bid}:
\begin{equation}
    \hat Y\stackrel{(e_4,e_5|e_7)}{\longrightarrow}Y_6\stackrel{(e_2,e_4|e_6)}{\longrightarrow}Y_5\stackrel{(e_2,e_3|e_5)}{\longrightarrow}Y_4\stackrel{(y,e_3|e_4)}{\longrightarrow}Y_3\stackrel{(x,e_2|e_3)}{\longrightarrow}Y_2\stackrel{(y,e_1|e_2)}{\longrightarrow}Y_1\stackrel{(x,y,s|e_1)}{\longrightarrow}Y\,.
\end{equation}
This resolution smooths out all singularities on $Y$ up to codimension
3. The exceptional divisors on $\hat Y$ are given by
\begin{align}
    D_1 &= (E_1-E_2)\cap \hat Y\,, \nonumber \\
    D_2 &= (-E_1+2E_2-E_3-E_5-E_6)\cap \hat Y\,, \nonumber \\
    D_3 &= (E_1-2E_2+E_3+2E_5+E_6-E_7)\cap \hat Y\,, \nonumber \\
    D_4 &= E_7\cap \hat Y\,, \nonumber \\
    D_5 &= (E_3-E_4-E_5)\cap \hat Y\,, \nonumber \\
    D_6 &= (-E_3+2E_4+E_5-E_6-E_7)\cap \hat Y\,, \nonumber \\
    D_7 &= (-E_1+2E_2-E_3-2E_5+E_7)\cap \hat Y\,.
\end{align}
Using the above information, we can then compute $c_2(\hat Y)$ using the
techniques in \cite{Esole:2017kyr,Jefferson:2021bid}. The computation
involves a pushforward formula from $\hat Y$ to $Y$ for the total Chern
class, and homology relations to relate all $D_i\cdot D_j$ to $D_i\cdot D_\alpha$. The result is
\begin{align}
[c_{2}(\hat Y)] & =  [c_{2}(B)]+11\pi^{*}K_{B}^{2}\nonumber \\
& +\left(-12D_{0}+14D_{1}+30D_{2}+48D_{3}+41D_{4}+28D_{5}+17D_{6}+27D_{7}\right)\cdot\pi^{*}K_{B}\nonumber \\
 & +\left(2D_{1}+6D_{2}+12D_{3}+12D_{4}+8D_{5}+6D_{6}+8D_{7}\right)\cdot\pi^{*}\Sigma\,.
\end{align}
It is known that the first row of the above is even \cite{Collinucci:2010gz}. Therefore, the potentially odd terms are
$D_i\cdot\pi^* K_B$ for $i=4,6,7$. To determine the parity of these terms, it is
more convenient to work with their pushforward
$\pi_*(D_i\cdot\pi^*K_B)=\Sigma\cdot K_B$.
For the model in \S\ref{sec:ExplicitConstruction}, we calculate
\begin{equation}
\Sigma\cdot K_{B}=K_{\Sigma}+N_{\Sigma}=-\sigma_{A}\cdot\left(9F_{\sigma}\cdot F_{E_{1}}+6F_{\sigma}\cdot F_{E_{2}}+11F_{\sigma}\cdot F_{F}+6F_{E_{1}}\cdot F_{F}\right)\,,
\end{equation}
which has odd coefficients. Notice that
\begin{equation}
\Sigma\cdot\left.(F_{E_1}+F_F)\right|_B=B\cdot\sigma_{A}\cdot\left(F_{E_{1}}+F_{F}\right)=\sigma_{A}\cdot\left(F_{\sigma}\cdot F_{E_{1}}+F_{\sigma}\cdot F_{F}+2F_{E_{1}}\cdot F_{F}\right)\,,
\end{equation}
has the same parity as $\Sigma\cdot K_B$, so the pullback
\begin{equation}
    (D_4+D_6+D_7)\cdot\pi^*\left.(F_{E_1}+F_F)\right|_B\,,
\end{equation}
has the same parity as $[c_2(\hat Y)]$. We see that flux quantization as
in Eq. (\ref{quantization}) can be satisfied by turning on half-integer $\phi_{iE_1}$ and $\phi_{iF}$ for $i=4,6,7$. From Eq. (\ref{eq:phi-n}),
this is the same as half-integer flux parameters $n_{E_1},n_F,\phi_{6E_1},\phi_{6F}$.

\bibliography{references}

\providecommand{\href}[2]{#2}\begingroup\raggedright\begin{thebibliography}{100}

\bibitem{TaylorWangVacua}
W.~Taylor and Y.-N. Wang, \emph{{The F-theory geometry with most flux vacua}},
  \href{https://doi.org/10.1007/JHEP12(2015)164}{\emph{JHEP} {\bfseries 12}
  (2015) 164} [\href{https://arxiv.org/abs/1511.03209}{{\ttfamily
  1511.03209}}].

\bibitem{Cvetic:2022fnv}
M.~Cvetic, J.~Halverson, G.~Shiu and W.~Taylor, \emph{{Snowmass White Paper:
  String Theory and Particle Physics}},
  \href{https://arxiv.org/abs/2204.01742}{{\ttfamily 2204.01742}}.

\bibitem{Marchesano:2022qbx}
F.~Marchesano, B.~Schellekens and T.~Weigand, \emph{{D-brane and F-theory Model
  Building}},  12, 2022, \href{https://arxiv.org/abs/2212.07443}{{\ttfamily
  2212.07443}}.

\bibitem{VafaF-theory}
C.~Vafa, \emph{{Evidence for F theory}},
  \href{https://doi.org/10.1016/0550-3213(96)00172-1}{\emph{Nucl. Phys.}
  {\bfseries B469} (1996) 403}
  [\href{https://arxiv.org/abs/hep-th/9602022}{{\ttfamily hep-th/9602022}}].

\bibitem{MorrisonVafaI}
D.~R. Morrison and C.~Vafa, \emph{{Compactifications of F theory on Calabi--Yau
  threefolds --- I}},
  \href{https://doi.org/10.1016/0550-3213(96)00242-8}{\emph{Nucl. Phys.}
  {\bfseries B473} (1996) 74}
  [\href{https://arxiv.org/abs/hep-th/9602114}{{\ttfamily hep-th/9602114}}].

\bibitem{MorrisonVafaII}
D.~R. Morrison and C.~Vafa, \emph{{Compactifications of F theory on Calabi--Yau
  threefolds --- II}},
  \href{https://doi.org/10.1016/0550-3213(96)00369-0}{\emph{Nucl. Phys.}
  {\bfseries B476} (1996) 437}
  [\href{https://arxiv.org/abs/hep-th/9603161}{{\ttfamily hep-th/9603161}}].

\bibitem{WeigandTASI}
T.~Weigand, \emph{{TASI Lectures on F-theory}},
  \href{https://arxiv.org/abs/[1806.01854]}{{\ttfamily [1806.01854]}}.

\bibitem{HeckmanReview}
J.~J. Heckman, \emph{{Particle Physics Implications of F-theory}},
  \href{https://doi.org/10.1146/annurev.nucl.012809.104532}{\emph{Ann. Rev.
  Nucl. Part. Sci.} {\bfseries 60} (2010) 237}
  [\href{https://arxiv.org/abs/1001.0577}{{\ttfamily 1001.0577}}].

\bibitem{Li:2021eyn}
S.~Y. Li and W.~Taylor, \emph{{Natural F-theory constructions of standard model
  structure from E7 flux breaking}},
  \href{https://doi.org/10.1103/PhysRevD.106.L061902}{\emph{Phys. Rev. D}
  {\bfseries 106} (2022) L061902}
  [\href{https://arxiv.org/abs/2112.03947}{{\ttfamily 2112.03947}}].

\bibitem{Li:2022aek}
S.~Y. Li and W.~Taylor, \emph{{Gauge symmetry breaking with fluxes and natural
  Standard Model structure from exceptional GUTs in F-theory}},
  \href{https://doi.org/10.1007/JHEP11(2022)089}{\emph{JHEP} {\bfseries 11}
  (2022) 089} [\href{https://arxiv.org/abs/2207.14319}{{\ttfamily
  2207.14319}}].

\bibitem{Bies:2021nje}
M.~Bies, M.~Cveti\v{c}, R.~Donagi, M.~Liu and M.~Ong, \emph{{Root bundles and
  towards exact matter spectra of F-theory MSSMs}},
  \href{https://doi.org/10.1007/JHEP09(2021)076}{\emph{JHEP} {\bfseries 09}
  (2021) 076} [\href{https://arxiv.org/abs/2102.10115}{{\ttfamily
  2102.10115}}].

\bibitem{Bies:2021xfh}
M.~Bies, M.~Cveti\v{c} and M.~Liu, \emph{{Statistics of limit root bundles
  relevant for exact matter spectra of F-theory MSSMs}},
  \href{https://doi.org/10.1103/PhysRevD.104.L061903}{\emph{Phys. Rev. D}
  {\bfseries 104} (2021) L061903}
  [\href{https://arxiv.org/abs/2104.08297}{{\ttfamily 2104.08297}}].

\bibitem{Bies:2022wvj}
M.~Bies, M.~Cveti\v{c}, R.~Donagi and M.~Ong, \emph{{Brill-Noether-general
  limit root bundles: absence of vector-like exotics in F-theory Standard
  Models}}, \href{https://doi.org/10.1007/JHEP11(2022)004}{\emph{JHEP}
  {\bfseries 11} (2022) 004}
  [\href{https://arxiv.org/abs/2205.00008}{{\ttfamily 2205.00008}}].

\bibitem{Bies:2023jqg}
M.~Bies, \emph{{Root bundles: Applications to F-theory Standard Models}},  3,
  2023, \href{https://arxiv.org/abs/2303.08144}{{\ttfamily 2303.08144}}.

\bibitem{Bies:2023sfm}
M.~Bies, M.~Cveti\v{c}, R.~Donagi and M.~Ong, \emph{{Improved statistics for
  F-theory standard models}},
  \href{https://arxiv.org/abs/2307.02535}{{\ttfamily 2307.02535}}.

\bibitem{Cvetic:2019sgs}
M.~Cvetic, L.~Lin, M.~Liu, H.~Y. Zhang and G.~Zoccarato, \emph{{Yukawa
  Hierarchies in Global F-theory Models}},
  \href{https://doi.org/10.1007/JHEP01(2020)037}{\emph{JHEP} {\bfseries 01}
  (2020) 037} [\href{https://arxiv.org/abs/1906.10119}{{\ttfamily
  1906.10119}}].

\bibitem{HalversonLongSungAlg}
J.~Halverson, C.~Long and B.~Sung, \emph{{Algorithmic universality in F-theory
  compactifications}},
  \href{https://doi.org/10.1103/PhysRevD.96.126006}{\emph{Phys. Rev.}
  {\bfseries D96} (2017) 126006}
  [\href{https://arxiv.org/abs/1706.02299}{{\ttfamily 1706.02299}}].

\bibitem{TaylorWangLandscape}
W.~Taylor and Y.-N. Wang, \emph{{Scanning the skeleton of the 4D F-theory
  landscape}}, \href{https://doi.org/10.1007/JHEP01(2018)111}{\emph{JHEP}
  {\bfseries 01} (2018) 111}
  [\href{https://arxiv.org/abs/1710.11235}{{\ttfamily 1710.11235}}].

\bibitem{twy}
W.~Taylor, Y.-N. Wang and Y.~Yu, \emph{{work in progress}}, .

\bibitem{Andriolo:2019gcb}
S.~Andriolo, S.~Y. Li and S.~H.~H. Tye, \emph{{String Landscape and Fermion
  Masses}}, \href{https://doi.org/10.1103/PhysRevD.101.066005}{\emph{Phys. Rev.
  D} {\bfseries 101} (2020) 066005}
  [\href{https://arxiv.org/abs/1902.06608}{{\ttfamily 1902.06608}}].

\bibitem{BraunWatariGenerations}
A.~P. Braun and T.~Watari, \emph{{Distribution of the Number of Generations in
  Flux Compactifications}},
  \href{https://doi.org/10.1103/PhysRevD.90.121901}{\emph{Phys. Rev.}
  {\bfseries D90} (2014) 121901}
  [\href{https://arxiv.org/abs/1408.6156}{{\ttfamily 1408.6156}}].

\bibitem{TaylorWangMC}
W.~Taylor and Y.-N. Wang, \emph{{A Monte Carlo exploration of threefold base
  geometries for 4d F-theory vacua}},
  \href{https://doi.org/10.1007/JHEP01(2016)137}{\emph{JHEP} {\bfseries 01}
  (2016) 137} [\href{https://arxiv.org/abs/1510.04978}{{\ttfamily
  1510.04978}}].

\bibitem{Donagi:2008ca}
R.~Donagi and M.~Wijnholt, \emph{{Model Building with F-Theory}},
  \href{https://doi.org/10.4310/ATMP.2011.v15.n5.a2}{\emph{Adv. Theor. Math.
  Phys.} {\bfseries 15} (2011) 1237}
  [\href{https://arxiv.org/abs/0802.2969}{{\ttfamily 0802.2969}}].

\bibitem{BeasleyHeckmanVafaI}
C.~Beasley, J.~J. Heckman and C.~Vafa, \emph{{GUTs and Exceptional Branes in
  F-theory - I}},
  \href{https://doi.org/10.1088/1126-6708/2009/01/058}{\emph{JHEP} {\bfseries
  01} (2009) 058} [\href{https://arxiv.org/abs/0802.3391}{{\ttfamily
  0802.3391}}].

\bibitem{BeasleyHeckmanVafaII}
C.~Beasley, J.~J. Heckman and C.~Vafa, \emph{{GUTs and Exceptional Branes in
  F-theory - II: Experimental Predictions}},
  \href{https://doi.org/10.1088/1126-6708/2009/01/059}{\emph{JHEP} {\bfseries
  01} (2009) 059} [\href{https://arxiv.org/abs/0806.0102}{{\ttfamily
  0806.0102}}].

\bibitem{DonagiWijnholtGUTs}
R.~Donagi and M.~Wijnholt, \emph{{Breaking GUT Groups in F-Theory}},
  \href{https://doi.org/10.4310/ATMP.2011.v15.n6.a1}{\emph{Adv. Theor. Math.
  Phys.} {\bfseries 15} (2011) 1523}
  [\href{https://arxiv.org/abs/0808.2223}{{\ttfamily 0808.2223}}].

\bibitem{Blumenhagen:2009yv}
R.~Blumenhagen, T.~W. Grimm, B.~Jurke and T.~Weigand, \emph{{Global F-theory
  GUTs}}, \href{https://doi.org/10.1016/j.nuclphysb.2009.12.013}{\emph{Nucl.
  Phys. B} {\bfseries 829} (2010) 325}
  [\href{https://arxiv.org/abs/0908.1784}{{\ttfamily 0908.1784}}].

\bibitem{Marsano:2009wr}
J.~Marsano, N.~Saulina and S.~Schafer-Nameki, \emph{{Compact F-theory GUTs with
  U(1) (PQ)}}, \href{https://doi.org/10.1007/JHEP04(2010)095}{\emph{JHEP}
  {\bfseries 04} (2010) 095} [\href{https://arxiv.org/abs/0912.0272}{{\ttfamily
  0912.0272}}].

\bibitem{Grimm:2009yu}
T.~W. Grimm, S.~Krause and T.~Weigand, \emph{{F-Theory GUT Vacua on Compact
  Calabi-Yau Fourfolds}},
  \href{https://doi.org/10.1007/JHEP07(2010)037}{\emph{JHEP} {\bfseries 07}
  (2010) 037} [\href{https://arxiv.org/abs/0912.3524}{{\ttfamily 0912.3524}}].

\bibitem{KRAUSE20121}
S.~Krause, C.~Mayrhofer and T.~Weigand, \emph{G4-flux, chiral matter and
  singularity resolution in f-theory compactifications},
  \href{https://doi.org/https://doi.org/10.1016/j.nuclphysb.2011.12.013}{\emph{Nuclear
  Physics B} {\bfseries 858} (2012) 1 }
  [\href{https://arxiv.org/abs/1109.3454}{{\ttfamily 1109.3454}}].

\bibitem{Braun:2013nqa}
V.~Braun, T.~W. Grimm and J.~Keitel, \emph{{Geometric Engineering in Toric
  F-Theory and GUTs with U(1) Gauge Factors}},
  \href{https://doi.org/10.1007/JHEP12(2013)069}{\emph{JHEP} {\bfseries 12}
  (2013) 069} [\href{https://arxiv.org/abs/1306.0577}{{\ttfamily 1306.0577}}].

\bibitem{Chen:2010ts}
C.-M. Chen, J.~Knapp, M.~Kreuzer and C.~Mayrhofer, \emph{{Global SO(10)
  F-theory GUTs}}, \href{https://doi.org/10.1007/JHEP10(2010)057}{\emph{JHEP}
  {\bfseries 10} (2010) 057} [\href{https://arxiv.org/abs/1005.5735}{{\ttfamily
  1005.5735}}].

\bibitem{Chen:2010tg}
C.-M. Chen and Y.-C. Chung, \emph{{On F-theory $E_{6}$ GUTs}},
  \href{https://doi.org/10.1007/JHEP03(2011)129}{\emph{JHEP} {\bfseries 03}
  (2011) 129} [\href{https://arxiv.org/abs/1010.5536}{{\ttfamily 1010.5536}}].

\bibitem{Callaghan:2012rv}
J.~C. Callaghan and S.~F. King, \emph{{E6 Models from F-theory}},
  \href{https://doi.org/10.1007/JHEP04(2013)034}{\emph{JHEP} {\bfseries 04}
  (2013) 034} [\href{https://arxiv.org/abs/1210.6913}{{\ttfamily 1210.6913}}].

\bibitem{Callaghan:2013kaa}
J.~C. Callaghan, S.~F. King and G.~K. Leontaris, \emph{{Gauge coupling
  unification in $E_6$ F-theory GUTs with matter and bulk exotics from flux
  breaking}}, \href{https://doi.org/10.1007/JHEP12(2013)037}{\emph{JHEP}
  {\bfseries 12} (2013) 037} [\href{https://arxiv.org/abs/1307.4593}{{\ttfamily
  1307.4593}}].

\bibitem{Mayrhofer:2013ara}
C.~Mayrhofer, E.~Palti and T.~Weigand, \emph{{Hypercharge Flux in IIB and
  F-theory: Anomalies and Gauge Coupling Unification}},
  \href{https://doi.org/10.1007/JHEP09(2013)082}{\emph{JHEP} {\bfseries 09}
  (2013) 082} [\href{https://arxiv.org/abs/1303.3589}{{\ttfamily 1303.3589}}].

\bibitem{Braun:2014pva}
A.~P. Braun, A.~Collinucci and R.~Valandro, \emph{{Hypercharge flux in F-theory
  and the stable Sen limit}},
  \href{https://doi.org/10.1007/JHEP07(2014)121}{\emph{JHEP} {\bfseries 07}
  (2014) 121} [\href{https://arxiv.org/abs/1402.4096}{{\ttfamily 1402.4096}}].

\bibitem{Buican:2006sn}
M.~Buican, D.~Malyshev, D.~R. Morrison, H.~Verlinde and M.~Wijnholt,
  \emph{{D-branes at Singularities, Compactification, and Hypercharge}},
  \href{https://doi.org/10.1088/1126-6708/2007/01/107}{\emph{JHEP} {\bfseries
  01} (2007) 107} [\href{https://arxiv.org/abs/hep-th/0610007}{{\ttfamily
  hep-th/0610007}}].

\bibitem{Braun:2014xka}
A.~P. Braun and T.~Watari, \emph{{The Vertical, the Horizontal and the Rest:
  anatomy of the middle cohomology of Calabi-Yau fourfolds and F-theory
  applications}}, \href{https://doi.org/10.1007/JHEP01(2015)047}{\emph{JHEP}
  {\bfseries 01} (2015) 047} [\href{https://arxiv.org/abs/1408.6167}{{\ttfamily
  1408.6167}}].

\bibitem{CveticEtAlQuadrillion}
M.~Cveti\v{c}, J.~Halverson, L.~Lin, M.~Liu and J.~Tian, \emph{{Quadrillion
  $F$-Theory Compactifications with the Exact Chiral Spectrum of the Standard
  Model}}, \href{https://doi.org/10.1103/PhysRevLett.123.101601}{\emph{Phys.
  Rev. Lett.} {\bfseries 123} (2019) 101601}
  [\href{https://arxiv.org/abs/1903.00009}{{\ttfamily 1903.00009}}].

\bibitem{KleversEtAlToric}
D.~Klevers, D.~K. Mayorga~Pena, P.-K. Oehlmann, H.~Piragua and J.~Reuter,
  \emph{{F-Theory on all Toric Hypersurface Fibrations and its Higgs
  Branches}}, \href{https://doi.org/10.1007/JHEP01(2015)142}{\emph{JHEP}
  {\bfseries 01} (2015) 142} [\href{https://arxiv.org/abs/1408.4808}{{\ttfamily
  1408.4808}}].

\bibitem{Raghuram:2019efb}
N.~Raghuram, W.~Taylor and A.~P. Turner, \emph{{General F-theory models with
  tuned $(\operatorname{SU}(3) \times \operatorname{SU}(2) \times
  \operatorname{U}(1)) / \mathbb{Z}_6$ symmetry}},
  \href{https://doi.org/10.1007/JHEP04(2020)008}{\emph{JHEP} {\bfseries 04}
  (2020) 008} [\href{https://arxiv.org/abs/1912.10991}{{\ttfamily
  1912.10991}}].

\bibitem{Jefferson:2022yya}
P.~Jefferson, W.~Taylor and A.~P. Turner, \emph{{Chiral spectrum of the
  universal tuned (SU(3) $\times$ SU(2) $\times$
  U(1))/\ensuremath{\mathbb{Z}}$_{6}$ 4D F-theory model}},
  \href{https://doi.org/10.1007/JHEP02(2023)254}{\emph{JHEP} {\bfseries 02}
  (2023) 254} [\href{https://arxiv.org/abs/2210.09473}{{\ttfamily
  2210.09473}}].

\bibitem{MorrisonTaylorClusters}
D.~R. Morrison and W.~Taylor, \emph{{Classifying bases for 6D F-theory
  models}}, \href{https://doi.org/10.2478/s11534-012-0065-4}{\emph{Central Eur.
  J. Phys.} {\bfseries 10} (2012) 1072}
  [\href{https://arxiv.org/abs/1201.1943}{{\ttfamily 1201.1943}}].

\bibitem{MorrisonTaylor4DClusters}
D.~R. Morrison and W.~Taylor, \emph{{Non-Higgsable clusters for 4D F-theory
  models}}, \href{https://doi.org/10.1007/JHEP05(2015)080}{\emph{JHEP}
  {\bfseries 05} (2015) 080} [\href{https://arxiv.org/abs/1412.6112}{{\ttfamily
  1412.6112}}].

\bibitem{MorrisonTaylorToric}
D.~R. Morrison and W.~Taylor, \emph{{Toric bases for 6D F-theory models}},
  \href{https://doi.org/10.1002/prop.201200086}{\emph{Fortsch. Phys.}
  {\bfseries 60} (2012) 1187}
  [\href{https://arxiv.org/abs/1204.0283}{{\ttfamily 1204.0283}}].

\bibitem{Candelas:2000nc}
P.~Candelas, D.-E. Diaconescu, B.~Florea, D.~R. Morrison and G.~Rajesh,
  \emph{{Codimension three bundle singularities in F theory}},
  \href{https://doi.org/10.1088/1126-6708/2002/06/014}{\emph{JHEP} {\bfseries
  06} (2002) 014} [\href{https://arxiv.org/abs/hep-th/0009228}{{\ttfamily
  hep-th/0009228}}].

\bibitem{Lawrie:2012gg}
C.~Lawrie and S.~Schäfer-Nameki, \emph{{The Tate Form on Steroids: Resolution
  and Higher Codimension Fibers}},
  \href{https://doi.org/10.1007/JHEP04(2013)061}{\emph{JHEP} {\bfseries 04}
  (2013) 061} [\href{https://arxiv.org/abs/1212.2949}{{\ttfamily 1212.2949}}].

\bibitem{Achmed-Zade:2018idx}
I.~Achmed-Zade, I.~n. Garc\'\i{}a-Etxebarria and C.~Mayrhofer, \emph{{A note on
  non-flat points in the SU(5) U(1)$_{PQ}$ F-theory model}},
  \href{https://doi.org/10.1007/JHEP05(2019)013}{\emph{JHEP} {\bfseries 05}
  (2019) 013} [\href{https://arxiv.org/abs/1806.05612}{{\ttfamily
  1806.05612}}].

\bibitem{Jefferson:2021bid}
P.~Jefferson, W.~Taylor and A.~P. Turner, \emph{{Chiral matter multiplicities
  and resolution-independent structure in 4D F-theory models}},
  \href{https://arxiv.org/abs/2108.07810}{{\ttfamily 2108.07810}}.

\bibitem{46}
P.~Jefferson, S.~Y. Li and W.~Taylor, \emph{{work in progress}}, .

\bibitem{DonagiWijnholtModelBuilding}
R.~Donagi and M.~Wijnholt, \emph{{Model Building with F-Theory}},
  \href{https://doi.org/10.4310/ATMP.2011.v15.n5.a2}{\emph{Adv. Theor. Math.
  Phys.} {\bfseries 15} (2011) 1237}
  [\href{https://arxiv.org/abs/0802.2969}{{\ttfamily 0802.2969}}].

\bibitem{Grimm:2010ks}
T.~W. Grimm, \emph{{The N=1 effective action of F-theory compactifications}},
  \href{https://doi.org/10.1016/j.nuclphysb.2010.11.018}{\emph{Nucl. Phys. B}
  {\bfseries 845} (2011) 48} [\href{https://arxiv.org/abs/1008.4133}{{\ttfamily
  1008.4133}}].

\bibitem{Grimm:2011tb}
T.~W. Grimm, M.~Kerstan, E.~Palti and T.~Weigand, \emph{{Massive Abelian Gauge
  Symmetries and Fluxes in F-theory}},
  \href{https://doi.org/10.1007/JHEP12(2011)004}{\emph{JHEP} {\bfseries 12}
  (2011) 004} [\href{https://arxiv.org/abs/1107.3842}{{\ttfamily 1107.3842}}].

\bibitem{Kodaira}
K.~Kodaira, \emph{On compact analytic surfaces: Ii}, {\emph{Annals of
  Mathematics} {\bfseries 77} (1963) 563}.

\bibitem{Neron}
A.~N\'eron, \emph{Mod\`eles minimaux des vari\'et\'es ab\'eliennes sur les
  corps locaux et globaux}, {\emph{Publications Math\'ematiques de l'IH\'ES}
  {\bfseries 21} (1964) 5}.

\bibitem{shioda1972}
T.~Shioda, \emph{On elliptic modular surfaces},
  \href{https://doi.org/10.2969/jmsj/02410020}{\emph{J. Math. Soc. Japan}
  {\bfseries 24} (1972) 20}.

\bibitem{Wazir}
R.~Wazir, \emph{Arithmetic on elliptic threefolds},
  \href{https://doi.org/10.1112/S0010437X03000381}{\emph{Compositio
  Mathematica} {\bfseries 140} (2004) 567–580}
  [\href{https://arxiv.org/abs/math/0112259}{{\ttfamily math/0112259}}].

\bibitem{Witten:1996md}
E.~Witten, \emph{{On flux quantization in M theory and the effective action}},
  \href{https://doi.org/10.1016/S0393-0440(96)00042-3}{\emph{J. Geom. Phys.}
  {\bfseries 22} (1997) 1}
  [\href{https://arxiv.org/abs/hep-th/9609122}{{\ttfamily hep-th/9609122}}].

\bibitem{Becker:1996gj}
K.~Becker and M.~Becker, \emph{{M theory on eight manifolds}},
  \href{https://doi.org/10.1016/0550-3213(96)00367-7}{\emph{Nucl. Phys. B}
  {\bfseries 477} (1996) 155}
  [\href{https://arxiv.org/abs/hep-th/9605053}{{\ttfamily hep-th/9605053}}].

\bibitem{Gukov:1999ya}
S.~Gukov, C.~Vafa and E.~Witten, \emph{{CFT's from Calabi-Yau four folds}},
  \href{https://doi.org/10.1016/S0550-3213(00)00373-4}{\emph{Nucl. Phys. B}
  {\bfseries 584} (2000) 69}
  [\href{https://arxiv.org/abs/hep-th/9906070}{{\ttfamily hep-th/9906070}}].

\bibitem{Sethi:1996es}
S.~Sethi, C.~Vafa and E.~Witten, \emph{{Constraints on low dimensional string
  compactifications}},
  \href{https://doi.org/10.1016/S0550-3213(96)00483-X}{\emph{Nucl. Phys. B}
  {\bfseries 480} (1996) 213}
  [\href{https://arxiv.org/abs/hep-th/9606122}{{\ttfamily hep-th/9606122}}].

\bibitem{Grimm:2011fx}
T.~W. Grimm and H.~Hayashi, \emph{{F-theory fluxes, Chirality and Chern-Simons
  theories}}, \href{https://doi.org/10.1007/JHEP03(2012)027}{\emph{JHEP}
  {\bfseries 03} (2012) 027} [\href{https://arxiv.org/abs/1111.1232}{{\ttfamily
  1111.1232}}].

\bibitem{Grimm:2011sk}
T.~W. Grimm and R.~Savelli, \emph{{Gravitational Instantons and Fluxes from
  M/F-theory on Calabi-Yau fourfolds}},
  \href{https://doi.org/10.1103/PhysRevD.85.026003}{\emph{Phys. Rev. D}
  {\bfseries 85} (2012) 026003}
  [\href{https://arxiv.org/abs/1109.3191}{{\ttfamily 1109.3191}}].

\bibitem{Dasgupta:1999ss}
K.~Dasgupta, G.~Rajesh and S.~Sethi, \emph{{M theory, orientifolds and G -
  flux}}, \href{https://doi.org/10.1088/1126-6708/1999/08/023}{\emph{JHEP}
  {\bfseries 08} (1999) 023}
  [\href{https://arxiv.org/abs/hep-th/9908088}{{\ttfamily hep-th/9908088}}].

\bibitem{Li:2022vfj}
S.~Y. Li and W.~Taylor, \emph{{Large U(1) charges from flux breaking in 4D
  F-theory models}}, \href{https://doi.org/10.1007/JHEP02(2023)186}{\emph{JHEP}
  {\bfseries 02} (2023) 186}
  [\href{https://arxiv.org/abs/2211.11768}{{\ttfamily 2211.11768}}].

\bibitem{Braun_2012}
A.~Braun, A.~Collinucci and R.~Valandro, \emph{G-flux in f-theory and algebraic
  cycles}, \href{https://doi.org/10.1016/j.nuclphysb.2011.10.034}{\emph{Nuclear
  Physics B} {\bfseries 856} (2012) 129–179}
  [\href{https://arxiv.org/abs/1107.5337}{{\ttfamily 1107.5337}}].

\bibitem{Marsano_2011}
J.~Marsano and S.~Schäfer-Nameki, \emph{Yukawas, g-flux, and spectral covers
  from resolved calabi-yau’s},
  \href{https://doi.org/10.1007/jhep11(2011)098}{\emph{Journal of High Energy
  Physics} {\bfseries 2011} (2011) }
  [\href{https://arxiv.org/abs/1108.1794}{{\ttfamily 1108.1794}}].

\bibitem{Bies:2017fam}
M.~Bies, C.~Mayrhofer and T.~Weigand, \emph{{Gauge Backgrounds and Zero-Mode
  Counting in F-Theory}},
  \href{https://doi.org/10.1007/JHEP11(2017)081}{\emph{JHEP} {\bfseries 11}
  (2017) 081} [\href{https://arxiv.org/abs/1706.04616}{{\ttfamily
  1706.04616}}].

\bibitem{Bies:2014sra}
M.~Bies, C.~Mayrhofer, C.~Pehle and T.~Weigand, \emph{{Chow groups, Deligne
  cohomology and massless matter in F-theory}},
  \href{https://arxiv.org/abs/1402.5144}{{\ttfamily 1402.5144}}.

\bibitem{Blumenhagen:2008zz}
R.~Blumenhagen, V.~Braun, T.~W. Grimm and T.~Weigand, \emph{{GUTs in Type IIB
  Orientifold Compactifications}},
  \href{https://doi.org/10.1016/j.nuclphysb.2009.02.011}{\emph{Nucl. Phys. B}
  {\bfseries 815} (2009) 1} [\href{https://arxiv.org/abs/0811.2936}{{\ttfamily
  0811.2936}}].

\bibitem{BershadskyEtAlSingularities}
M.~Bershadsky, K.~A. Intriligator, S.~Kachru, D.~R. Morrison, V.~Sadov and
  C.~Vafa, \emph{{Geometric singularities and enhanced gauge symmetries}},
  \href{https://doi.org/10.1016/S0550-3213(96)90131-5}{\emph{Nucl. Phys.}
  {\bfseries B481} (1996) 215}
  [\href{https://arxiv.org/abs/hep-th/9605200}{{\ttfamily hep-th/9605200}}].

\bibitem{HeckmanMorrisonVafa}
J.~J. Heckman, D.~R. Morrison and C.~Vafa, \emph{{On the Classification of 6D
  SCFTs and Generalized ADE Orbifolds}},
  \href{https://doi.org/10.1007/JHEP05(2014)028}{\emph{JHEP} {\bfseries 05}
  (2014) 028} [\href{https://arxiv.org/abs/1312.5746}{{\ttfamily 1312.5746}}].

\bibitem{Apruzzi:2018oge}
F.~Apruzzi, J.~J. Heckman, D.~R. Morrison and L.~Tizzano, \emph{{4D Gauge
  Theories with Conformal Matter}},
  \href{https://doi.org/10.1007/JHEP09(2018)088}{\emph{JHEP} {\bfseries 09}
  (2018) 088} [\href{https://arxiv.org/abs/1803.00582}{{\ttfamily
  1803.00582}}].

\bibitem{Witten:1996bn}
E.~Witten, \emph{{Nonperturbative superpotentials in string theory}},
  \href{https://doi.org/10.1016/0550-3213(96)00283-0}{\emph{Nucl. Phys. B}
  {\bfseries 474} (1996) 343}
  [\href{https://arxiv.org/abs/hep-th/9604030}{{\ttfamily hep-th/9604030}}].

\bibitem{Blumenhagen:2006xt}
R.~Blumenhagen, M.~Cvetic and T.~Weigand, \emph{{Spacetime instanton
  corrections in 4D string vacua: The Seesaw mechanism for D-Brane models}},
  \href{https://doi.org/10.1016/j.nuclphysb.2007.02.016}{\emph{Nucl. Phys. B}
  {\bfseries 771} (2007) 113}
  [\href{https://arxiv.org/abs/hep-th/0609191}{{\ttfamily hep-th/0609191}}].

\bibitem{Ibanez:2006da}
L.~E. Ibanez and A.~M. Uranga, \emph{{Neutrino Majorana Masses from String
  Theory Instanton Effects}},
  \href{https://doi.org/10.1088/1126-6708/2007/03/052}{\emph{JHEP} {\bfseries
  03} (2007) 052} [\href{https://arxiv.org/abs/hep-th/0609213}{{\ttfamily
  hep-th/0609213}}].

\bibitem{Blumenhagen:2007zk}
R.~Blumenhagen, M.~Cvetic, D.~Lust, R.~Richter and T.~Weigand,
  \emph{{Non-perturbative Yukawa Couplings from String Instantons}},
  \href{https://doi.org/10.1103/PhysRevLett.100.061602}{\emph{Phys. Rev. Lett.}
  {\bfseries 100} (2008) 061602}
  [\href{https://arxiv.org/abs/0707.1871}{{\ttfamily 0707.1871}}].

\bibitem{Banks:2010zn}
T.~Banks and N.~Seiberg, \emph{{Symmetries and Strings in Field Theory and
  Gravity}}, \href{https://doi.org/10.1103/PhysRevD.83.084019}{\emph{Phys. Rev.
  D} {\bfseries 83} (2011) 084019}
  [\href{https://arxiv.org/abs/1011.5120}{{\ttfamily 1011.5120}}].

\bibitem{Harlow:2018tng}
D.~Harlow and H.~Ooguri, \emph{{Symmetries in quantum field theory and quantum
  gravity}}, \href{https://doi.org/10.1007/s00220-021-04040-y}{\emph{Commun.
  Math. Phys.} {\bfseries 383} (2021) 1669}
  [\href{https://arxiv.org/abs/1810.05338}{{\ttfamily 1810.05338}}].

\bibitem{Tatar:2006dc}
R.~Tatar and T.~Watari, \emph{{Proton decay, Yukawa couplings and underlying
  gauge symmetry in string theory}},
  \href{https://doi.org/10.1016/j.nuclphysb.2006.04.025}{\emph{Nucl. Phys. B}
  {\bfseries 747} (2006) 212}
  [\href{https://arxiv.org/abs/hep-th/0602238}{{\ttfamily hep-th/0602238}}].

\bibitem{Workman:2022ynf}
{\scshape Particle Data Group} collaboration, \emph{{Review of Particle
  Physics}}, \href{https://doi.org/10.1093/ptep/ptac097}{\emph{PTEP} {\bfseries
  2022} (2022) 083C01}.

\bibitem{Ibanez:2012zg}
L.~E. Ibanez, F.~Marchesano, D.~Regalado and I.~Valenzuela, \emph{{The
  Intermediate Scale MSSM, the Higgs Mass and F-theory Unification}},
  \href{https://doi.org/10.1007/JHEP07(2012)195}{\emph{JHEP} {\bfseries 07}
  (2012) 195} [\href{https://arxiv.org/abs/1206.2655}{{\ttfamily 1206.2655}}].

\bibitem{Hebecker:2014uaa}
A.~Hebecker and J.~Unwin, \emph{{Precision Unification and Proton Decay in
  F-Theory GUTs with High Scale Supersymmetry}},
  \href{https://doi.org/10.1007/JHEP09(2014)125}{\emph{JHEP} {\bfseries 09}
  (2014) 125} [\href{https://arxiv.org/abs/1405.2930}{{\ttfamily 1405.2930}}].

\bibitem{Palti:2016kew}
E.~Palti, \emph{{Vector-Like Exotics in F-Theory and 750 GeV Diphotons}},
  \href{https://doi.org/10.1016/j.nuclphysb.2016.04.026}{\emph{Nucl. Phys. B}
  {\bfseries 907} (2016) 597}
  [\href{https://arxiv.org/abs/1601.00285}{{\ttfamily 1601.00285}}].

\bibitem{Heckman:2008qa}
J.~J. Heckman and C.~Vafa, \emph{{Flavor Hierarchy From F-theory}},
  \href{https://doi.org/10.1016/j.nuclphysb.2010.05.009}{\emph{Nucl. Phys. B}
  {\bfseries 837} (2010) 137}
  [\href{https://arxiv.org/abs/0811.2417}{{\ttfamily 0811.2417}}].

\bibitem{Cecotti:2009zf}
S.~Cecotti, M.~C.~N. Cheng, J.~J. Heckman and C.~Vafa, \emph{{Yukawa Couplings
  in F-theory and Non-Commutative Geometry}},
  \href{https://arxiv.org/abs/0910.0477}{{\ttfamily 0910.0477}}.

\bibitem{Font:2012wq}
A.~Font, L.~E. Ibanez, F.~Marchesano and D.~Regalado, \emph{{Non-perturbative
  effects and Yukawa hierarchies in F-theory SU(5) Unification}},
  \href{https://doi.org/10.1007/JHEP03(2013)140}{\emph{JHEP} {\bfseries 03}
  (2013) 140} [\href{https://arxiv.org/abs/1211.6529}{{\ttfamily 1211.6529}}].

\bibitem{Font:2013ida}
A.~Font, F.~Marchesano, D.~Regalado and G.~Zoccarato, \emph{{Up-type quark
  masses in SU(5) F-theory models}},
  \href{https://doi.org/10.1007/JHEP11(2013)125}{\emph{JHEP} {\bfseries 11}
  (2013) 125} [\href{https://arxiv.org/abs/1307.8089}{{\ttfamily 1307.8089}}].

\bibitem{Marchesano:2015dfa}
F.~Marchesano, D.~Regalado and G.~Zoccarato, \emph{{Yukawa hierarchies at the
  point of E$_{8}$ in F-theory}},
  \href{https://doi.org/10.1007/JHEP04(2015)179}{\emph{JHEP} {\bfseries 04}
  (2015) 179} [\href{https://arxiv.org/abs/1503.02683}{{\ttfamily
  1503.02683}}].

\bibitem{Carta:2015eoh}
F.~Carta, F.~Marchesano and G.~Zoccarato, \emph{{Fitting fermion masses and
  mixings in F-theory GUTs}},
  \href{https://doi.org/10.1007/JHEP03(2016)126}{\emph{JHEP} {\bfseries 03}
  (2016) 126} [\href{https://arxiv.org/abs/1512.04846}{{\ttfamily
  1512.04846}}].

\bibitem{Taylor:1996ik}
W.~Taylor, \emph{{D-brane field theory on compact spaces}},
  \href{https://doi.org/10.1016/S0370-2693(97)00033-6}{\emph{Phys. Lett. B}
  {\bfseries 394} (1997) 283}
  [\href{https://arxiv.org/abs/hep-th/9611042}{{\ttfamily hep-th/9611042}}].

\bibitem{Martin:1997ns}
S.~P. Martin, \emph{{A Supersymmetry primer}},
  \href{https://doi.org/10.1142/9789812839657_0001}{\emph{Adv. Ser. Direct.
  High Energy Phys.} {\bfseries 18} (1998) 1}
  [\href{https://arxiv.org/abs/hep-ph/9709356}{{\ttfamily hep-ph/9709356}}].

\bibitem{Blumenhagen:2008aw}
R.~Blumenhagen, \emph{{Gauge Coupling Unification in F-Theory Grand Unified
  Theories}}, \href{https://doi.org/10.1103/PhysRevLett.102.071601}{\emph{Phys.
  Rev. Lett.} {\bfseries 102} (2009) 071601}
  [\href{https://arxiv.org/abs/0812.0248}{{\ttfamily 0812.0248}}].

\bibitem{Demirtas:2018akl}
M.~Demirtas, C.~Long, L.~McAllister and M.~Stillman, \emph{{The Kreuzer-Skarke
  Axiverse}}, \href{https://doi.org/10.1007/JHEP04(2020)138}{\emph{JHEP}
  {\bfseries 04} (2020) 138}
  [\href{https://arxiv.org/abs/1808.01282}{{\ttfamily 1808.01282}}].

\bibitem{Danilov_1987}
V.~I. Danilov and A.~G. Khovanskii, \emph{Newton polyhedra and an algorithm for
  computing hodge–deligne numbers},
  \href{https://doi.org/10.1070/IM1987v029n02ABEH000970}{\emph{Mathematics of
  the USSR-Izvestiya} {\bfseries 29} (1987) 279}.

\bibitem{Jefferson:2022ssj}
P.~Jefferson and M.~Kim, \emph{{On the intermediate Jacobian of M5-branes}},
  \href{https://arxiv.org/abs/2211.00210}{{\ttfamily 2211.00210}}.

\bibitem{batyrev1993dual}
V.~V. Batyrev, \emph{Dual polyhedra and mirror symmetry for calabi-yau
  hypersurfaces in toric varieties},
  \href{https://arxiv.org/abs/alg-geom/9310003}{{\ttfamily alg-geom/9310003}}.

\bibitem{Esole:2017kyr}
M.~Esole, P.~Jefferson and M.~J. Kang, \emph{{Euler Characteristics of Crepant
  Resolutions of Weierstrass Models}},
  \href{https://doi.org/10.1007/s00220-019-03517-1}{\emph{Commun. Math. Phys.}
  {\bfseries 371} (2019) 99}
  [\href{https://arxiv.org/abs/1703.00905}{{\ttfamily 1703.00905}}].

\bibitem{Bhardwaj_2019}
L.~Bhardwaj and P.~Jefferson, \emph{{Classifying $5d$ SCFTs via $6d$ SCFTs:
  Rank one}}, \href{https://doi.org/10.1007/JHEP07(2019)178}{\emph{JHEP}
  {\bfseries 07} (2019) 178}
  [\href{https://arxiv.org/abs/1809.01650}{{\ttfamily 1809.01650}}].

\bibitem{Collinucci:2010gz}
A.~Collinucci and R.~Savelli, \emph{{On Flux Quantization in F-Theory}},
  \href{https://doi.org/10.1007/JHEP02(2012)015}{\emph{JHEP} {\bfseries 02}
  (2012) 015} [\href{https://arxiv.org/abs/1011.6388}{{\ttfamily 1011.6388}}].

\end{thebibliography}\endgroup
\bibliographystyle{JHEP}

\end{document}